\documentclass[12pt,a4paper,fleqn]{book}
\usepackage{graphicx}
\DeclareGraphicsExtensions{.jpg}
\usepackage[tbtags]{amsmath}
\usepackage{amssymb}
\usepackage{mathrsfs}
\usepackage{appendix}
\usepackage[pdftex,hyperindex,bookmarks=true,colorlinks=true,citecolor=blue,linkcolor=blue]{hyperref}

\usepackage{fancyhdr}
\makeatletter
\def\cleardoublepage{\clearpage\if@twoside \ifodd\c@page\else%
    \hbox{}%
    \thispagestyle{empty}
    \newpage%
    \if@twocolumn\hbox{}\newpage\fi\fi\fi}
\makeatother

\newcommand{\fillblank}{\textsf}

\begin{document}


\setlength{\parskip}{1.0ex plus 0.5ex minus 0.5ex}
\frontmatter 

\begin{titlepage}
\begin{center}

 {\bf \uppercase{\huge F\Large ield \huge T\Large heory \huge A\Large spects of \huge C\Large osmology \\and \\
 \vspace{0.4 cm} \huge B\Large lack \huge H\Large oles}}
\vfill

\normalsize
{\Large Thesis submitted for the degree of}\\[2.2ex]
\textbf{\Large Doctor of Philosophy (Science)}\\[2ex]
{\Large of}\\[2ex]
\textbf{\Large Jadavpur University, Kolkata}

\vfill

{\Large August 2009}

\vfill

\textbf{{\Large Kulkarni Shailesh Gajanan}}\\[2ex]
{\large \mbox{Satyendra Nath Bose National Centre for Basic Sciences}}\\
{\large JD Block, Sector 3, Salt Lake, Kolkata 700098, India}

\end{center}
\end{titlepage}


\begin{figure}[t]
  \begin{center}
    \includegraphics[width=\textwidth]{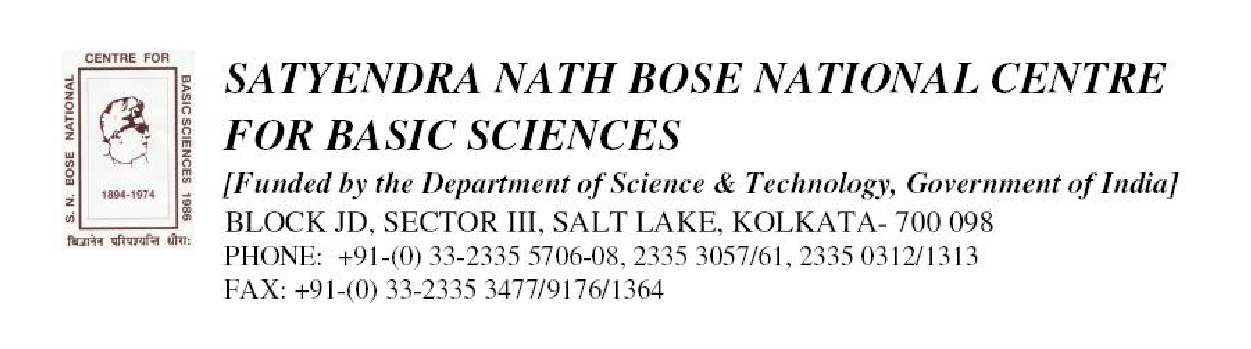}
  \end{center}

\end{figure}
\section*{\begin{center}Certificate from the supervisor\end{center}}

\thispagestyle{empty}


This is to certify that the thesis entitled \fillblank{``Field theory aspects of cosmology and black holes''} submitted by Sri. \fillblank{Kulkarni Shailesh Gajanan}, who got his name registered on \fillblank{October 8, 2007} for the award of \fillblank{Ph.D.~(Science)} degree of \fillblank{Jadavpur University}, is absolutely based upon his own work under the supervision of \fillblank{Professor Rabin Banerjee} at \fillblank{S.N.~Bose National Centre for Basic Sciences, Kolkata, India}, and that neither this thesis nor any part of it has been submitted for any degree/diploma or any other academic award anywhere before.

\vspace{3.0cm}

\hfill \begin{tabular}{@{}l@{}}
\fillblank{Rabin Banerjee}\\
Professor\\
S.N.~Bose National Centre for Basic Sciences\\
JD Block, Sector 3, Salt Lake\\
Kolkata 700098, India
\end{tabular}


\thispagestyle{empty}
\vspace*{19.5 cm}




\hspace*{8.6 cm} $\mathcal{TO}$\\ \hspace*{10 cm}$\mathcal{MY \ MOTHER}$

\chapter*{Acknowledgements}
\thispagestyle{empty}

This thesis is the culminative outcome of five years work, which has been
 made possible by the blessings and support of many individuals. I take this
 opportunity to express my sincere gratitude to all of them.

First and foremost, I would like to thank Prof. Rabin Banerjee, my thesis supervisor.
 His uncanny ability to select a particular problem, a keen and strategic analysis of it
 and deep involvement among the students makes him something special.
 Thank you Sir for giving me all that could last my entire life. 

 I thank Prof. Subir Ghosh, ISI Kolkata. It was a quite nice experience to work with
 him. I would also like to express my gratitude to Dr. Amitabha Lahiri, Dr. Samir Kumar Paul, Dr. Biswajit Chakraborty and Dr. Pradip Mukherjee for helping me in my academics at
 Satyendra Nath National Center for Basic Sciences (SNBNCBS). 

  Prof. Jayanta Kumar Bhattacharjee and Prof. Binayak Dutta Roy have 
 always been available to clarify very elementary but at same time subtle concepts
 in physics. I sincerely thank them for giving me their valuable attention. 

  I would like to thank Prof. Sushanta Dattagupta, ex. Director of SNBNCBS for allowing me
  to carry out my research here. I thank Prof. Arup K. Raychaudhuri, present
 Director, for providing an excellent academic atmosphere.

 I am also thankful to the Library staff of SNBNCBS for their support.

 I am indebted to Prof. Anil Gangal, University of Pune, for giving me 
 crucial suggestions and guidance, whenever I required. I would also like to thank
 Prof. Dileep Kanhere, University of Pune and my college teachers
 Prof. S. Deshpande, Prof. S. Bapat and Prof. G. Pujari. 
   
  My life during this voyage has been made cherishable and interesting by some
 colourful and memorable personalities to whom I would like to express my heartiest feelings.
 I am indebted to my seniors - Abhishek Chowdhury, Mani, Ankush, Kuldeep, Chandrashekhar and Sunandan.
 Things, which I gained from them, turn out to be most important in my thesis work.
 I would like to thank all of them for their brotherly support. Saurav Samanta, a senior,
 a friend and an ``$8\times 8$ squares'' addicted personality deserves special mention for
 those strategic discussions, on both the academic as well as non-academic fronts. 
  Then come my brilliant and helpful group mates,
 ranging from and enthusiastic (+ violant, under {\it specific} circumstances) Bibhas to poignant Debraj and 
 `innocent' Sujoy in between, making the whole set complete and orthonormal. I am very much grateful
 to you comrades for helping me in difficult situations. My other group, founded by Niraj, consists
 of many bright minds like, Hemant, Mrinal, Debu, Venkat, Navin, Irfan and the deadly duo of Sudip-Rajiv.
 This group is almost incomplete if I don't mention the names - Abhishek Pandey, Kapil and Chinmay, who understand better the {\it sacredness} and true {\it colours} of {\it GH three O seven}. I am indebted to you guys for giving me those
 precious moments. I am grateful to Himadri for his {\it sometimes} meaningless but still delightful gossips.
 I would also like to thank all the cricket players of SNBNCBS for making my
 stay enjoyable. How can I forget those {\it saturdays} and {\it sundays} and my {\it backfoot punch}? 
   
 I would like to mention my {\it Sahyadri group} friends who have made
 my each trip to Pune, memorable - Bhalchandra, Manish, Chaitanya, Sandeep, Hemant, Dhananjay,
 Pramod and Sukumaran.  All of them  have something special, which is only visible 
 from the {\it cliffs} of {\it western ghats}. I am also thankful to Hrishikesh and Smita,
 my school friends, for their friendly attitude on many occasions.          

I owe my deepest gratitude to each and every member of my of family. Especially,
 I would like to thank my elder sister Sachu and my Jijaji for understanding, and
 helping me, on many occasions. I am indebted to my elder brother Satyendra and my sister-in-law 
 Archana for their help during my stay in Pune. I thank Kedar (my brother and friend),
 with whom I shared some {\it sweet} moments of my graduate life and uncountable
 visits to that {\it stop}.
      
 Finally, I would like to say sorry to my {\it mother}, my late {\it father} and my friend
 {\it nandu}; I could not find a proper word for you! 


\chapter*{List of publications}
\thispagestyle{empty}


\begin{enumerate} \raggedright
\item \textbf{Deformed symmetry in Snyder space and relativistic particle dynamics.}\\
    Rabin Banerjee, Shailesh Kulkarni and Saurav Samanta\\
    {\em JHEP} {\bf 0605}, 077 (2006) [hep-th/0602151].

\item \textbf{Remarks on the generalized Chaplygin gas.}\\
    Rabin Banerjee, Subir Ghosh and Shailesh Kulkarni\\
    {\em Phys. Rev.} {\bf D75}, 025008 (2007) [gr-qc/0611131].
     
\item \textbf{Hawking radiation and covariant anomalies.}\\
    Rabin Banerjee and Shailesh Kulkarni \\
    {\em Phys. Rev.} {\bf D77}, 024018 (2008) arXiv:0707.2449 [hep-th].
    
\item \textbf{Hawking Radiation, Effective Actions and Covariant Boundary Conditions.}\\
  Rabin Banerjee and Shailesh Kulkarni \\
  {\em Phys. Lett.} {\bf B659}, 827 (2008) arXiv:0709.3916 [hep-th].

\item \textbf{Hawking radiation in GHS and non-extremal D1-D5 blackhole via covariant
  anomalies.}\\
  Sunandan Gangopadhyay and Shailesh Kulkarni \\
  {\em Phys. Rev.} {\bf D77}, 024038 (2008) arXiv:0710.0974 [hep-th].

\item \textbf{Hawking Fluxes, Back reaction and Covariant Anomalies}\\
  Shailesh Kulkarni\\
  {\em Class. Quant. Grav.} {\bf 25}, 225023 (2008) arXiv:0802.2456 [hep-th].

\newpage

\item \textbf{Hawking Radiation, Covariant Boundary Conditions and Vacuum States}\\
  Rabin Banerjee and Shailesh Kulkarni \\
  {\em Phys. Rev.} {\bf D79}, 084035 (2009) arXiv:0810.5683 [hep-th].
\end{enumerate}
This thesis is based on the papers numbered by [2,3,4,5,6,7] whose reprints are attached at
 the end of the thesis.



\chapter*{}
\pagenumbering{roman}
\thispagestyle{empty}

\begin{center}
\uppercase{Field theory aspects of cosmology\\ and black holes}
\end{center}



\tableofcontents



\mainmatter
\chapter{\label{chap:introduction} Introduction}
\section{\label{Overview}Overview}
Recently, there has been significant development in the field of 
cosmology. Detail study of cosmology enable us to understand the
 origin and ultimate fate of our universe. 
Research in cosmology has become astonishingly lively in the early 1980s.
 An idea of the cosmic inflation \cite{starobinski,guth1,linde1} (for review see \cite{guth2,linde2})
 offered a way to understand some outstanding
 cosmological puzzles and provided a mechanism for the origin of large-scale structure.
 Results predicted by
 the theory of inflation could be tested by observations of anisotropies in
 the cosmic microwave background. In the late 1990, observations of
 Type Ia supernovae led to the discovery that the expansion of the universe
 is accelerating \cite{obs}. Further, CMB data \cite{barlett} and cluster mass distribution
 \cite{carlberg} seem to favor models in which the energy density contributed by the negative
 pressure component should be roughly twice as much as the energy density of the matter, thus
 leading to the flat universe i.e  the fraction of total density  $\Omega_{tot}=1$ with
 $\Omega_{M} \sim 0.4$ and $\Omega_{\Lambda} \sim 0.6$. Therefore the universe
 should be presently dominated by a smooth component with effective negative pressure;
 this is infact the most general requirement in order to explain the observed 
 accelerated expansion of the universe.

Recently, it has been suggested that the change of behavior
of the missing energy density, responsible for the
 accelerated expansion of the Universe, might be governed 
 by a change in the equation of state of the background
fluid instead of the form of the potential, thereby avoiding
the fine-tuning problems present in the above approaches. This is
achieved via introduction of an exotic background fluid,
 the Chaplygin gas \cite{kam}, described by the equation of state
\begin{equation}
P=-\frac{B}{\rho^{\alpha}}~. \label{intro1}
\end{equation}
Where $P$ and $\rho$ are the pressure and density of the fluid in comoving frame, respectively
 with $\rho >0$ and $B$ is some constant. The exponent $\alpha$ is bounded by $ 0 < \alpha \leq 1 $. 
When $\alpha=1$, the model (\ref{intro1}) is referred as standard Chaplygin gas.

S. Chaplygin introduced this equation of state \cite{chap} as
a convenient soluble model to study the lifting force
on a plane wing in aerodynamics. Later on, the
same equations were rediscovered in \cite{tsien}, again in
an aerodynamical context.


    Apart from its applications in cosmology, the Chaplygin gas model 
 has drawn considerable interest  because of its many remarkable
and intriguingly unique features. In the action formulation, the standard
 $\alpha=1$ Chaplygin gas has very deep connection with string theory \cite{jac1,bazeia,jac2}.
Indeed, it was shown in \cite{jac2} that, in the light cone parameterization
 there is a one to one correspondence between the reparameterization invariant 
 Nambu-Goto action for $d$-brane in ($d+1,1$) dimensions and the 
  Galileo invariant (nonrelativistic) action for ($d,1$) dimensional Chaplygin gas.
 While, in the Cartesian parameterization, the reparameterization invariant 
 Nambu-Goto action for $d$-brane in ($d+1,1$) dimensions is dual to the
 Poincare invariant (relativistic) action for Born-Infeld model in ($d,1$) dimensional
 spacetime \cite{bord}. In the nonrelativistic limit, ($d,1$) dimensional Born-Infeld
 action reduces to  action for ($d,1$) dimensional Chaplygin gas. In addition
 to this, in the nonrelativistic decent from the Born-Infeld theory to the Chaplygin gas,
 there exists a mapping of one system to another, and between solutions of one system
 to another, because both, with the certain choice of parameterizations,
 reduces to the Nambu-Goto action \cite{jac2}. Also,
the Chaplygin gas is the only fluid which, up to now,
admits a supersymmetric generalization \cite{jac3,hopp}.   
 All this analysis was done for $\alpha=1$ Chaplygin gas model. 
 Thus, it is clear that the field theoretical techniques play a vital role
 in the understanding of some issues in the modern cosmology.     
 
    Now we focus our attention on the applications of field theory to black hole 
 physics. Black holes are among the most fascinating predictions
 of Einstein's theory of gravitation. One of the basic results of general relativity
 is that matter affects the spacetime geometry. The gravitational field produced
 by the matter could become so strong as to substantially modify the causal structure
 of spacetime and eventually produce a region from which nothing can escape.
 A boundary of such region of spacetime is called an event horizon. Black hole
 is characterized by certain parameters like mass, charge, angular momentum.
 Classical black hole mechanics can be summarized by the following three basic
 laws,
\begin{enumerate}
 \item  Zeroth law : The surface gravity $\kappa$ of a black hole is constant
on the horizon.
\item First law :  The variations in the black hole parameters, i.e  mass $M$, area $A$, angular momentum
$L$, and charge $Q$, obey 
\begin{equation}
\delta M = \frac{\kappa}{8\pi} \delta A + \Omega \delta L - V \delta Q \label{intro3}
\end{equation}
where $\Omega$ and $V$ are the angular velocity and the electrostatic potential, respectively 
\item Second law : The area of a black hole horizon $A$ is nondecreasing in time \cite{hawkingarea},
\begin{equation}
 \delta A \geq 0~. \label{intro4}
\end{equation}
\end{enumerate}
These laws have a close resemblance to the corresponding
 laws of thermodynamics. The zeroth law of thermodynamics
says that the temperature $T$ is constant throughout a system in thermal
equilibrium. The first law states that in small variations between equilibrium
configurations of a system, the changes in the energy $M$ and entropy $S$
of the system obey equation (\ref{intro3}), if the surface gravity $\kappa$
 is replaced by a term proportional to $T$ (other terms on the right hand side are interpreted as work terms).
 The second law of thermodynamics states that, for a closed system, entropy always increases
in any (irreversible or reversible) process, i.e $\delta S \geq 0$. Jacob Bekenstein in 1973 \cite{bekenstein1} 
suggested that a physical identification does hold between
 the laws of thermodynamics and the laws of black hole mechanics. The
 surface gravity $\kappa$ and horizon area $A$ are identified with multiple of temperature $T$ and
 entropy $S$, respectively. However, he was unable to elevate this analogy to a more formal level.
 For example, if the correspondence among the laws of black hole mechanics and thermodynamics were true
 then black holes must radiate. However black holes do not radiate. 

  This discrepancy was successfully removed by Stephen Hawking.
 In 1975 he published his famous paper ``Particle Creation by Black Holes'' \cite{hawking}
 where he explicitly showed that black holes do radiate if one takes
 into the account the quantum mechanical nature of matter fields in the spacetime.
 The key idea behind quantum particle production in curved spacetime
is that the definition of a particle is observer dependent. It depends on the
choice of reference frame. Since the theory is generally
covariant, any time coordinate, possibly defined only locally within
a patch, is a legitimate choice with which to define positive and negative
 frequency modes. Hawking considered a massless quantum scalar field moving in the background of a collapsing
 star. If the quantum field was initially in the vacuum state (no particle state)
 defined in the asymptotic past, then at late times it will appear as
 if particles are present in that state. Hawking showed \cite{hawking}, by explicit computation of
 the Bogoliubov coefficients (see also  \cite{birrell, ford} for detail calculation of Bogoliubov coefficients) 
 between the two sets of vacuum states defined at asymptotic past and future respectively, that the spectrum
 of the emitted particles is identical to that of black body with the temperature
 \begin{equation}
 T_{H} = \frac{\hbar \kappa}{2\pi}, \label{intro6}
\end{equation}
 known as the Hawking temperature \cite{hawking}. This astonishing result is obtained using the approximation that the
 matter field behaves quantum mechanically but the gravitational field (metric) satisfy
 the classical Einstein equation. This semiclassical approximation holds good for energies below the
Planck scale \cite{hawking}. Although it is a semiclassical result, Hawking's 
computation is considered an important clue in the search for a theory of quantum
gravity. Any theory of quantum gravity that is proposed must predict black
hole evaporation.

 Apart from Hawking's original calculation, this effect has been studied by different
 methods. S. Hawking and G. Gibbons, in 1977 \cite{gibbons} developed an approach
 based on the Euclidean quantum gravity. In this approach they computed
 an action for gravitational field, including the boundary term, on the
 complexified spacetime. The purely imaginary values of this
 action gives a contribution of the metrics to the partition function for a grand
 canonical ensemble at Hawking temperature $T_{H}$. Using this, they were able to 
 show that the entropy associated with these metrics is always equal to $\frac{A}{4}$, where $A$
 is an area of the event horizon. In the same year Christensen and Fulling \cite{fulling}, by 
 exploiting the structure of trace anomaly, were able to obtain the expectation value 
 for each component of the stress tensor $\langle T_{\mu\nu} \rangle$, which eventually
 lead to the Hawking flux. This approach is exact in $(1+1)$ dimensions, however in $3+1$ dimensions, the requirements
 of spherical symmetry, time independence and covariant conservation are not sufficient
 to fix completely the flux of Hawking radiation in terms of the trace anomaly \cite{birrell,fulling}.
 There is an additional arbitrariness in the expectation values of the angular components of 
 the stress tensor. Another intuitive way to understand the Hawking effect was proposed independently
 by T. Padmanabhan, K. Srinivasan  \cite{paddy} and  F. Wilczek, M. Parikh \cite{parikh}.
 This approach is based on the quantum tunneling.
  The essential idea is that a particle-antiparticle pair forms
 close to the event horizon which is similar to pair formation in an external electric field.
 The ingoing mode is trapped inside the horizon while the outgoing mode can quantum mechanically
tunnel through the event horizon. It is observed at infinity as a Hawking flux. Within the tunneling
 mechanism the expressions for temperature and entropy for a black hole in presence
 of gravitational back reaction were also computed by  R. Banerjee and B. Majhi \cite{bibhas}.
 
 Recently, S. Robinson and F. Wilczek \cite{robwilczek} gave a new approach to compute the Hawking flux
 from a black hole. This approach is based on gravitational or diffeomorphism anomaly. Basic and  
 essential fact used in their analysis is that the theory of matter fields
 (scalar or fermionic) in the $3+1$ dimensional static black hole background 
 can effectively be represented, in the vicinity of event horizon, by an
 infinite collection of free massless $1+1$ dimensional fields, each propagating
 in the background of an effective metric given by the $r-t$ sector of full $3+1$ dimensional
 metric \footnote{Such a dimensional reduction of matter fields 
 has been already used in the analysis of \cite{carlip1,solodukhin} to compute the entropy of
 $2+1$ dimensional $BTZ$ black hole.}. By definition the horizon is null surface and hence
 the region inside it is causally disconnected from the exterior. Thus, in the region near
 to the horizon the modes which are going into the black hole do not affect the 
 physics outside the horizon. In other words, the theory near the event horizon acquires 
   a definite chirality. Any two dimensional chiral theory in general curved background
 possesses gravitational anomaly \cite{witten}. This anomaly is manifested in the nonconservation
 of the stress tensor. The theory far away from the event horizon
 is $3+1$ dimensional and anomaly free and the stress tensor in this region satisfies
 the usual conservation law. Consequently, the total energy-momentum tensor, which is a sum of two contribution
 from the two different regions, is also anomalous. However, it becomes anomaly free once we 
 take into account the contribution from classically irrelevant ingoing modes. This imposes 
 restrictions on the structure of the energy-momentum tensor and is ultimately responsible
 for the Hawking radiation \cite{robwilczek}. The expression for energy-momentum flux
 obtained by this anomaly cancellation approach is in exact agreement with the 
 flux from the perfectly black body kept at Hawking temperature \cite{robwilczek}.
 Soon this analysis was extended to compute Hawking fluxes 
 from the Reissner-Nordstrom (charged) black hole \cite{isowilczek} and Kerr (rotating)
 black hole \cite{isowilczekPRD}.
 
  It is worth to note that there are certain similarities between the trace \cite{fulling}
 and the gravitational \cite{robwilczek} anomaly method. Both the approaches uses two inputs: the 
 usual conservation law,  and the trace \cite{fulling} or gravitational 
 \cite{robwilczek} anomaly. Further, since the structure of trace as well as gravitational 
 anomaly, apart from the overall multiplicative factor, is identical for different field species
 (e.g scalar,fermionic etc. ) the methodology of two approaches would not alter for different field
 species. However, the analysis of \cite{fulling} is
 restricted to $1+1$ dimensional conformal fields. In this sense the anomaly cancellation 
 method \cite{robwilczek} is more appealing compared to the trace anomaly approach \cite{fulling}
 \footnote{Comparison among these two approaches and their connection with the $W-infinity$ algebra
 has been discussed in detail by L. Bonora and collaborators \cite{bonora1,bonora2}.}.
\section{\label{structure} Outline of the thesis}
This thesis, based on the work \cite{shailesh0,shailesh1,shailesh2,shailesh3,shailesh4,shailesh5}, is focussed towards the applications of field theory, classical as well as quantum,
 in the context of cosmology and black holes.
 On the cosmology side, we study some theoretical aspects of generalized
 Chaplygin gas, a strong candidate for explaining the origin of 
 accelerated expansion of the Universe. In the remaining part of the thesis, we discuss thoroughly,
 the relationship between the quantum gauge and gravitational anomalies and Hawking effect.
 We propose two different approaches, based on the covariant anomalies \cite{shailesh1} and
 chiral effective actions \cite{shailesh2}, to compute the fluxes of Hawking radiation. Further,
  we provide a way to understand the covariant boundary condition used in the analysis of \cite{isowilczek,shailesh1,shailesh2}. A connection of this boundary condition
  with the various vacuum states defined in the black hole spacetime is also elucidated. 
 
  Chapter wise summary of the thesis is given below.

In chapter-\ref{chap:chaplygin}, we focus our
attention on generalized Chaplygin gas model, which is considered as an alternative model for
explaining the accelerated expansion of the Universe. In the nonrelativistic regime, we give
 the most general action for the generalized Chaplygin gas. This construction has been
done in two versions. In one case the action involves the density and the velocity potential.
Elimination of density is possible leading to the second version involving the velocity potential
only. The form for density independent action is similar to the Born-Infeld type action in the
nonrelativistic limit.
In the case of relativistic generalized Chaplygin gas a Born-Infeld action involving only velocity
 potential is proposed which has the correct nonrelativistic limit. We  also provide
a form for the action of a relativistic generalized Chaplygin gas involving both density and velocity potential
which also has a proper nonrelativistic limit. Our whole analysis of the generalized Chaplygin
 gas is consistent in the  $\alpha = 1$ limit which corresponds to the standard Chaplygin gas model.

 In chapter-\ref{chap:coanomaly} we provide a derivation of Hawking radiation using 
 covariant gauge and gravitational anomalies. Our derivation is essentially linked with the 
 approach given in \cite{robwilczek,isowilczek}, but with important distinctions.
 A crucial ingredient in the analysis of \cite{robwilczek} is that quantum field theory 
 in the region near the event horizon becomes two dimensional and chiral. A two dimensional
 chiral theory is anomalous with respect to gauge and general coordinate transformation. Such
  theories admit two types of anomalous currents and energy-momentum tensors - the consistent and
 the covariant. The covariant divergence of these currents and energy-momentum tensors
 yields  either the consistent or the covariant form of the gauge and gravitational anomaly, respectively
 \cite{fujikawabook,bardeenzumino}. The consistent current and anomaly satisfy the Wess-Zumino
 condition but do not transform covariantly under the gauge transformation. Expressions for
 covariant current and anomaly, on the other hand, transform covariantly under the gauge transformation
 but do not satisfy the Wess-Zumino condition. The covariant and consistent structures are connected
 by a local counterterm \cite{fujikawabook,bertlmannbook}. In fact this difference between the 
 covariant and consistent currents is the germ of the anomaly. For usual (anomaly free) theory
 the covariant and consistent expressions are identical. Similar conclusions also hold for the gravitational case.
 In \cite{robwilczek,isowilczek} the fluxes of Hawking radiation were obtained by cancellation of
 consistent gauge and gravitational anomalies. However, the analysis of \cite{robwilczek,isowilczek}
 raises several issues, both technically and conceptually. The Hawking flux is obtained from
 the consistent expression for gauge and gravitational anomaly but the boundary condition, necessary to fix the 
 form of current and energy-momentum tensor, involves the covariant form. Note that
 the Hawking flux is measured at infinity where there is no anomaly, so that covariant and consistent
 structures are identical. Hence, one can also obtain the flux from the covariant expressions. 
 In our derivation  we completely reformulate the analysis of \cite{robwilczek,isowilczek} 
 totally in terms of covariant expressions leading to a simple and conceptually clean way to understand
 the Hawking effect. 

 We begin this chapter by giving a brief discussion on some aspects of gauge and
 gravitational anomalies highlighting the peculiarities of two dimensional spacetime.
 Then we compute  Hawking charge and energy-momentum 
 flux by using the covariant gauge and gravitational anomalies. Since the boundary condition
 involves the vanishing of covariant current and energy-momentum tensor at event horizon, all
 calculations involve only covariant expressions. We discuss essential differences among the consistent and
 covariant anomaly based methods, emphasizing the utility of our approach. Also, we show that
 the analysis of \cite{robwilczek,isowilczek} is resilient and the results are unaffected
 by taking  more general expressions for the consistent gauge and gravitational anomalies, which occur
 due to peculiarities of two dimensional spacetime. We then implement our covariant
 anomaly approach to compute the Hawking radiation from non-trivial 
 black hole geometries arising in the string theory. Finally, we provide an appendix discussing the 
 dimensional reduction of real and complex scalar fields.

 In chapter-\ref{chap:chiralaction} we present a new formalism, based on the chiral
 effective action, to compute the Hawking fluxes from generic spherically symmetric
 static black hole. The expressions for current and energy-momentum tensor are obtain from
 the chiral effective action, suitably modified by a local counterterm. The covariant divergence
 of this current and energy-momentum tensor satisfy covariant gauge and gravitational anomaly, respectively.
 The role of chirality in imposing constraints on the structure of current and energy-momentum
 tensor is elucidated. The arbitrary constants appearing in the current and energy-momentum tensor are fixed by imposing the covariant boundary condition. Since the covariant gauge and gravitational anomaly vanish
 in the asymptotic infinity limit, we can obtain the Hawking charge and energy-momentum flux
 by appropriately taking the asymptotic limit of the covariant anomalous current and energy-momentum tensor.
 Since this approach uses only  the near horizon structure of effective action and covariant boundary condition,
 splitting of spacetime into two different regions and consequently the use of discontinuous step functions,
 as required in the earlier approaches \cite{robwilczek,isowilczek,shailesh1}, are not mandatory in the 
 computation of Hawking flux. As an application of this chiral effective action approach, we compute
 the correction to the Hawking flux due to the effect of one loop back reaction.
  
 One of the most important and crucial step in the anomaly \cite{robwilczek,isowilczek,shailesh1} or
  in the chiral effective action \cite{shailesh2} approach of computing the Hawking flux 
 was the implementation of covariant boundary condition; namely, the vanishing of the covariant
 current and energy-momentum tensor at event horizon. Apart from the fact that it is covariant under
 the gauge or general coordinate transformation, there was no other justification or physical
 interpretation in favor of this boundary condition. In chapter-\ref{chap:boundarycondition} we 
 address this issue by giving a detailed explanation for the covariant boundary condition.  
 
 We begin this excersise by giving a derivation of Hawking charge and energy-momentum flux from generic 
 spherically symmetric static black hole. Here we adopt the technique developed in \cite{rabinessay}.
  This method \cite{rabinessay}, like the chiral effective action approach,  uses only
 the near horizon structures for the covariant currents and energy-momentum tensors and the covariant boundary
 condition. 

   Next, we use the structures of covariant current/energy-momentum tensor
 derived earlier from the chiral effective action, appropriately modified by the
 local counterterm. In order to make the chiral nature of the theory more transparent
 we transform the various components of current and energy-momentum tensor to null coordinates.
 Only one component of the covariant chiral current and energy-momentum tensor
 is independent, while other components get fix by chirality and trace anomaly. 
 The independent components of current and energy-momentum tensor involve
 arbitrary constants, which are then fixed by imposing the condition that, a freely falling observer
 must see a finite amount of flux across the event horizon. This is the regularity condition and
 it implies that the current and energy-momentum tensor in Kruskal coordinates must be regular at future event horizon. For the chiral theory however, this is the same condition on outgoing modes in either
 the Unruh vacuum \cite{unruh} or Hartle-Hawking vacuum \cite{hartle}.   
  The structures for currents and energy-momentum tensors obtained from
 this analysis are seen to be in 
exact agreement with that found by solving the anomaly equations subjected to
 the covariant boundary condition.
 The fact that the Unruh and Hartle-Hawking vacua gives identical
 results is a consequence of chirality. This provides a clear
justification for the covariant boundary condition used in determining the Hawking flux form
the gauge/gravitational anomalies. Further, we compare our findings with the results derived from
 the conventional analysis of the various vacua states \cite{fullingvacuum}. 
 For $1+1$ dimensional chiral theory it is possible to give a connection between
 the trace and gravitational anomaly. This is explained in the appendix-\ref{appendix3A}.

 Finally, in chapter-\ref{chap:conclusions} we present our conclusion and outlook.

\chapter{\label{chap:chaplygin}Generalized Chaplygin gas}

 The recent observation of accelerated expansion of the universe,
concluded \cite{obs} from the study of luminosity of type Ia
distant supernova, has put Cosmology at the center stage. Our
inability to explain the origin of this expansion has led to the
naming of this  phenomenon as "Dark Energy" effect. The coinage 
obviously matches the
other fuzzy area in Cosmology, {\it{i.e.}} the existence of
"Dark Matter".There exist several plausible models at hand that attempt to explain the
astronomical data \cite{obs}. The traditional one - vacuum energy
or non-zero cosmological constant - fits well with the
observational data. Unfortunately it is plagued with serious
conceptual difficulties: smallness of the value of the
cosmological constant in comparison with Planck mass scale and the
coincidence problem, (that questions the reason for the near
equality between energy densities of Dark Energy and dust-like
matter in the present epoch), to name a few. The latter is
circumvented by introducing scalar field (or Quintessence) models
\cite{quint} inducing a dynamical vacuum energy, but only at the
expense of fine tuning the scalar potential parameters. 

 An alternative dynamical model \cite{kam} for Dark Energy, featuring
Chaplygin Gas \cite{chap} or its generalization \cite{gcg} - the
Generalized Chaplygin Gas (GCG) - has created some interest in
recent times. Conventional analysis of the model \cite{ch1}
 allows a smooth interpolation between a dust dominated era (at early times) to the
Cosmological constant dominated era (at present times). A further
generalization \cite{bilic} to inhomogeneous GCG model allows one to
address the issue of Dark Matter  as well. The GCG model has
passed several experimental tests of various nature, such as high
precision Cosmic Microwave Background Radiation data \cite{exp1},
supernova data \cite{exp2} and gravitational lensing \cite{exp3}.
Naive analysis \cite{teg} seemed to suggest  a disturbing
phenomenon in the GCG model: possible existence of unphysical
oscillations or even an exponential blow-up in the matter power
spectrum at present. However, this problem has been solved in
\cite{sol} by taking into account the interaction between Dark
Matter, Dark Energy and phantom-type Dark Energy \cite{Xin}. (For a detailed
exposition of these issues, see \cite{orf}.).

The above mentioned ideal fluid system  was introduced long ago by
Chaplygin \cite{chap} as an effective model in computing the
lifting force on a wing of an airplane. It obeys an  exotic
equation of state,
\begin{equation}
P=-\frac{B}{\rho} \label{chaply1}
\end{equation}
where $P(x)$ and $\rho (x)$ denote pressure and density
respectively and $B$  is a constant parameter.  However, the interest in Chaplygin gas model
 actually goes beyond Cosmology (see \cite{jac1,bazeia,jac2} for a review, oriented towards the High Energy Physics community).
 It has a deep connection with the $D$-branes in a higher dimensional Nambu-Goto formulation in light-cone
 parameterization \cite{bord}. It is also unique in admitting a supersymmetric generalization \cite{jac1} for a fluid. The dynamical role of Chaplygin gas
in cosmology has been shown in \cite{bg}.  The above discussion clearly underlines the relevance
 of GCG models in Cosmology and High Energy Physics.
 
 In the present work, we follow the same theme and focus our attention on the GCG models, where the generalization amounts to
 postulating the Chaplygin equation of state as,
\begin{equation}
P = -\frac{B}{\rho^{\alpha}};~~B>0;~~0<\alpha <1  \label{chaply2}
\end{equation}
where the standard Chaplygin pressure equation (\ref{chaply1}) is recovered for
$\alpha =1$. In the nonrelativistic regime, we have constructed the most 
general action for GCG consistent with (\ref{chaply2}). This construction has been done in two versions. In one case,
the action involves the density $\rho$ and the velocity potential $\theta$.
Elimination of $\rho$ is possible leading to the second version involving only 
$\theta$. This may be interpreted as a Born-Infeld type action in the nonrelativistic limit \cite{jac2}.

    Next, the construction of relativistic action for GCG is discussed.
Here a Born-Infeld action involving only $\theta$ is proposed which
has the correct nonrelativistic limit. Also for $\alpha =1$, it reproduces
the standard Born-Infeld action for the Chaplygin gas. We may mention
that our action is different from the one given in the literature \cite{gcg}.
Since the density plays an important role it becomes worthwhile to write
the relativistic action for GCG involving both $\rho$ and $\theta$ analogous
to the usual $\alpha =1$ case \cite{jac2}. However here we are faced with certain problems.
Our suggested form for GCG action has the correct nonrelativistic and
$\alpha =1$ limits. But for $\alpha \ne 1$ it is relativistic only for large $\rho$.
 This is found by an explicit check of the Poincare algebra.

The chapter is organized as follows: In Section-\ref{chaplygin1}  we provide
a brief review of the non-relativistic and  relativistic action
formulations of the normal Chaplygin gas ($\alpha =1$). This helps
us to fix the notation and charts the course of our subsequent
analysis. Section-\ref{chaplygin2} comprises an analogous study
for the GCG ($\alpha \ne 1$) and introduces new expressions for the
nonrelativistic GCG action. Section-\ref{chaplygin3} is devoted to the construction and
subsequent  analysis of the relativistic GCG. In Section-\ref{chaplygin4} we provide our conclusion and
propose avenues for future study.

\section{\label{chaplygin1}Normal ($\alpha =1$) Chaplygin gas : a brief review}
Before concentrating on the Chaplygin gas, let us discuss some basic notions of fluid dynamics, in the
Eulerian formulation \cite{jac2}.

 We start with the non-relativistic scenario. The equations of motion, governing an ideal fluid in  arbitrary space dimensions,  are given by
\begin{eqnarray}
\partial_t \rho(t, {\bf x}) + \nabla \cdot (\rho(t,{\bf x}){\bf v}(t,{\bf x})) &=& 0 \label{chaply3},\\
\partial_t {\bf v}(t,{\bf x}) +  {\bf v}(t,{\bf x})\cdot \nabla {\bf v}(t,{\bf x}) &=& {\bf f}(t,{\bf x}) \label{chaply4}
\end{eqnarray}
where  $\rho(t,{\bf x})$ and ${\bf v}(t,{\bf x})$ are the  matter
density  and velocity     fields respectively. The first identity
reflects the matter conservation and the second is the Euler equation of
motion. We consider the motion of fluid to be isentropic. Hence the force $f$ is
 derived from a $\rho $-dependent potential $V(\rho)$,
\begin{equation}
{\bf f} = -\frac{1}{\rho}{\bf \nabla}P ~.\label{chaply5}
\end{equation}
Here  $P$ is the pressure.
For isentropic motion $P$ is a function of $\rho$ only. Hence we can write (\ref{chaply5}) as
 \begin{equation}
{\bf f} =  -{\bf \nabla}V'(\rho).\label{chaply6}
\end{equation}
 prime denotes derivative with respect to $\rho$. Note that $V'(\rho)$ is the enthalpy, given by,
\begin{equation}
P(\rho) = \rho V'(\rho) - V(\rho). \label{chaply7}
\end{equation}
In the case of irrotational fluid further simplification occurs. For this case,
 the vorticity vanishes, which implies
\begin{equation}
{\bf v} = {\nabla} \theta.\label{chaply8}
\end{equation}
where $\theta(x,t)$ is some scalar field.
The (non-relativistic) Hamiltonian for irrotational motion is just
the sum of kinetic and potential energy,
\begin{equation}
H = \int dr (\frac{1}{2}\rho (\partial_i \theta)^2 + V(\rho)).
\label{chaply9}
\end{equation}
  Now, the first order form of the Lagrangian $L$, corresponding to (\ref{chaply9})
 is given by,
\begin{equation}
L=\int dr (\theta \dot \rho -\frac{1}{2}\rho (\partial_i \theta)^2
- V(\rho)). \label{chaply10}
\end{equation}
From the symplectic structure it is clear that $\rho$ and $\theta$ are conjugate variables,
 satisfying the the canonical Poisson bracket,
\begin{equation}
\{\theta({\bf x}), \rho({\bf y})\} = \delta({\bf x}- {\bf y}). \label{chaply11}
\end{equation}
The nature of the potential function $V(\rho )$ will specify the particular fluid model under study. For Chaplygin
gas the potential profile is given by,
\begin{equation}
V(\rho) = \frac{\lambda}{\rho } \label{chaply12}
\end{equation}
where $\lambda$ is the interaction strength. Using (\ref{chaply12}) the Lagrangian for the Chaplygin gas model is
given by
\begin{equation}
L=\int dr (\theta \dot \rho -\frac{1}{2}\rho (\partial_i \theta)^2
- \frac{\lambda}{\rho})~.  \label{chaply13}
\end{equation}
 Varying $L$ with respect to $\rho$, yields 
\begin{equation}
\dot \theta + \frac{1}{2}(\partial_i \theta)^2 = \frac{\lambda}{\rho ^2}.\label{chaply14}
\end{equation}
This the Bernoulli equation. 

It is possible to  eliminate  $\rho$ from the Lagrangian to obtain a non-relativistic Born-Infeld like structure
in  $\theta$,
\begin{equation}
L(\theta) = - 2\sqrt{\lambda}\int dr \sqrt{\left(\dot \theta +
\frac{1}{2}(\partial_i \theta)^2\right)} \label{chaply15}
\end{equation}
with the equation of motion,
\begin{equation}
\partial_t\left(\dot \theta + \frac{1}{2}(\partial_i \theta)^2\right)^{-\frac{1}{2}}
+ \partial_i \left[\partial_i \theta \left(\dot \theta + \frac{1}{2}(\partial_i \theta)^2\right)^{-\frac{1}{2}}\right]=0~. \label{chaply16}
\end{equation}

  Now we come to the relativistic generalization of Chaplygin gas\cite{jac2}. A Lagrangian has been suggested for the normal $(\alpha =1)$ Chaplygin gas in \cite{jac2},
\begin{equation}
L = \int dr (\theta \dot\rho - \sqrt{(\rho^2c^2 + a^2)}\sqrt{c^2 + (\partial_i \theta)^2}), \label{chaply17}
\end{equation}
where $a$ is a interaction strength.  Although (\ref{chaply17}) does not have a manifestly relativistic form, its
Poincare invariance has been demonstrated explicitly in \cite{bg} in a Hamiltonian framework (see section-\ref{chaplygin3} of the present chapter, also). On the other hand, one can eliminate $\rho $ once again from (\ref{chaply17}) to obtain
the Born-Infeld form,
\begin{equation}
L= -a\int dr \sqrt{c^2 - \partial_{\mu}\theta\partial^{\mu}\theta}
\label{chaply18}
\end{equation}
which is manifestly relativistic.

To get the correct  non-relativistic limit, one has to consider the map \cite{jac2},
\begin{equation}
\theta \rightarrow  \theta - tc^2. \label{chaply19}
\end{equation}
Under this transformation, the relativistic model (\ref{chaply17}) will reduce to the non-relativistic one in
(\ref{chaply13}) with the  identification $\lambda \equiv \frac{a^2}{2}$. This concludes our review \cite{jac2} of action
formulation of Chaplygin gas.

\section{\label{chaplygin2}Nonrelativistic generalized ($\alpha \ne 1$) Chaplygin gas}
In GCG the equation of state (\ref{chaply1}) is replaced by a more flexible one,
 given in (\ref{chaply2}). In order to incorporate
this generalization in the action formulation, our starting  point is to find a suitable potential $V(\rho)$
compatible with (\ref{chaply2}). To find the general solution for $V(\rho)$
 we start from an ansatz,  
\begin{equation}
 V(\rho) = \left(\frac{B}{\alpha +1}\right) \frac{1}{\rho^{\alpha}} + u(\rho) \label{chaply20}
\end{equation}
where $u(\rho)$ is such that $V(\rho)$ satisfies
\begin{equation}
-\frac{B}{\rho^{\alpha}} = \rho \frac{dV(\rho)}{d\rho} - V(\rho)~.\label{chaply21}
\end{equation}
This follows from (\ref{chaply2}) and the enthalpy relation (\ref{chaply7}). This implies $u(\rho)$ must satisfy
\begin{equation}
\rho \frac{du}{d\rho} - u = 0~.\label{chaply22}
\end{equation}
The solution for above equation is
\begin{equation}
u(\rho) = I \rho~. \label{chaply23} 
\end{equation}
where $I$ is an integration constant.

    Hence the most general form of the potential $V(\rho)$  is
\begin{equation}
V(\rho) = \left(\frac{B}{\alpha +1}\right)\frac{1}{\rho^{\alpha}} + I\rho.\label{chaply24}
\end{equation}

 For the irrotational fluid we can write the Hamiltonian (\ref{chaply9}), with  $V(\rho)$
 as given in (\ref{chaply24})
\begin{equation}
H = \int dr \left(\frac{1}{2}\rho (\partial_i \theta)^2
+\frac{B}{(\alpha +1)\rho^{\alpha}} + I\rho\right)~. \label{chaply25}
\end{equation}
Now the first order form of the Lagrangian follows from (\ref{chaply25}),
\begin{equation}
L^{\alpha} = \int dr \left[\theta \dot \rho -\frac{1}{2}\rho
(\partial_i \theta)^2 - \frac{B}{(\alpha +1)\rho^{\alpha}} - I\rho\right]
\label{chaply26}
\end{equation}
where the superscript $\alpha$ on $L$ reveals the fact that we are
dealing with GCG. 

  Variation of $\rho$ yields the Bernoulli equation,
\begin{equation}
\dot \theta + \frac{1}{2}(\partial_i \theta)^2 =
\frac{B\alpha}{(\alpha+1)\rho^{\alpha +1}} - I ~. \label{chaply27}
\end{equation}

To obtain the $\rho$ independent Lagrangian for GCG, one can  use the Bernoulli equation to reexpress   $\rho$ in terms of  $\theta$,
\begin{equation}
\rho = \left[\frac{\alpha B}{\alpha+1}(\dot \theta
+\frac{1}{2}(\partial_i\theta)^2 + I )^{-1}
\right]^{\frac{1}{\alpha+1}}. \label{chaply28}
\end{equation}
It is very convenient to rewrite the Lagrangian given in
(\ref{chaply26}) in the following form
\begin{equation}
L^{\alpha} = -\int dr \left( \dot \theta \rho +
\frac{\rho}{2}(\partial_i \theta)^2 + \frac{B}{(\alpha
+1)\rho^{\alpha}} + I\rho\right). \label{chaply29}
\end{equation}
In the above equation we have omitted  total derivative terms. Substituting
 $\rho$ from (\ref{chaply28}) in (\ref{chaply29}) we find, 
\begin{equation}
L^{\alpha}(\theta) = -
\left(\frac{\alpha}{\alpha +1}\right)^{\frac{\alpha}{\alpha+1}}
B^{\frac{1}{\alpha+1}}\int dr \sqrt{\left(\dot\theta +
\frac{1}{2}(\partial_i
\theta)^2 + I\right)^{\frac{2\alpha}{\alpha+1}}}~. \label{chaply30}
\end{equation}
This is the most general form of GCG Lagrangian and is a central result of
our analysis. It is the Born-Infeld version of nonrelativistic GCG.

    The  equation of motion for $\theta $ turns out to be,
\begin{equation}
\partial_t\left(\dot \theta + \frac{1}{2}(\partial_i \theta)^2\right)^{\frac{-1}{\alpha+1}}
+ \partial_i \left[\partial_i \theta \left(\dot \theta
+ \frac{1}{2}(\partial_i \theta)^2 + I\right)^{\frac{-1}{\alpha+1}}\right]=0~. \label{chaply31}
\end{equation}
A definite simplification occurs by setting $I=0$. Then the Lagrangians
 (\ref{chaply26}) and (\ref{chaply30}) reduce to  
\begin{equation}
L^{\alpha} = \int dr \left(\theta \dot \rho -\frac{1}{2}\rho
(\partial_i \theta)^2 - \frac{B}{(\alpha +1)\rho^{\alpha}}\right)
\label{chaply32}
\end{equation}
and
\begin{equation}
L^{\alpha}(\theta) = -
\left(\frac{\alpha}{\alpha +1}\right)^{\frac{\alpha}{\alpha+1}}
B^{\frac{1}{\alpha+1}}\int dr \sqrt{\left(\dot\theta +
\frac{1}{2}(\partial_i
\theta)^2 \right)^{\frac{2\alpha}{\alpha+1}}}. \label{chaply33}
\end{equation}
Putting $\alpha =1$ in (\ref{chaply32}) and (\ref{chaply33}) reproduces 
the expressions for the usual Chaplygin gas \cite{jac2}.

\section{\label{chaplygin3}Relativistic generalized Chaplygin gas}
Now we turn to the relativistic form of GCG. Any relativistic version of GCG 
must satisfy two conditions: it should have the correct nonrelativistic limit
 (\ref{chaply26}) or (\ref{chaply30}), secondly, for $\alpha =1$ it should reduce to
 (\ref{chaply13}) or (\ref{chaply15}).
 
To begin with we suggest a manifestly Poincare invariant model for GCG, given by\begin{equation}
L^{\alpha} = -(a')^{\frac{1}{\alpha+1}} \int dr \sqrt{(c^2 - \partial_{\mu}\theta\partial^{\mu}\theta)^{\frac{2\alpha}{\alpha+1}}}\label{chaply34} 
\end{equation}
In the nonrelativistic  limit it agrees with (\ref{chaply33}).To show this we exploit (\ref{chaply19}) and use the fact that  $\partial_{\mu}\theta\partial^{\mu}\theta = \frac{\dot \theta^2}{c^2} -\partial_i\theta^2$, to simplify the above Lagrangian, 
\begin{eqnarray}
L^{\alpha} &=& -(a')^{\frac{1}{\alpha+1}}\int dr
\left[-\frac{\dot\theta}{c^2}+2\dot\theta
+ (\partial_i \theta)^2 \right]^{\frac{\alpha}{\alpha+1}}.\label{chaply35}
\end{eqnarray}
Now taking the large $c$ limit we get
\begin{eqnarray}
\lim_{c\rightarrow \infty} L^{\alpha} = -
(2)^{\frac{\alpha}{\alpha+1}}(a')^{\frac{1}{\alpha+1}}\int dr \sqrt{\left[\dot \theta+(\frac{\partial_i \theta)^2}{2}\right]^{\frac{2\alpha}{\alpha+1}}}\label{chaply36}.
\end{eqnarray}
After making the identification
\begin{equation}
 a' \equiv (\frac{\alpha}{2(\alpha+1)})^{\alpha}B \label{chaply37}
\end{equation}
 we see that (\ref{chaply36}) agrees with  (\ref{chaply33}).
Also, in the $\alpha =1$ limit our Lagrangian (\ref{chaply34}) reduces to that of usual relativistic Chaplygin gas (\ref{chaply18}). This shows that it is possible to interpret (\ref{chaply34}) as a viable form for the relativistic GCG Lagrangian.

  At this point we should mention that there exists in the literature a Poincare invariant form for GCG \cite{gcg}
\begin{equation}
L_b = - A^{\frac{1}{1+\alpha}} \int dr \left[c^2 -
(\partial_{\mu}\theta
\partial^{\mu}\theta)^{\frac{1+\alpha}{2\alpha}}\right]^{\frac{\alpha}{\alpha+1}}.\label{chaply38}
\end{equation}
Note that, for $\alpha\ne 1$ the above Lagrangian is different from (\ref{chaply34}). However for $\alpha =1$ it agrees with normal Chaplygin gas Lagrangian \cite{jac2}. GCG of similar nature \cite{gcg}  coupled to gravity has been considered in \cite{Hess}.  
 
   Now consider the nonrelativistic limit of (\ref{chaply38}). Following 
the same procedure as discussed above we get in this limit 
\begin{equation}
L_b = -A^{\frac{1}{1+\alpha}}(2Y)^{\frac{\alpha}{\alpha+1}}\sqrt{\left[\frac{Z}{2Y} + (\dot\theta+\frac{1}{2}\nabla\theta^2)\right]^{\frac{2\alpha}{1+\alpha}}}.
\label{chaply39}
\end{equation}
where
\begin{eqnarray}
Y&=& \frac{1+\alpha}{2\alpha}c^{\frac{1+\alpha}{\alpha}}, \label{chaply40}\\
Z&=& c^2 - c^{\frac{1+\alpha}{\alpha}}. \label{chaply41}
\end{eqnarray}
This Lagrangian is same as that of (\ref{chaply30}) provided we identify $I$ with $\frac{Z}{2Y}$. Thus both (\ref{chaply34}) and (\ref{chaply38}) are valid forms for the relativistic GCG whose nonrelativistic limits correspond to different
 parameterizations of the general form for nonrelativistic GCG given in (\ref{chaply30}). 
 
 Let us next attempt to construct the relativistic GCG model by including
 the density field $\rho$. Also, since the  density field plays an 
important role in the observational analysis of GCG it is worthwhile to have 
 a relativistic version for GCG involving $\rho$ and the velocity
 potential $\theta$.

   To this end, we consider the following  Lagrangian for relativistic GCG:
\begin{equation}
L^{\alpha} = \int dr \left(\theta \dot\rho - \sqrt{(\rho^2c^2 +
\frac{a^2}{\rho^{\alpha-1}})}\sqrt{c^2 + (\partial_i
\theta )^2}\right) \label{chaply42}
\end{equation}
where  $a$ is a constant parameter. To ensure the correct  nonrelativistic limit
we use the same  map as (\ref{chaply19}), and  explicitly check that in the  $c\rightarrow \infty$ limit, the nonrelativistic GCG model (\ref{chaply32}) is reproduced, provided we identify, 
\begin{equation}
a = \sqrt{\frac{2B}{\alpha+1}}. \label{chaply43}
\end{equation}
We put $c =1$ and obtain the equations of motion, 
      \begin{equation}
\dot \rho + \partial_i \left( \frac{\sqrt{(\rho^2 +
\frac{a^2}{\rho^{\alpha-1}})}}{\sqrt{1 + (\partial_i \theta)^2}}
\partial_i \theta\right) = 0, \label{chaply44}
\end{equation}
\begin{equation}
\dot \theta = -\frac{\sqrt{1 + (\partial_i
\theta)^2}}{\sqrt{(\rho^2 + \frac{a^2}{\rho^{\alpha-1}})}}[\rho
c^2 - (\frac{\alpha -1}{2})\frac{a^2}{\rho^{\alpha}}]~.\label{chaply45}
\end{equation}
They also have the correct  $\alpha=1$ limit \cite{jac2}.

As we have pointed out before, the Lagrangian (\ref{chaply42}) has been posited by us in analogy with the relativistic Lagrangian given in (\ref{chaply17}) \cite{jac2}. Also we show that (\ref{chaply42}) has the correct $\alpha =1$ limit. Since the model (\ref{chaply42}) is not manifestly Lorentz invariant, it becomes imperative to check the Poincare algebra. To this end, we follow the method discussed in \cite{bg} and compute the canonical energy-momentum tensor $T_{\mu\nu}$ (in  the Noether prescription),
\begin{equation}
T_{\mu\nu} =
\frac{\partial\mathcal{L}}{\partial(\partial^{\mu}\psi_{i})}\partial_{\nu}\psi^{i}
- g_{\mu\nu}\mathcal{L}.\label{chaply46}
\end{equation}
  Using the above definition, the explicit form of the components of $T_{\mu\nu}$ are given by,
\begin{eqnarray}
T_{00}&=& \sqrt{(\rho^2 + \frac{a^2}{\rho^{\alpha-1}})}\sqrt{1 + (\partial_i \theta)^2},\label{chaply47}\\
T_{0i}&=& \theta \  \partial_{i} \rho,\label{chaply48}\\
T_{i0}&=& -\frac{\sqrt{\rho^2 + \frac{a^{2}}{\rho^{\alpha-1}}}}{\sqrt{1+(\partial_{k}\theta)^2}} (\partial_{i}\theta) \dot \theta = \left(\rho({\bf x}) +
\left(\frac{1-\alpha}{2}\right)\frac{a^2}{\rho^{\alpha}({\bf
x})}\right)\partial_{i}\theta ,\label{chaply49}\\
T_{ij}&=& -\frac{\sqrt{\rho^2 +
\frac{a^{2}}{\rho^{\alpha-1}}}}{\sqrt{1+(\partial_{k}\theta)^2}}
(\partial_{i}\theta) (\partial_{j}\theta) -
g_{ij}\mathcal{L}^{\alpha}.\label{chaply50}
\end{eqnarray}
Notice that $T_{0i} \ne T_{i0}$. Using the equations of motion (\ref{chaply44}), (\ref{chaply45}) one can  explicitly  verify the conservation law,
\begin{equation}
\partial^{\mu}T_{\mu\nu}= 0. \label{chaply51}
\end{equation}
Hence $T_{\mu\nu}$ is a conserved but non-symmetric energy-momentum tensor.

    Once we have the forms of $T_{00}$ and $T_{0i}$ we can easily obtain the expression for the momenta $P_{\mu}$ and the angular momenta $M_{\mu\nu}$. They are related to the components of the energy-momentum tensor as
\begin{eqnarray}
P_{\mu}&=&\int d^3 x \ T_{0\mu},\label{chaply52}\\
M_{\mu\nu}&=&\int d^3 x \  (T_{0\mu}x_{\nu} - T_{0\nu}x_{\mu}).
\label{chaply53}
\end{eqnarray}
By using (\ref{chaply47}) and (\ref{chaply48}) we get,
\begin{eqnarray}
P_{0}&=& \int d^3 x \ \sqrt{(\rho^2+ \frac{a^2}{\rho^{\alpha-1}})}\sqrt{1 + (\partial_i \theta)^2}, \label{54}\\
P_{i}&=& \int d^3 x \  \theta \ \partial_{i} \rho,\label{chaply55} \\
M_{0i}&=& \int d^3 x \  \sqrt{(\rho^2 + \frac{a^2}{\rho^{\alpha-1}})}\sqrt{1 + (\partial_i \theta)^2} \  x_{i} - \theta \ \partial_{i} \rho \ x_{0},\label{chaply56}\\
M_{ij}&=& \int d^3 x \  (\theta \ \partial_{i}\rho \  x_{j} -
\theta \  \partial_{j} \rho \ x_{i}). \label{chaply57}
\end{eqnarray}
 Using the Poisson bracket (\ref{chaply11}) we are able to compute the following algebra, 
\begin{eqnarray}
\{M_{ij},M_{kl}\}&=&  (g_{jk}M_{il} - g_{ik}M_{jl} - g_{il}M_{kj} + g_{jl}M_{ki}),\label{chaply58}\\
\{M_{oi},M_{kl}\}&=& (g_{ik}M_{0l} - g_{il}M_{0k}),\label{chaply59}\\
\{M_{0i},M_{oj}\}&=& - g_{00}\int d^3x \ (\theta \
\partial_{i}\rho \  x_{j} - \theta \  \partial_{j} \rho \ x_{i})
\left(1 -\alpha\left(\frac{1-\alpha}{2}\right)\frac{a^2}{\rho^{\alpha+1}}\right).\label{chaply60}
\end{eqnarray}
Similarly the algebra between $P_{\mu}$-$M_{\mu\nu}$ is given by
\begin{eqnarray}
\{M_{0i},P_{j}\}&=& P_{0}g_{ij},\label{chaply61}\\
\{M_{ij},P_{k}\}&=& g_{jk}P_{i} - g_{ik}P_{j},\label{chaply62}\\
\{M_{0i},P_{0}\}&=& -g_{00} \int d^3 x \  \theta \ \partial_{i}
\rho \ \left(1-\alpha\left(\frac{1-\alpha}{2}\right)\frac{a^2}{\rho^{\alpha+1}}\right).\label{chaply63}
\end{eqnarray}
Finally, the algebra between $P_{\mu}$-$P_{\nu}$ is found out to be,
\begin{eqnarray}
\{P_{\mu},P_{\nu}\}&=& 0.\label{chaply64}
\end{eqnarray}
Concentrate on the two Poisson brackets (\ref{chaply60}) and (\ref{chaply63}). We find that for $\alpha =1$ the complete Poincare algebra is satisfied. This corresponds to the usual Chaplygin Poincare algebra \cite{bg}. However for $\alpha\ne 1$
(which corresponds to the GCG model) the Poincare algebra closes only in the large density limit $(\rho>>1)$. It is, however, reassuring to note that the Schwinger condition,      
\begin{equation}
\{T_{00}({\bf x}),T_{00}({\bf y})\} = (T_{i0}({\bf x}) +
T_{i0}({\bf y}))\partial^{x}_{i}\delta({\bf x}-{\bf
y}),\label{chaply65}
\end{equation}
is satisfied for any $\alpha$ and $\rho$.
\section{\label{chaplygin4}Discussions}
To conclude, we have studied various aspects of the Generalized Chaplygin Gas (GCG) models. In the nonrelativistic regime, we have constructed a general form
of the Lagrangian for GCG, that obeys the generalized equation of state. Different parameterizations of this master Lagrangian yield different {\it{inequivalent}} models for GCG, such as the one studied here and the one in \cite{gcg}.
In this sense the construction of nonrelativistic GCG is not unique.
Naturally, the same conclusion extends for a relativistic formulation 
 of GCG. 

  For the relativistic scenario, we have proposed a Born-Infeld like
 model for GCG, which in the nonrelativistic limit, reduces to the conventional GCG. However, unlike the usual $\alpha=1$ Chaplygin gas case, the construction
 of a relativistic GCG model, including both density field and velocity potential is nontrivial. In this context, our model reduces to the usual one, quoted
 in literature \cite{jac2} for $\alpha =1$ and also has the correct nonrelativistic limit. However the Poincare algebra closes only in the limit of large 
density.

\chapter{\label{chap:coanomaly}Hawking fluxes from covariant gauge and gravitational anomalies}
 Hawking effect provides an important step 
towards understanding the quantum aspects of black holes.
 Specifically, it arises in a background spacetime with event horizons. 
Hawking studied quantum effects of matter in the black hole background
 formed during collapse and concluded that the black hole emits thermal radiation
 as if it was a black body at a temperature proportional to the surface gravity
 of the black hole \cite{hawking}. The radiation emitted from the black holes
has a spectrum with Planck distribution. Hawking's original derivation \cite{hawking}
 was based on the computation of Bogoliubov coefficients between 'in' and 'out'
 states. Apart from this derivation there are several ways 
to compute the flux of thermal radiation emitted by the black hole
\cite{gibbons,paddy, parikh, fulling}, each having its 
own merits and dimerits. This has led to open problems leading
to alternative approaches with fresh insights. In this chapter we 
would discuss a different approach to derive  the Hawking flux from 
 a black hole. This approach is based on the covariant gauge and gravitational anomalies. 

   A relationship between gravitational anomalies and  
Hawking radiation was first noted by S. Robinson and F. Wilczek
\cite{robwilczek}. They considered quantum
scalar fields propagating on the $(3+1)$ dimensional Schwarzschild 
black hole background. Their analysis rests on the fact that 
quantum field theory in the region near to the event horizon can
 effectively be described by a $(1+1)$ dimensional \cite{carlip1,solodukhin}
 chiral \cite{robwilczek} theory. 
Any $2$-dimensional chiral theory 
on a curved background possesses gravitational anomaly \cite{witten}. In the
region away from the horizon, the theory is still $3+1$ dimensional
and also usual (anomaly free). 
The  energy flux of the Hawking radiation, which is necessary to
cancel the anomaly present near the horizon, was then computed by  
solving the anomalous as well as usual conservation laws in the
respective regions together with implementation of certain boundary conditions.
 This method is expected to hold in any dimensions. In this sense it is distinct
from the approach given by Christiansen and Fulling \cite{fulling}, where 
 the form for energy-momentum tensor of massless quantum field in a 
$(1+1)$ dimensional black hole background was obtained by exploiting
 the structure of trace anomaly. The flux obtained via trace anomaly 
is in quantitative agreement with Hawking's original result. However,
the masslessness of the quantum fields and also limitation to $(1+1)$ dimensional
black holes are quite essential ingredients in this analysis. 
 The approach of \cite{robwilczek} was also applied to compute the Hawking
charge and energy flux coming from the Reissner-Nordstrom black hole 
\cite{isowilczek}. Further advances and application
of this anomaly cancellation approach may be found in a host of papers
(\cite{muratasoda1}-\cite{saurya}).

   However, an unpleasant feature of \cite{robwilczek,isowilczek} was that 
whereas the expressions for chiral anomalies were taken to be consistent,
the boundary conditions required to fix the arbitrary constants were covariant.
In this chapter we present a derivation that is solely based on covariant expressions. 
The expressions for the covariant anomalous currents and energy-momentum tensors,
in the region near to the event horizon, are obtained by solving the covariant anomalous
 gauge and gravitational Ward identities, respectively. The arbitrary constants appearing 
 in these expressions are fixed by imposing the covariant boundary condition: namely, the vanishing
 of covariant current and energy-momentum tensor at the event horizon.
 On the other hand the corresponding expressions for the current
 and energy-momentum in the region far away from the horizon are derived by solving the 
 usual conservation laws. The charge/energy-momentum flux is then obtained by
 demanding that the total current/energy-momentum tensor of the theory must be
 anomaly free. The charge
and energy fluxes obtained by our anomaly approach matches with the standard
expression of Hawking flux \cite{hawking,birrell}. Further, as a side
 calculation, we show that
the analysis of \cite{robwilczek,isowilczek} is resilient and 
the results are unaffected by taking more general expressions for the 
consistent anomaly which occur due to peculiarities of two dimensional
spacetime. 

   This chapter is organized in the following manner.
 Section-\ref{coanomaly1} discusses the general 
aspects of anomalies in quantum field theory. 
Since the anomalies in $(1+1)$ dimensions are particularly relevant in 
deriving the Hawking flux, we simply restrict the discussion of
anomalies  to $(1+1)$ dimensions. 
Distinction among the two types of anomalies -
covariant and consistent - is also emphasized. In section-\ref{coanomaly2}
 we derive the 
Hawking charge and energy flux by using the covariant gauge and 
gravitational anomalies. A comparison between our covariant 
 anomaly based with the consistent one \cite{isowilczek} is given
 in section-\ref{coanomaly3}. In section-\ref{coanomaly4} we comment
 on the robustness in the analysis of \cite{robwilczek,isowilczek} by
 considering more general expressions for the consistent anomaly which occur
 due to peculiarities of two dimensional spacetime.
 Some applications of the covariant anomaly cancellation mechanism
 for more nontrivial black hole geometries is provided in section-\ref{coanomaly5}.
 There we compute  the fluxes of Hawking radiation comming from the 
 Garfinkle-Horowitz-Strominger (GHS) and $D1-D5$ nonextremal 
 black holes. 
Our concluding remarks are given in section-\ref{coanomaly6}. 
Finally,  we provide an appendix containing a detailed discussion of 
the dimensional reduction procedure, for the neutral and charged scalar  
fields propagating on the Schwarzschild or Reissner-Nordstrom black hole background.
\section{\label{coanomaly1}General discussion on covariant and consistent anomalies}
 Symmetries play an important role in physics in general and in quantum field theory in particular.
A continuous symmetry of the classical action is a transformation of the fields that leaves the action invariant.
Corresponding to each such symmetry operation there exist a conserved charge. This is the
 Noether's theorem. Standard examples are Lorentz, or more generally Poincare transformations, and gauge transformations in gauge theories. In the functional integral formulation of quantum field theory, symmetries of the classical action are easily seen to translate into the Ward identities for the correlation functions computed from the quantum effective action. Naturally, it becomes important 
to know whether a certain classical symmetry is still valid in the quantum theory. 

      An anomaly in quantum field theory is a breakdown of some classical symmetry 
 due to the process of quantization. This surprising feature of
quantum theory plays a fundamental role in physics 
 (for reviews, see \cite{fujikawabook,bertlmannbook,adler,jackiwanomaly1,jackiwanomaly2,billal}).  
 Specifically, for instance, a gauge  anomaly is an anomaly in gauge symmetry, taking the form of nonconservation  of the gauge current. Such  anomalies characterize a theoretical 
 inconsistency, leading to problems with the probabilistic interpretation
 of quantum mechanics. The cancellation of gauge anomalies gives strong
 constraints on model building. Likewise, a gravitational anomaly 
\cite{witten,bardeenzumino} is an anomaly in general coordinate invariance, taking the
 form of nonconservation of the energy-momentum tensor. There are other
 types of anomalies but here we shall be concerned with only gauge 
 and gravitational anomalies. The simplest case for these anomalies,
 which is also relevant for the present analysis, occurs for $1+1$ dimensional
 chiral fields.        

      In this section we would discuss some important aspects of 
gauge and gravitational anomalies in  ($1+1$) dimensions. In general, the 
 anomalous theories admit two types of currents and energy-momentum tensors;
 the consistent and the covariant \cite{fujikawabook,bardeenzumino,bertlmannbook,rabin1}.
 The covariant divergence of these currents and energy-momentum tensors yields either
 consistent or covariant gauge and gravitational anomalies, respectively
 \cite{witten,fujikawabook,bardeenzumino,bertlmannbook,bertlmann,rabin1,rabin2}. 
 The consistent current and anomaly satisfy the Wess-Zumino integrability condition but
 do not transform covariantly under a gauge transformation. Expressions for covariant
 current and anomaly, on contrary, transform covariantly under gauge the transformation
 but do not satisfy the Wess-Zumino integrability condition. Similar conclusions also
 hold for the gravitational case, except that currents are now replaced by
 energy-momentum tensors and gauge transformations by general coordinate transformations.
The consistent and covariant of currents (energy-momentum tensors) are interrelated by
 a local counterterm. 

 For simplicity, we mainly focus our attention on the gauge anomaly. For the 
 gravitational case  we just quote some basic results which shall be important in 
 discussing the Hawking effect.  
 
     Consider the chiral (Weyl) fermions moving in the presence of
 external abelian gauge field $A_{\mu}$ on a flat $1+1$ dimensional
 spacetime. The action for this  chiral theory is 
 \begin{equation}
 S = -\int d^{2}x \ \bar\Psi \gamma^{\mu}\left(\partial_{\mu} - iA_{\mu}
 \frac{1\pm\gamma_{5}}{2}\right)\Psi \label{1.1}
\end{equation}
where $+(-)$ corresponds to the left and right moving fermions, respectively. 
In the Minkowski space the Dirac matrices satisfy 
\begin{equation}
 (i\gamma^{0})^{\dagger} = i\gamma^{0} \ ; \ (\gamma^{1})^{\dagger} = \gamma^{1} \ ; \ 
\gamma_{5}^{\dagger} = \gamma_{5} ~.\label{1.1a}
\end{equation}
 The chiral gauge current, derived from (\ref{1.1}) is  
\begin{equation}
 J_{\mu} = i\bar\Psi \gamma_{\mu}\frac{1\pm\gamma_{5}}{2} \Psi ~.\label{1.2}
\end{equation}
On using the equations of motion for $\Psi$ and $\bar\Psi$
 we can easily show that the chiral current
given above is conserved
\begin{equation}
 \partial_{\mu}J^{\mu} = 0~. \label{1.2a}
\end{equation}
Also, it transform covariantly under the chiral gauge transformations,
\begin{eqnarray}
 \Psi(x)&\rightarrow& \exp\left(i\alpha(x)\frac{1\mp\gamma_{5}}{2}\right)\Psi(x) \label{1.6'}\\
 \bar\Psi(x)&\rightarrow& \bar\Psi(x) \exp\left(-i\alpha(x)\frac{1\pm\gamma_{5}}{2}\right)\label{1.6}\\
 A_{\mu}(x) &\rightarrow& \exp(i\alpha(x)) [A_{\mu} - i\partial_{\mu}]\exp(-i\alpha(x))~, \label{1.6''}
\end{eqnarray}

 However, when we quantize the theory
described by (\ref{1.1}), the regularized current $\langle J_{\mu}(x)\rangle$ does not
 conserve. In fact, it satisfies
\begin{equation}
\partial_{\mu}\langle J^{\mu}(x)\rangle = \pm G \label{1.3}
\end{equation}
where $G$ is the chiral abelian gauge anomaly. The explicit form for
$G$ depends upon how we regularize $\langle J_{\mu}\rangle$.  Let $W$ be
 the quantum effective action for the chiral theory, defined as 
\begin{equation}
 e^{ iW[A]} = \int [D\Psi D\bar\Psi] e^{iS[\Psi,\bar\Psi,A_{\mu}]} \label{1.4}
\end{equation}
where $S$ is the classical action (\ref{1.1}) . 
The current $\langle \tilde J_{\mu}(x)\rangle$ obtained by taking the 
functional derivative of the quantum effective action, i.e
\begin{equation}
 \langle \tilde J^{\mu}(x)\rangle = \frac{\delta}{\delta A_{\mu}(x)} W \label{1.5}
\end{equation}
does not transform covariantly under the chiral gauge transformations (\ref{1.6'}, \ref{1.6}, \ref{1.6''}).
Rather, it satisfies the Wess-Zumino integrability condition \cite{rabin1,rabin2}. 
\begin{equation}
 \frac{\delta \langle\tilde J_{\mu}(x)\rangle}{\delta A^{\nu}(x')} = \frac{\delta\langle\tilde J_{\nu}(x')\rangle}{\delta A^{\mu}(x)}~. \label{1.7}
\end{equation}
This current is called the consistent current \cite{fujikawabook,rabin1}. 
Divergence of the consistent current yields the consistent 
anomaly 
\begin{equation}
 \partial_{\mu}\langle \tilde J^{\mu} \rangle = \pm \frac{e^2}{4\pi}\epsilon^{\mu\nu}\partial_{\mu}A_{\nu}~.
 \label{1.8}
\end{equation}
where  $\epsilon^{\mu\nu}$ is the numerical antisymmetric tensor with 
\begin{equation}
 \epsilon^{01}=-\epsilon^{10}=1~. \label{1.8a}
\end{equation}
 The relation (\ref{1.8}) can be easily extended to a general curved background defined by the 
 metric $g_{\mu\nu}$ by replacing partial derivative with the covariant derivative compatible
 with the metric $g_{\mu\nu}$ and  $\epsilon^{\mu\nu}$ by $\bar\epsilon^{\mu\nu}$
\begin{equation}
 \nabla_{\mu}\langle \tilde J^{\mu} \rangle = \pm \frac{e^2}{4\pi}\bar\epsilon^{\mu\nu}\partial_{\mu}A_{\nu}~.
 \label{1.9}
\end{equation}
 where
\begin{equation}
\bar\epsilon^{\mu\nu}=\frac{\epsilon^{\mu\nu}}{\sqrt{-g}}~. \label{1.9a}
\end{equation}
   
 The structure appearing in (\ref{1.9}) is the minimal form, since only
odd parity terms occur.  However it is possible that normal parity 
terms appear in (\ref{1.9}). Indeed, as we now argue, such a term is a natural
 consequence of two dimensional properties. To emphasize this point we note that
 in $1+1$ dimensions, $\gamma_{\mu}$ satisfy
\begin{equation}
\gamma_{5}\gamma^{\mu} = -\frac{\epsilon^{\mu\nu}}{\sqrt{-g}} \gamma_{\nu}~.
\label{1.10}
\end{equation}
Using this it is found that the gauge field $A_{\mu}$ couples as a chiral combination
$(g^{\mu\nu}\pm \epsilon^{\mu\nu})A_{\nu}$. Hence the expression for the consistent 
anomaly in (\ref{1.9}) generalizes to
\begin{equation}
 \nabla_{\mu}\langle \tilde J'^{\mu} \rangle = \pm \frac{e^2}{4\pi}\nabla_{\alpha}
[(\epsilon^{\alpha\beta}\pm g^{\alpha\beta})A_{\beta}]~.\label{1.11}
\end{equation} 
This is the nonminimal form for the consistent gauge anomaly
dictated by the symmetry of the Lagrangian, and has already
 appeared earlier in the literature \cite{bardeenzumino}. The current
 $\langle \tilde J'^{\mu}\rangle$ is related to $\langle \tilde J^{\mu}\rangle$  
 as
\begin{equation}
 \langle \tilde J'^{\mu}\rangle = \langle \tilde J^{\mu}\rangle + \frac{e^{2}}{4\pi}A_{\mu}
\label{1.12}
\end{equation}
so that the covariant divergence of $\langle \tilde J'^{\mu}\rangle$ yields the 
nonminimal form of the consistent gauge anomaly (\ref{1.11}). Also, note that
the $\langle \tilde J'^{\mu}\rangle$ is the consistent current since
 the extra piece satisfies the Wess-Zumino integrability condition (\ref{1.7}). 

     As mentioned earlier, value of the anomaly (\ref{1.3}) depends upon
 the regularization prescription.  If the current $ \langle J^{\mu}(x)\rangle$  
is defined  by a gauge invariant regularization then it transform covariantly
 under the chiral gauge transformations (\ref{1.6'}, \ref{1.6}, \ref{1.6''}).
 Let us denote this current by $\langle \hat J^{\mu} \rangle$
 \footnote{In this section all the covariantly regularized objects are indicated by
 the hatted variables. From the next section onwards, we would denote them by usual 
(unhatted) variables.}, then 
\begin{equation}
\langle \hat J^{\mu}(x) \rangle \rightarrow e^{-i\alpha}\langle \hat J^{\mu}(x) \rangle e^{i\alpha}~.
\label{1.13}
\end{equation} 
The current defined in such a way is called the covariant current. Divergence of the 
covariant current yields the covariant chiral gauge anomaly \cite{fujikawabook,rabin1,rabin2} 
\begin{equation} 
\partial_{\mu}\langle \hat J^{\mu} \rangle = \pm \frac{e^2}{4\pi}\epsilon^{\alpha\beta}F_{\alpha\beta}~. 
\label{1.14}
\end{equation}
As before, the curved space generalization of this is given by
\begin{equation}
 \nabla_{\mu}\langle \hat J^{\mu} \rangle = \pm \frac{e^2}{4\pi}\bar\epsilon^{\alpha\beta}F_{\alpha\beta} 
\label{1.15}
\end{equation}
It is possible to modify the consistent current (\ref{1.5}), by adding
 a local counterterm, so that it becomes covariant,
\begin{equation}
 \langle \hat J^{\mu} \rangle = \langle \tilde J^{\mu} \rangle \mp \frac{e^2}{4\pi} A_{\alpha}\bar\epsilon^{\alpha\mu} \label{1.16}
\end{equation}
 Note that the covariant current (\ref{1.16}) does not satisfy the Wess-Zumino
 consistency condition since the counterterm violates the integrability
 condition (\ref{1.7}). Moreover the gauge covariant anomaly (\ref{1.14}) or
 its curved space generalization (\ref{1.15}) has a unique form 
 dictated by the gauge transformation properties. This
 is contrary to the consistent anomaly which may have a minimal (\ref{1.9})
or non-minimal (\ref{1.11}) structure. 

     Now we will concentrate our attention on the gravity sector. It
was shown by Alvarez-Gaume and E. Witten \cite{witten} that in $4k+2 \ (k=0,1,2 \cdots)$ 
dimensions, Einstein's general coordinate transformation can contain an anomaly
 in the chiral sector (see for a review \cite{fujikawabook,bertlmannbook,bertlmann}).  
The breakdown of general coordinate invariance is manifested 
in the nonconservation of the energy-momentum tensor. As in the
case of $U(1)$ gauge current, the energy-momentum tensor for the chiral theory 
on the general curved background is either consistent or covariant depending
 on the choice of regularization adopted to quantize the theory. We denote
 the consistent energy-momentum tensor by $\tilde T_{\mu\nu}$. The covariant 
 divergence of the consistent energy-momentum tensor yields the 
consistent anomaly \cite{witten,bardeenzumino,bertlmann}. In $1+1$ dimensions
 the form of consistent gravitational anomaly, for right moving fermions, is given by  
\begin{equation}
 \nabla_{\mu}\langle{\tilde T}^{\mu}{}_{\nu}\rangle = \frac{1}{96\pi}\bar\epsilon^{\beta\delta}\partial_{\delta}\partial_{\alpha}\Gamma^{\alpha}_{\nu\beta}~.
\label{1.17}
\end{equation}
It is worthwhile to point out that consistent gravitational anomaly and the consistent
gauge anomaly are analogous satisfying similar consistency conditions (\ref{1.7}). 
This is easily observed here by comparing (\ref{1.17}) with (\ref{1.9}) where
 the affine connection plays the role of the gauge potential. We therefore omit the details
 and write the generalized gravitational consistent anomaly by an inspection of (\ref{1.11})
 on how to include the normal parity term. The result is
\begin{equation}
\nabla_{\mu}\langle{\tilde T}^{'\mu}{}_{\nu}\rangle = \frac{1}{96\pi}\partial_{\delta}\partial_{\alpha}\left[(\epsilon^{\beta\delta} + g^{\beta\delta})\Gamma^{\alpha}_{\nu\beta}\right] =  \mathcal{\tilde A'_{\nu}}.\label{1.18} 
\end{equation}
The covariant energy-momentum tensor $\hat T_{\mu\nu}$, on the other hand, has the divergence anomaly,
\begin{equation}
 \nabla_{\mu}\langle \hat T^{\mu}{}_{\nu}\rangle = \frac{1}{96\pi}\bar\epsilon_{\nu\mu}\nabla^{\mu}R = \mathcal{A_{\nu}}
\label{1.19} 
\end{equation}
where $R$ is the Ricci scalar corresponding to the metric $g_{\mu\nu}$. This is
 the covariant form of the gravitational anomaly. Note that the right hand side of
 (\ref{1.19}) is manifestly covariant, since it contains terms proportional to the 
derivative of the Ricci scalar. However this is not true
 for the consistent anomaly, (\ref{1.17}). The consistent and covariant
energy-momentum tensors are related via local counterterm  
\begin{equation}
 \langle \hat T_{\mu\nu}\rangle = \langle\tilde T_{\mu\nu}\rangle + \mathcal{P_{\mu\nu}} \label{1.20}
\end{equation}
where $\mathcal{P_{\mu\nu}}$ satisfies \cite{bardeenzumino} 
\begin{equation}
 \nabla^{\mu}\mathcal{P_{\mu\nu}} = \frac{1}{96\pi}[\bar\epsilon_{\nu\sigma}\nabla^{\sigma}R - \bar\epsilon
^{\beta\delta}\partial_{\delta}\partial_{\alpha}\Gamma^{\alpha}_{\nu\beta}]~.\label{1.21}
\end{equation}

\section{\label{coanomaly2}Covariant anomalies and Hawking fluxes}
Now we discuss the relationship between Hawking radiation
and the gauge and gravitational anomalies. In \cite{robwilczek},
 it was proposed that the flux of 
Hawking radiation can be obtain by a knowledge of the gravitational anomaly
at the horizon. An essential observation in  \cite{robwilczek} is that
quantum fields near the horizon of $(d+1)$ dimensional black hole behave as
an infinite collection of two dimensional massless fields propagating on $r-t$ sector
of the full $(d+1)$ dimensional black hole metric. 
Then, by transforming into the null coordinate and using the 
 equations of motion for fields under consideration, we can easily decompose the field
 into two parts, propagating either `in' to the horizon or `out' from the horizon. 
 We interpret ingoing modes as left moving and outgoing modes as right moving. Once
 the left moving modes fall into the black hole, they never come out classically and cannot
 affect the physics outside the black hole. Classically, modes inside the event horizon
 are causally disconnected from the outer region. Consequently, the effective field theory
 near the horizon is two dimensional and chiral. If we then integrate over the relevant right 
 moving modes to obtain the quantum effective action in the exterior region, it becomes anomalous
 with respect to gauge or general coordinate symmetry. However, the original theory is 
 of course gauge and diffeomorphism invariant. Therefore the anomalies, which are present near the
 horizon, must be cancelled by the quantum effects of classically irrelevant left moving modes. 
 This fixes the flux of the outgoing modes which is interpreted as the Hawking flux as
 measured by an observer at the asymptotic infinity. 
 Also, since the source of anomaly is
 located in an arbitrary small region near the event horizon, the  fluxes of radiation are universally determined by the properties of black holes at the horizon.

    To put the above considerations in a proper perspective, it is important 
 to realize that $(1+1)$ dimensional chiral theories admit two types of anomalous currents and energy-
 momentum tensors- the consistent and the covariant (section-\ref{coanomaly1}).
 The analysis of \cite{robwilczek,isowilczek} uses the consistent form for the
 gauge and gravitational anomalies 
 to obtain the form for the current and energy-momentum tensor in the vicinity of 
 the event horizon. However, the boundary conditions,
 required to fix the arbitrary constants appearing in the 
 expressions for current and energy-momentum tensor, were covariant. This raises 
 several issues, both technically and conceptually. Note that the flux is measured
 at infinity. Since, in the region far away from the horizon, the theory is anomaly free,
 covariant and consistent structures are identical. Note that the mismatch between the
 covariant and consistent currents (energy-momentum tensors) is the germ of the anomaly
 \cite{rabin1,rabin2}. Hence if the anomaly approach
 is viable the fluxes of Hawking radiation should equally well be obtainable from the
 covariant expressions. Also, as we shall demonstrate below,
 the use of covariant expressions entails considerable technical simplification.  For example the shift between
 covariant and consistent expressions through local counterterms, as is mandatory in
 \cite{robwilczek,isowilczek}, is not necessary if we use the covariant expressions
 for the gauge and gravitational anomalies. We now elaborate step by step the covariant 
 anomaly method to compute the flux of Hawking radiation. We do this analysis for
 the Reissner-Nordstrom black hole. 

      Consider the Einstein-Maxwell theory represented by the action
\begin{equation} 
 S_{EM} = \int \ d^{4}x \ \sqrt{-\gamma}\left[R_{(4)} - F_{ab}F^{ab}\right] \label{1.2.1}
\end{equation}
where $\gamma_{ab}$ is the metric on
 $(3+1)$ dimensional spacetime\footnote{Latin indices $a,b$, unless otherwise stated, represent 
 $(3+1)$ dimensional spacetime while the Greek indices $\mu,\nu$ are reserved for $(1+1)$ dimensions.}
 while $\gamma = det\gamma_{ab}$. $R_{(4)}$ is the curvature scalar associated with $\gamma_{ab}$. The electromagnetic field strength tensor $F_{ab}$ is defined in terms of gauge potential $A_{a}$ as   
\begin{equation}
F_{ab} = \nabla_{a}A_{b} - \nabla_{b}A_{a}~. \label{1.2.2}
\end{equation}
Variation of (\ref{1.2.1}) with respect to the metric and gauge potential gives the coupled
Einstein and Maxwell equations, respectively. The solutions
for Einstein equations are given by
 \begin{equation}
 ds^2 = \left(1-\frac{2M}{r}+\frac{Q^2+P^2}{r^2}\right)dt^2 - \left(1-\frac{2M}{r}+\frac{Q^2+P^2}{r^2}\right)^{-1}dr^2 - r^2(d\theta^2 + \sin^2\theta d\phi^2)~. \label{1.2.3} 
 \end{equation}
The metric (\ref{1.2.3}) is static and it is know as the Reissner-Nordstrom metric.
 The solutions for the Maxwell's equations are given by
 \begin{equation}
 E_{r}=F_{rt} = \frac{Q}{r^{2}} \ ; B_{r}=\frac{F_{\theta\phi}}{r^{2}\sin\theta} = \frac{P}{r^2}~. \label{1.2.4}
 \end{equation}
Here $M$ is the mass of the black hole and $Q$, $P$ are electric and magnetic charges respectively. 
For our purpose we consider only electrically charged black hole (i.e $P=0$).
 Also, we shall work in the gauge $A_{r}=0$. Then the metric (\ref{1.2.3}) becomes
\begin{equation}
 ds^2 = \gamma_{ab}dx^{a}dx^{b}=f(r)dt^2 - \frac{1}{f(r)}dr^2 - r^2(d\theta^2 + \sin^2\theta d\phi^2) \label{1.2.5} 
\end{equation}
with 
\begin{equation}
 f(r) = 1 - \frac{2M}{r} + \frac{Q^2}{r^2} = \frac{(r-r_{+})(r-r_{-})}{r^2}~, \label{1.2.6}
\end{equation}
 while the gauge field $A_{a}$ is given by
\begin{equation}
 A_{t}(r) = -\frac{Q}{r} \ ; \ A_{r}=A_{\theta}=A_{\phi} =0~. \label{1.2.7}
\end{equation}
The location of inner ($r_{-}$) and outer ($r_{+}$) horizons are given by  
\begin{equation}
r_{\pm} = M \pm \sqrt{M^2 - Q^2}~. \label{1.2.8}
\end{equation}
Now we consider charged (complex) scalar fields moving on the background 
 represented by the metric given in (\ref{1.2.5}). The  metric $\gamma_{ab}$ 
and the gauge field $A_{a}$ serves as external fields. In the region near the 
 outer event horizon, upon transforming to $r_{*}$ (tortoise) coordinate and performing 
 the partial wave decomposition, we can show that the effective radial potential corresponding
 to each partial wave mode is proportional to the metric function $f(r(r_{*}))$ which decay
 exponentially  fast near the event horizon. The same reasoning holds for the mass terms (in the matter
 Lagrangian). Hence, the matter action in the region near the event horizon can be described by
 an infinite collection of $(1+1)$ dimensional free massless partial wave modes, each propagating
 in a spacetime with the effective metric given by  $r-t$ sector of the full spacetime metric
 (\ref{1.2.5}), i.e
 \begin{equation}
 ds^2 =g_{\mu\nu}dx^{\mu} dx^{\nu}= f(r)dt^2 -\frac{1}{f(r)}dr^2 \label{1.2.9} 
\end{equation}
with $\mu,\nu = t,r$. This is a kind of dimensional reduction, of the field theory under consideration, from $(3+1)$ to $(1+1)$ dimensions\footnote{See appendix of this chapter for a detailed discussion of dimensional reduction.}.
Now we further split each partial wave  into left moving and right moving parts. This splitting
is always possible. Consider for example the free massless complex scalar field $\Phi(t,r)$ satisfying
 the Klein-Gordon equation
\begin{equation}
 \nabla_{\mu}\nabla^{\mu}\Phi(t,r) = 0 \label{1.2.10}
\end{equation}
and a similar equation with $\Phi$ replaced by its complex conjugate $\Phi^{*}$. Upon
transforming to null coordinates, defined as 
\begin{eqnarray}
 u &=& t - r_{*} \ ; \ v = t+r_{*} \nonumber\\
 \frac{dr}{dr_{*}} &=& f(r) \label{1.2.11}
\end{eqnarray}
the equation (\ref{1.2.10}) becomes 
\begin{equation}
 \partial_{u}\partial_{v}\Phi(u,v) = 0~. \label{1.2.12}
\end{equation}
The general solution for (\ref{1.2.12}) can be taken as 
\begin{equation}
 \Phi(u,v) = \Phi^{R}(u) + \Phi^{L}(v) \label{1.2.13}
\end{equation}
where $\Phi^{R}(u)$ and $\Phi^{L}(v)$ are right moving and left moving modes, satisfying
\begin{eqnarray}
 \partial_{v}\Phi^{R} &=& 0 \ ; \ \partial_{u}\Phi^{R} \ne 0 \nonumber \\
 \partial_{u}\Phi^{L} &=& 0 \ ; \ \partial_{v}\Phi^{L} \ne 0~.\label{1.2.14} 
\end{eqnarray}
 Similar analysis holds for $\Phi^{*}$. 
  
     Since the horizon is a null hypersurface, all the left moving ($\Phi^{L}$) modes at the horizon
 cannot classically affect the theory outside the horizon.  In other
 words, in equation (\ref{1.2.13}) we have $\Phi^{L}(v) =0$ and hence the field $\Phi(u,v)$
 possesses definite handedness (in our case it is right handed). \\
{\it Hawking fluxes for the Reissner-Nordstrom black hole :}\\

  We now compute the Hawking charge and energy fluxes coming from the Reissner-
 Nordstrom black hole by using the covariant expressions for the gauge and gravitational anomalies.
 First, let us consider the charge flux. We denote the expectation value of
 the covariant current very near the outer event horizon by
 $\langle J^{\mu}_{(H)} \rangle$. This covariant current satisfies the $(1+1)$ dimensional chiral covariant gauge anomaly  \cite{fujikawabook,bertlmannbook}. For the right-handed fields it is given by (\ref{1.15})
\begin{equation}
 \nabla_{\mu}\langle J^{\mu}_{(H)}\rangle = -\frac{e^2}{4\pi} \bar\epsilon^{\mu\nu}F_{\mu\nu} \label{1.2.15}
\end{equation}
where $\bar\epsilon^{\mu\nu}$ is an antisymmetric tensor defined in (\ref{1.9a}).
 For the effective $(1+1)$ dimensional
Reissner-Nordstrom metric (\ref{1.2.9}) the left hand side of (\ref{1.2.15}) becomes
\begin{eqnarray}
 \nabla_{\mu}\langle J^{\mu}_{(H)} \rangle &=& \frac{1}{\sqrt{-g}}\partial_{\mu}(\sqrt{-g}\langle J^{\mu}_{(H)}\rangle)= \partial_{r} \langle J^{r}_{(H)}\rangle \label{1.2.16}
\end{eqnarray}
while the right hand side of (\ref{1.2.15}) is 
\begin{equation}
 -\frac{e^2}{4\pi}\bar\epsilon^{\mu\nu}F_{\mu\nu} = \frac{e^2}{2\pi} \partial_{r}A_{t}~.\label{1.2.17}
\end{equation}
Here we have used the fact that $\sqrt{-g}=1$ for the metric (\ref{1.2.9})\footnote{An example where $\sqrt{-g}\ne1 $ is discussed in the section-\ref{coanomaly5}.}. Hence the equation (\ref{1.2.15}) now reads
\begin{equation}
 \partial_{r}\langle J^{r}_{(H)}\rangle = \frac{e^2}{2\pi}\partial_{r}A_{t}~.\label{1.2.18}
\end{equation}
 The solution for the above equation is given by
\begin{equation}
 \langle J^{r}_{(H)}(r) \rangle = c_{H} + \frac{e^2}{2\pi}[A_{t}(r)-A_{t}(r_{+})] \label{1.2.19}
\end{equation}
where $c_{H}$ is an integration constant. This is the expression for the chiral covariant 
 current. By construction, (\ref{1.2.19}) is valid only in the region near to the event horizon. 

     Next, we consider the theory away from the event horizon. Theory away from the horizon is 
 still $(3+1)$ dimensional (since the dimensional reduction procedure is  valid only in the vicinity
 of the horizon). Also, in this region, both the modes, left and right handed, are present which
 makes the theory anomaly free. Consequently, the $(3+1)$ dimensional current, denoted by
 $\langle J^{a}_{(4)} \rangle$ satisfy the usual conservation law
\begin{equation}
 \nabla_{a}\langle J^{a}_{(4)} \rangle =\frac{1}{\sqrt{-\gamma}}\partial_{a}(\sqrt{-\gamma}
\langle J^{a}_{(4)}\rangle)= 0 \label{1.2.20}
\end{equation}
where $\gamma$ is determinant of the full Reissner-Nordstrom metric $\gamma_{ab}$ given in 
(\ref{1.2.5}) and $a \in {t,r,\theta,\phi}$. Now since the the current $\langle J^{a}_{(4)}\rangle$
 only  depends upon the radial coordinate (since the metric is static and spherically symmetric), equation (\ref{1.2.20}) becomes
\begin{equation}
 \partial_{r}(r^2 \sin\theta \langle J^{r}_{(4)}(r)\rangle) + [\langle J^{\theta}_{(4)}(r)\rangle +\langle J^{\phi}_{(4)}(r)\rangle]r^2 \cos\theta  = 0~. \label{1.2.21}
\end{equation}
We now define the effective $(1+1)$ dimensional current $\langle J^{a}_{(o)}\rangle$ corresponding to
$3+1$ dimensional one \cite{isowilczek,isowilczekPRD} as 
\begin{equation}
 \langle J^{a}_{(o)}\rangle  = \int \  d\theta d\phi \ \sin\theta \ r^2 \langle J^{a}_{(4)}
\rangle ~.\label{1.2.22} 
\end{equation}
Then by integrating (\ref{1.2.21}) over the angular degrees of freedom and using
(\ref{1.2.22}), we arrive at
\begin{equation}
 \partial_{r}\langle J^{r}_{(o)}\rangle = 0~. \label{1.2.23}
\end{equation}
The solution of (\ref{1.2.23}) is given by
\begin{equation}
 \langle J^{r}_{(o)}\rangle = c_{o} \label{1.2.24} 
\end{equation}
where $c_{o}$ is an integration constant. We would like to point out that by construction $\langle J^{r}_{(o)} \rangle$
is an integrated current and hence it gives the amount of current passing through the spatial 
hypersurface defined by $\theta$ and $\phi$. As we shall see below the Hawking charge flux 
is related to the radial component $\langle J^{r}_{(o)}\rangle$. 

    Now we write the total current $\langle J^{\mu}\rangle$ as a sum of two contributions from the two regions - the
 region near to the horizon ranging from  $[r_{+},r_{+}+\epsilon]$ and the other region ranging from
 $[r_{+}+\epsilon,\infty]$. Then we have  
\begin{equation}
 \langle J^{\mu}\rangle = \langle J^{\mu}_{(o)}\rangle \Theta(r-r_{+}-\epsilon) + 
\langle J^{\mu}_{(H)}\rangle H(r) \label{1.2.25}
\end{equation}
where $\Theta(r-r_{+}-\epsilon)=1$ for $r>r_{+}+\epsilon$ and otherwise zero while, 
$H(r)$ is the top hat function given by $H(r)=1-\Theta(r-r_{+}-\epsilon)$. 
Taking the covariant divergence of $\langle J^{\mu} \rangle$ the Ward identity becomes
\begin{eqnarray}
 \nabla_{\mu}\langle J^{\mu}\rangle&=& \partial_{r} \langle J^{r} \rangle \nonumber\\
&=& \partial_{r}\langle J^{r}_{(o)}\rangle \Theta(r-r_{+}-\epsilon) + 
\partial_{r}\langle J^{r}_{(H)}\rangle  H(r)\label{1.2.26}\\
&& + [\langle J^{r}_{(o)}\rangle - \langle J^{r}_{(H)}\rangle]\delta(r-r_{+}-\epsilon)~. \nonumber 
\end{eqnarray}
By using the conservation relations for $\langle J^{r}_{(H)}\rangle$ and $\langle J^{r}_{(o)}\rangle$ given in 
(\ref{1.2.18}) and (\ref{1.2.23}) respectively, we get
\begin{eqnarray}
 \partial_{r}\langle J^{r}\rangle &=& \partial_{r}\left(\frac{e^2}{2\pi}A_{t}H\right) + [\langle J^{r}_{(o)} \rangle - \langle J^{r}_{(H)}\rangle + \frac{e^2}{2\pi}A_{t}]\delta(r-r_{+}-\epsilon)~. \label{1.2.27}
\end{eqnarray}
To make the the current anomaly free the first term must be cancelled by quantum effects of
the classically irrelevant left moving modes. This is the Wess-Zumino term induced by
these modes near the horizon. Effectively it implies a redefinition of the current
 as 
\begin{equation}
\langle J'^{r}\rangle = \langle J^{r}\rangle - \frac{e^2}{2\pi}A_{t}H(r)~. \label{1.2.28} 
\end{equation}
Since the cancellation of the anomaly occurs in the arbitrary small region near the horizon,
it is expected that the redefinition of $\langle J^{r}\rangle$(\ref{1.2.28}) should not
affect the current conservation (\ref{1.2.23}) valid far away from the horizon. 
Indeed, the explicit appearance of the top hat function $H(r)$ in (\ref{1.2.28}) assures
that in the region far away from the horizon  $\langle J'^{r}(r)\rangle = \langle J^{r}(r)\rangle$. 
Once we take into account quantum effects of the left moving modes the current (\ref{1.2.28}) becomes anomaly free provided the coefficient of the delta function in (\ref{1.2.27})
 vanishes,  leading to the relation
\begin{equation}
 \langle J^{r}_{(o)}\rangle = \langle J^{r}_{(H)} \rangle - \frac{e^2}{2\pi}A_{t}(r)~. \label{1.2.29} 
\end{equation}
Substituting (\ref{1.2.19}) and (\ref{1.2.24}) in (\ref{1.2.29}) gives
a relation among the integration constants $c_{o}$ and $c_{H}$
\begin{equation}
 c_{o} = c_{H} - \frac{e^2}{2\pi}A_{t}(r_{+})~.\label{1.2.30}
\end{equation}
The coefficient $c_{H}$ is fixed by imposing a boundary condition
 \footnote{In chapter-\ref{chap:boundarycondition} we will elaborate on
 the meaning and interpretation of this boundary condition.}
 requiring the vanishing of the covariant current
 (\ref{1.2.19}) at the horizon i.e
\begin{equation}
 \langle J^{r}_{(H)}(r=r_{+})\rangle = 0 ~.\label{1.2.31}
\end{equation}
 Using (\ref{1.2.31}) in (\ref{1.2.19}) we get $c_{H} = 0$. 
The other constant $c_{o}$ is now obtained from (\ref{1.2.30})
\begin{equation}
 c_{o} = - \frac{e^2}{2\pi}A_{t}(r_{+})~.\label{1.2.32}
\end{equation}
Thus, the integrated flux $\langle J^{r}_{(o)}\rangle$ given in (\ref{1.2.24}) now reads
\begin{equation}
 \langle J^{r}_{(o)}\rangle = c_{o} = - \frac{e^2}{2\pi}A_{t}(r_{+}) = \frac{e^2 Q}{2\pi r_{+}}~.\label{1.2.33} 
\end{equation}

     Now we focus our attention on the gravity sector. In the region near to the horizon
the theory is $(1+1)$ dimensional and chiral. Such a $(1+1)$ dimensional chiral theory 
is anomalous with respect to the general coordinate invariance. Consequently, in the region
near the horizon, the covariant divergence of the energy-momentum tensor will satisfy
either the consistent or the covariant gravitational anomaly. We denote the covariant 
chiral energy-momentum tensor for the charged scalar field by $\langle {T^{\mu}}_{\nu{(H)}}\rangle$. 
This energy-momentum tensor satisfies the covariant gravitational anomaly \cite{witten,bertlmann}
 and for the right moving fields it is given by (\ref{1.19})
\begin{equation}
 \nabla_{\mu}\langle {T^{\mu}}_{\nu{(H)}}\rangle = \frac{1}{96\pi}\bar\epsilon_{\nu\mu}\nabla^{\mu}R = \mathcal{A}_{\nu}~.
\label{1.2.34}
\end{equation}
Here $R$ is the Ricci scalar corresponding to the $(1+1)$ dimensional metric (\ref{1.2.9}).
 For the Reissner-Nordstrom black hole background, however, we have to take into account
 the effect of $U(1)$ gauge field leading to the modification in (\ref{1.2.34}).
 The corresponding anomalous Ward identity for the covariantly regularized energy-momentum
 tensor is then given by
\begin{equation}
 \nabla_{\mu}\langle{T^{\mu}}_{\nu {(H)}}\rangle = \mathcal{A}_{\nu} +  F_{\mu\nu}\langle J^{\mu}_{(H)}\rangle \label{1.2.35} 
\end{equation}
where $\langle J^{\mu}_{(H)}\rangle$ is given by (\ref{1.2.19}). The first term in the above expression 
 represents the covariant gravitational anomaly and it is purely a quantum effect while  
 the second one is the classical Lorentz force term which arises due to the effect of gauge field
  on the charged matter. Here we would like to point out that since the current 
 $\langle J^{\mu}_{(H)}\rangle$ itself is anomalous one might envisage the possibility of an additional term in (\ref{1.2.35}) proportional to the  gauge anomaly. 
 Indeed this happens in the Ward identity for consistently regularized objects
 \cite{isowilczek}. Such a term is ruled out here because there is no such covariant
 piece with the correct dimensions, having single a free index \cite{shailesh1}. Further, we observe that
 for the metric (\ref{1.2.9}) radial component of $\mathcal{A}_{\nu}$ vanishes. Thus    
 the covariant gravitational anomaly (\ref{1.2.34}) is purely timelike.
 Substituting $\nu=t$ in (\ref{1.2.34}) we obtain
 \begin{equation}
 \mathcal{A}_{t} = \frac{1}{96\pi}f \partial_{r}R ~.
 \label{1.2.36}
\end{equation}
For the metric (\ref{1.2.9}) the explicit form for the Ricci scalar is 
\begin{equation}
 R = \partial_{r}^{2}f = f''~. \label{1.2.37}
\end{equation}
 Substituting (\ref{1.2.37}) in (\ref{1.2.36}), we get
\begin{equation}
 \mathcal{A}_{t} = \partial_{r}N^{r}_{t} \label{1.2.38} 
\end{equation}
where $N^{r}_{t}$ is given by
\begin{equation}
 N^{r}_{t} = \frac{1}{192 \pi}\left(2ff'' - f'^{2}\right)~. \label{1.2.39}
\end{equation}
Taking $\nu=t$ component of (\ref{1.2.35}) and then using (\ref{1.2.19}, \ref{1.2.38}), yields 
\begin{eqnarray}
 \nabla_{\mu}\langle {T^{\mu}}_{t(H)}\rangle &=& \partial_{r}\langle {T^{r}}_{t(H)}\rangle \nonumber\\
 &=& \partial_{r}\left[\frac{e^2}{4\pi}(A_{t}^2(r) - 2A_{t}(r)A_{t}(r_{+})) + N^{r}_{t}(r)\right]~. \label{1.2.40}
\end{eqnarray}
Solving the above equation we get the form for energy-momentum tensor in the vicinity of the horizon 
 \begin{equation}
 \langle {T^{r}}_{t(H)}\rangle = a_{H} + \left[\frac{e^2}{4\pi}(A_{t}^2(r) - 2A_{t}(r)A_{t}(r_{+})) + N^{r}_{t}(r)\right]
\Big |_{r_{+}}^{r} \label{1.2.41}
\end{equation} 
where $a_{H}$ is an integration constant. 

    Now let us consider the theory away from the event horizon. In this region of spacetime,
 the theory is $3+1$ dimensional and anomaly free. Consequently, the covariant 
 divergence of  $3+1$ dimensional energy-momentum tensor, denoted as $\langle T^{ab}_{(4)}\rangle$
 satisfies the usual Lorentz force law
\begin{equation}
 \nabla_{a}\langle {T^{a}}_{b(4)}\rangle = F_{ab}\langle J^{a}_{(4)}\rangle ~.\label{1.2.42}
\end{equation}
For the static spherically symmetric metric (\ref{1.2.5}), $b=t$ component of the above equation 
is given by
\begin{equation}
 \partial_{r}(r^2 \sin\theta \langle {T^{r}}_{t(4)}\rangle) + r^2 \cos\theta \langle {T^{\theta}}_{t(4)}\rangle
 = F_{rt}\langle J^{r}_{(4)}\rangle~.
\label{1.2.43}
\end{equation}
Further using (\ref{1.2.22}) and its tensorial analog
\begin{equation}
 \langle {T^{a}}_{b(o)}\rangle = \int \ d\theta d\phi \  r^2 \sin\theta \ \langle {T^{a}}_{b(4)}\rangle \label{1.2.44}
\end{equation}
the equation (\ref{1.2.43}), after performing angular integrations, reduces to
\begin{equation}
 \partial_{r}\langle {T^{r}}_{t(o)}\rangle = (\partial_{r}A_{t})\langle J^{r}_{(o)}\rangle~. \label{1.2.45} 
\end{equation}
By substituting the known expression for $\langle J^{r}_{(o)}\rangle$, given by (\ref{1.2.33}), in the above 
 equation and then performing the integral, yields the solution   
\begin{equation}
 \langle {T^{r}}_{t(o)}(r)\rangle = a_{o} - \frac{e^2}{2\pi}A_{t}(r_{+})A_{t}(r)~. \label{1.2.46}
\end{equation}
As before, writing the energy-momentum tensor as a sum of two combinations
\begin{equation}
\langle {T^{r}}_{t}\rangle = \langle {T^{r}}_{t(o)}\rangle \Theta(r-r_{+}-\epsilon) +
\langle {T^{r}}_{t(H)}\rangle H(r) \label{1.2.47}
\end{equation}
we find
\begin{eqnarray}
 \nabla_{\mu}\langle {T^{\mu}}_{t}\rangle &=&\partial_{r}\langle {T^{r}}_{t}\rangle = -\frac{e^2}{2\pi}A_{t}(r_{+})\partial_{r}A_{t}(r)
+ \partial_{r}\left[\left(\frac{e^2}{4\pi}A_{t}^2+N^{r}_{t}\right)H\right] \nonumber\\
&& + \left(\langle {T^{r}}_{t(o)}\rangle - \langle {T^{r}}_{t(H)}\rangle +\frac{e^2}{4\pi}A_{t}^2 + N^{r}_{t}\right)\delta(r-r_{+}-\epsilon)~.
\label{1.2.48}
\end{eqnarray}
The first term is a classical effect coming from the Lorentz force. The 
 second term has to be cancelled
by the quantum effect of the left moving modes. As before, it implies the existence of a 
Wess-Zumino term modifying the energy-momentum tensor as 
\begin{equation}
 \langle {T'^{\mu}}_{t}\rangle = \langle {T^{\mu}}_{t}\rangle - \left[\left(\frac{e^2}{4\pi}A_{t}^2+N^{r}_{t}\right)H\right]
\label{1.2.49}
\end{equation}
which is anomaly free provided the coefficient of the delta function vanishes. This gives a
relation among the integration constants $a_{o}$ and $a_{H}$
\begin{equation}
 a_{o} = a_{H} + \frac{e^2}{4\pi}A_{t}^2(r_{+}) - N^{r}_{t}(r_{+}) \label{1.2.50}
\end{equation}
where $a_{H}$ is now fixed by boundary condition\footnote{See earlier footnote.}
 requiring that the covariant energy-momentum tensor
vanishes at the horizon, i.e
\begin{equation}
 \langle {T^{r}}_{t(H)}(r=r_{+})\rangle = 0 \label{1.2.51} ~.
\end{equation}
Using (\ref{1.2.51}) in (\ref{1.2.41}) gives $a_{H} = 0$. Then from (\ref{1.2.50}),
we have
\begin{equation}
 a_{o} = \frac{e^2}{4\pi}A_{t}^2(r_{+}) - N^{r}_{t}(r_{+})~. \label{1.2.52}
\end{equation}
 The Hawking flux is given by the asymptotic $(r\rightarrow \infty)$ 
limit of anomaly free energy momentum tensor. Substituting $a_{o}$ (\ref{1.2.50})
 in $\langle {T^{r}}_{t(o)}\rangle$ (\ref{1.2.46}) and then taking its asymptotic limit,
 we get the expression for energy-momentum flux of the charge particles emitted from the
 horizon 
\begin{equation}
 \langle {T^{r}}_{t(o)}(r\rightarrow \infty)\rangle = a_{o} = \frac{e^2}{4\pi}A_{t}^2(r_{+}) - N^{r}_{t}(r_{+})~.\label{1.2.53}
\end{equation}
Since $f(r_{+})=0$ we find from (\ref{1.2.39}) that
\begin{equation}
 N^{r}_{t}(r_{+}) = -\frac{f'^{2}(r_{+})}{192\pi}~. \label{1.2.54}
\end{equation}
Further, by using the known expressions 
\begin{equation}
 \kappa = \frac{2\pi}{\beta} = \frac{f'(r_{+})}{2} \label{1.2.55}
\end{equation}
for the surface gravity $\kappa$ and $\beta$ the inverse of Hawking temperature $T_{H}$,
we write (\ref{1.2.53}) into a more familiar form
\begin{equation}
 \langle {T^{r}}_{t(o)}(r\rightarrow \infty)\rangle = a_{o} = \frac{e^2 Q^2}{4\pi r_{+}^2} + \frac{\pi}{12 \beta^2}~. \label{1.2.56}  
\end{equation}
This is the expression for the energy-momentum flux obtained
 from the covariant anomaly method. 

 Further, since the basic structure of the covariant
 anomalous gauge and gravitational Ward identities 
(\ref{1.2.15},\ref{1.2.35}),
 apart from the coupling constant, is identical both 
for complex scalar field and fermionic field, 
the results given in (\ref{1.2.33}) and (\ref{1.2.56}) 
would remain unchanged if one uses the fermionic field instead of 
 the complex scalar field.  

    Now we compare our findings (\ref{1.2.33}) and (\ref{1.2.56}),
with the fluxes of Hawking radiation coming from the Reissner-Nordstrom black hole. 
Hawking radiation spectrum  is given by the Bose distribution 
 \begin{equation}
  N^{\pm}_{b}(\omega) = \frac{1}{e^{\beta(\omega\pm \mu)} - 1} \label{1.2.57}
 \end{equation}
in the case of bosons, and the Fermi-Dirac distribution
\begin{equation}
 N^{\pm}_{f}(\omega) = \frac{1}{e^{\beta(\omega\pm \mu)} + 1} \label{1.2.58}
\end{equation}
for fermions. Here $\mu = \frac{eQ}{r_{+}}$ is the chemical potential \cite{hawking}.
 $N^{\pm}_{b}$ and $N^{\pm}_{f}$ correspond to the distribution of particles with charge $\pm e$.
In the following we calculate the flux in the case of fermions in order to avoid the problem
of superradiance present in the bosonic case \cite{gibbonssuprad}.

     The charge flux of fermionic particles is given by
\begin{equation}
 \langle J^{r}\rangle = e \int_{0}^{\infty} \ \frac{d\omega}{2\pi}\left[N^{-}_{f}(\omega) - N^{+}_{f}(\omega)\right]~.\label{1.2.59} 
\end{equation}
After substituting (\ref{1.2.58}) in the above and performing the integral,
we get
\begin{equation}
 \langle J^{r}\rangle = \frac{e}{2\pi\beta}\ln \Big |\frac{1+e^{\beta \mu}}{1+e^{-\beta \mu}}\Big |~. 
\label{1.2.60}
\end{equation}
Expanding the right hand side of (\ref{1.2.60}) about $\beta=0$,  finally yields 
\begin{equation}
\langle J^{r}\rangle = \frac{e^2 Q}{2\pi r_{+}} \label{1.2.61}~. 
\end{equation}
Similarly, the flux of energy-momentum tensor is given by
\begin{equation}
 \langle {T^{r}}_{t}\rangle = \int_{0}^{\infty} \ \frac{d\omega}{2\pi} \omega \left[N^{-}_{f}(\omega) + N^{+}_{f}(\omega)\right]~.
\label{1.2.62}
\end{equation}
Using (\ref{1.2.58}) in (\ref{1.2.62}) and following similar steps as before,
yields
\begin{equation}
 \langle {T^{r}}_{t}\rangle = \frac{e^2 Q^2}{4\pi r_{+}^{2}} + \frac{\pi}{12 \beta^2}~. \label{1.2.63}
\end{equation}
The results (\ref{1.2.33}, \ref{1.2.56}), derived from the covariant anomaly
cancellation mechanism coincide with (\ref{1.2.61}, \ref{1.2.63}). 
Anomalies in the covariant current and energy-momentum tensor, present in the region near 
to the horizon, are compensated by the charge and energy-momentum flux emitted from
the Reissner-Nordstrom black hole. Note that the flux computed either from the 
 cancellation of covariant anomaly or from the consistent anomaly \cite{robwilczek,isowilczek} 
matches with the pure thermal flux of the blackbody radiation. 
The actual Hawking spectrum is obtained by propagating the emission, originated from
 the horizon, through the centrifugal barrier. Effectively, the particles 
 emitted from the horizon back scatter before reaching to spatial infinity. 
 This leads to the corrections in the standard Hawking flux (\ref{1.2.61}, \ref{1.2.63}).
The resulting radiation observed at infinity is that of a $(3+1)$ dimensional gray body at the 
Hawking temperature \cite{zelnikov}. These gray body factors are not accounted in our 
derivation of Hawking flux. 
\section{\label{coanomaly3}Comparison between consistent and covariant \\
 anomaly approach}
In the last section we saw that conditions imposed by the 
vanishing of covariant gauge and gravitational anomalies 
 are capable of giving the expressions for Hawking charge
 and energy-momentum flux. 
Similar results, based on the use of consistent anomalies,
 were already derived in \cite{isowilczek}. Therefore it is 
important to compare the efficiency of both, the consistent \cite{isowilczek}
 and the covariant anomaly approaches. 
     
   We would emphasize this point by considering the gauge sector of the total
Hawking radiation. In the analysis of \cite{isowilczek}, 
the explicit form for the universal component of the consistent current 
$\langle \tilde J^{r}_{(H)}\rangle$, in the region near to 
the horizon, was obtained by solving the consistent gauge anomaly
\cite{fujikawabook,bertlmannbook} for the right handed fields (\ref{1.9}) 
\begin{equation}
 \nabla_{\mu}\langle \tilde J^{\mu}_{(H)}\rangle = \partial_{r}\langle \tilde J^{r}\rangle= \frac{e^2}{2\pi}\partial_{r}A_{t}~.
\label{1.3.1}
\end{equation}
Solution of (\ref{1.3.1}) gives us the form for consistent gauge current 
\begin{equation}
 \langle\tilde J^{r}_{(H)}\rangle = \tilde c_{H} + \frac{e^2}{4\pi}[A_{t}(r)-A_{r}(r_{+})]~. \label{1.3.2}
\end{equation}
Like before, the integration constant $\tilde c_{H}$ is fixed by demanding 
the vanishing of the covariant current $\langle J^{\mu}_{(H)}\rangle$ at the horizon.
 However, since the current (\ref{1.3.2}) is 
the consistent current, knowledge of the local counterterm which
connects the consistent and covariant current becomes necessary. 
The explicit relation among these two anomalous currents is given by 
(\ref{1.16})
\begin{equation}
 \langle J^{r}_{(H)}\rangle = \langle\tilde J^{r}_{(H)}\rangle + \frac{e^2}{4\pi}A_{t}\bar\epsilon^{tr}~.
\label{1.3.3}
\end{equation}
Now by implementing the covariant boundary condition $(\langle J^{r}_{(H)}(r=r_{+})\rangle=0)$ in the above
expression and then using (\ref{1.3.2}), yields
\begin{equation}
 \tilde c_{H} = -\frac{e^2}{4\pi}A_{t}(r_{+}) \label{1.3.4}~.
\end{equation}
This fixes the form for the consistent current (\ref{1.3.2}) 
completely \cite{isowilczek}. On the other hand, in the region away 
from the horizon there is no anomaly in the gauge current. 
Hence the issue of covariant and consistent current 
would not arise. Consequently, the Hawking charge flux, which is measured 
 at asymptotic infinity, computed either from the covariant 
or  consistent anomaly, agrees. It is therefore clear from the above discussion
 that unlike in the covariant anomaly based approach, where only the 
boundary condition on the covariant current and the expression for covariant 
gauge anomaly were essential inputs, the consistent anomaly method, apart from
the boundary condition and the consistent gauge anomaly, also requires the knowledge of local
counterterms relating the different currents. The computation of this local counterterm,
 although quite straightforward for the case of gauge current, becomes cumbersome in the case
of higher rank tensors.  For example, the universal component of 
 the covariant energy-momentum tensor $\langle {T^{r}}_{t(H)}\rangle$ 
is related to its consistent counterpart $\langle \tilde {T}^{r}{}_{t(H)}\rangle$ 
by the Bardeen polynomial (\ref{1.20}) \cite{bardeenzumino,bertlmann}. For
 the metric (\ref{1.2.9}) the relation among the two types of energy-momentum
 tensor is given by 
\begin{equation}
 \langle {T^{r}}_{t(H)}\rangle = \langle \tilde {T}^{r}{}_{t(H)}\rangle + 
\frac{1}{192\pi}[ff'' -2f'^{2}]~. \label{1.3.5}
\end{equation}
All these computations become essential in  dealing with the consistent 
anomaly cancellation approach to compute the fluxes of Hawking radiation 
\cite{isowilczek}. In this sense, therefore, the covariant anomaly cancellation 
approach, presented in the last section, is more efficient and lucid compared 
to the one given in \cite{isowilczek}.   
\section{\label{coanomaly4}Generalized consistent anomaly and flux}
Here we show that the conclusions of \cite{robwilczek,isowilczek} 
remain unaffected by taking the general form of the consistent anomaly
 (\ref{1.11}) and (\ref{1.18}), rather than the minimal
 expressions (\ref{1.8}, \ref{1.17}) considered in \cite{robwilczek,isowilczek}.
 Instead of repeating the analysis
 of \cite{robwilczek,isowilczek} we just point out the reasons for this robustness.

   For static configuration and for the specific choice of the gauge
potential $(A_{r}=0)$, it is clear that the normal parity terms
in (\ref{1.11}) vanishes. Likewise the normal parity term in the counterterm
 relating the generalized consistent current $\langle \tilde J'^{\mu}\rangle$
(\ref{1.11}) and the covariant current $\langle J^{\mu}\rangle$ would
 also vanish since only the $\mu=r$ component in $\langle\tilde J'^{\mu}\rangle$ is relevant.
Hence, effectively the same structures of the consistent gauge anomaly
 and the counterterm relating the consistent and covariant currents,
 as used in \cite{isowilczek}, are valid. Since these were the two basic
 inputs, the results concerning the charge flux associated with Hawking radiation
 remain intact.

     Identical conclusions also hold for the gravity sector. Although
 not immediately obvious, a little algebra shows that the normal parity
 term in $\mathcal{\tilde A'}_{t}$ (\ref{1.18}) vanishes. Hence the energy-momentum
 flux obtained by solving the generalized consistent anomaly (\ref{1.18})
 agrees with the one given in \cite{isowilczek}.
\section{\label{coanomaly5}Application to stringy black holes}
In section-\ref{coanomaly2}  we gave a derivation, based on the 
 treatment of covariant gauge and gravitational anomalies,
 of Hawking radiation from the Reissner-Nordstrom black
 hole. As discussed earlier, such an approach can be applicable to
 a variety of black hole spacetimes. In this section we will adopt our
 covariant anomaly mechanism to discuss Hawking radiation from the 
 black hole backgrounds that arise in string theory. 

 Many interesting properties and physics of black holes
 can be acquired by studying other types of black hole solutions
 that may appear in theories aiming to generalize Einstein's
 theory of gravity. Of particularly interest are considering those
 black hole configurations that emerge as classical solutions of
 the low energy limit of supersting theory \cite{polchinski}. 
These black hole solutions  were discovered when the dilaton 
scalar field was included
 in the Einstein-Maxwell theory \cite{gibbonsGHS2,ghs}. 
This dilaton field couples in
 a nontrivial fashion to the metric and the gauge field.
 When the electromagnetic (gauge) field vanishes, the only static and spherically
 symmetric black hole solution is the Schwarzschild black hole with a 
 constant dilaton field. However, the dilaton field cannot remain constant
 in the presence of a gauge field (i.e in the case of charged black holes).
 This fact separates the stringy black holes from the Reissner-Nordstrom
 black holes. Nevertheless, these black holes also satisfy the usual 
 laws of black hole thermodynamics.

 Another example, motivated by string theory, is that of the charged non-extremal five-dimensional
black hole in string theory. This black hole solution 
is obtained from a specific $D$-brane configuration 
and often called the non-extremal $D1-D5$ black hole \cite{D1D5-1}
 (for review see \cite{stringBHreview}). This background is
particularly interesting since it is related to various black hole solutions by taking
different limits on parameters appearing in the background of five-dimensional
 Reissner-Nordstrom and Schwarzschild solutions, 
six-dimensional black string solution \cite{D1D5-2}, black
five-brane solution \cite{D1D5-3}, dyonic black string solution \cite{D1D5-4}.
All these black hole solutions possess similar thermodynamical
properties. Naturally, it would be an interesting exercise to implement
the covariant anomaly cancellation approach to these stringy black holes.
First we will study the Hawking radiation from the Garfinkle-Horowitz-Strominger
 (GHS) \cite{ghs} black hole. In the extremal limit of this black hole solution,
 we compute the flux of Hawking radiation by using our covariant anomaly approach.
 Next, we calculate the Hawking flux for D1-D5 nonextremal black holes.  
\subsection{\label{coanomaly5.1}Hawking Fluxes from GHS black hole}
The Garfinkle-Horowitz-Strominger (GHS) black hole is a member of a family of solutions
to low-energy string theory described by the $3+1$ dimensional action \cite{ghs}
 (in the string frame)
\begin{equation}
S_{GHS} = \int d^{4}x \sqrt{-\gamma} e^{-2\phi}
\left[ -R -4(\nabla\phi)^2 + F^{2}\right]
\label{2.1.1}
\end{equation}
where $\phi$ is the dilaton field and $F_{\mu\nu}$ is the Maxwell field
associated with a $U(1)$ subgroup of $E_{8}\times E_{8}$ 
or ${\it{Spin}(32)}/Z_{2}$. Its charged black hole solution is given by
\begin{equation}
ds^{2}_{string} = \gamma_{ab}dx^{a}dx^{b}= f(r)dt^{2} - \frac{1}{h(r)}dr^{2} - r^{2}d\Omega
\label{2.1.2} 
\end{equation}
where,
\begin{eqnarray}
f(r) &=& \left( 1- \frac{2Me^{\phi_{0}}}{r}\right)
\left(1 - \frac{Q^{2}e^{3\phi_{0}}}{Mr}\right)^{-1}\nonumber\\
h(r) &=& \left(1-\frac{2Me^{\phi_{0}}}{r}\right)
\left(1- \frac{Q^{2}e^{3\phi_{0}}}{Mr}\right)
\label{2.1.3}
\end{eqnarray}
with $\phi_{0}$ being the asymptotic constant value of the dilaton field,
and $Q$ the magnetic charge. We consider the case when 
$Q^{2}<2e^{-2\phi_{0}}M^{2}$ for which the above 
metric describes a black hole with an event horizon 
situated at \cite{ghs}  
\begin{eqnarray}
r_{H}&=&2Me^{\phi_{0}}~.
\label{2.1.4}
\end{eqnarray}
 Now  consider scalar fields propagating on the background (\ref{2.1.2}).
In the near horizon region, with the aid of dimensional reduction procedure, we 
can effectively describe scalar fields with a metric given by the by the 
``$r-t$" sector of the full spacetime metric (\ref{2.1.2}), i.e
\begin{equation}
 ds^2 = f(r)dt^2 - \frac{1}{h(r)}dr^2 \label{2.1.5}
\end{equation}
 where the metric functions $f(r)$ and $h(r)$ are given in (\ref{2.1.3}).
It is important to realize that, unlike the Reissner-Nordstrom black hole, the $1+1$
dimensional effective metric for GHS black hole (\ref{2.1.5}) has nontrivial 
determinant, i.e $\sqrt{-g} = \sqrt{-g_{tt}g_{rr}}\ne 1$. In fact
(\ref{2.1.2}) represents the most general spherically symmetric
metric. We shall see below that the  anomaly method works
without any difficulty for the GHS black hole also. 

As mentioned earlier, the theory near the horizon is $1+1$ dimensional
 and chiral and the energy-momentum tensor $\langle {T^{\mu}}_{\nu(H)}\rangle$ 
in this region satisfies the covariant gravitational anomaly 
(\ref{1.2.34}). For the metric (\ref{2.1.5}) the radial 
 component of the covariant anomaly $\mathcal{A}_{\nu}$ vanishes.
 Consequently, the covariant anomaly (\ref{1.2.34}) is timelike. This  
 feature of the covariant anomaly is common for all stationary
 black holes. The point is that for stationary black holes,
 the Ricci scalar $R$ corresponding to the $r-t$ sector of the 
full $3+1$ dimensional metric is time independent. On the 
 other hand, the radial component of the covariant anomaly
 $\mathcal{A}_{\nu}$ (\ref{1.2.34}), due to
 the presence of $\bar\epsilon_{\nu\mu}$, would  always be
 proportional to the time derivative of $R$. Hence 
 for the stationary black holes we have $\mathcal{A}_{r} =0$.
 Our task  now is to compute $\mathcal{A}_{t}$. 
 For the metric (\ref{2.1.5}), expression for the Ricci scalar
 $R$ is given by 
\begin{equation}
 R = \frac{f'' h}{f} + \frac{f'h'}{2f} - \frac{f'^{2}h}{2f^2}~. \label{2.1.6}
\end{equation}
Then taking the $\nu=t$ component of (\ref{1.2.34}), we find
\begin{equation}
 \nabla_{\mu}\langle{T^{\mu}}_{t(H)}\rangle = \frac{1}{\sqrt{-g}}\partial_{r}(\sqrt{-g}
\langle{T^{r}}_{t(H)}\rangle) = \frac{1}{\sqrt{-g}}\partial_{r}N^{r}_{t} \label{2.1.9} 
\end{equation}
with 
\begin{equation}
N^{r}_{t} = \frac{1}{96\pi}\left( hf'' + \frac{f'h'}{2} - \frac{f'^{2}h}{f}\right). \label{2.1.8}
\end{equation}
 Solution of (\ref{2.1.9}) yields,
\begin{equation}
\langle{T^{r}}_{t(H)}\rangle = \frac{1}{\sqrt{-g}}
\left[a_{H} + N^{r}_{t}(r) - N^{r}_{t}(r_{H})\right]
\label{2.1.10}
\end{equation}
where, $a_{H}$ is an integration constant. 
 
In the region far away from the horizon the theory is $3+1$ dimensional
 and anomaly free. Hence the energy-momentum tensor $\langle{T^{a}}_{b(4)}\rangle$
 is conserved, i.e 
\begin{equation}
 \nabla_{a}\langle{T^{a}}_{b(4)}\rangle = 0~.\label{2.1.11}
\end{equation}
The effective $1+1$ dimensional anomaly free energy-momentum 
 tensor, defined in (\ref{1.2.44}), then satisfies,
\begin{equation}
\partial_{r}(\sqrt{-g}\langle{T^{r}}_{t(o)}\rangle) = 0~, \label{2.1.12}
\end{equation}
which, after integrating, yields
\begin{equation}
\langle{T^{r}}_{t(o)}\rangle = \frac{a_{o}}{\sqrt{-g}}
\label{2.1.13}
\end{equation}
where $a_{o}$ is an integration constant. 

Writing the total energy-momentum tensor $\langle {T^{r}}_{t}\rangle$ 
as a sum of two combinations and following the same reasoning given
 in section-\ref{coanomaly2}, we arrive at
\begin{equation}
 \langle{T^{r}}_{t(o)}\rangle - \langle{T^{r}}_{t(H)}\rangle + \frac{N^{r}_{t}(r)}{\sqrt{-g}} = 0~.
\label{2.1.16}
\end{equation}
Substituting (\ref{2.1.10}) and (\ref{2.1.13}) in the above equation, yields 
\begin{equation}
a_{o} = a_{H} - N^{r}_{t}(r_{H})
\label{2.1.17}
\end{equation}
 The integration constant $a_{H}$ can be fixed by imposing the covariant
 boundary condition (\ref{1.2.51}).
This gives $a_{H}=0$. 
Hence the total flux of the energy-momentum tensor is given by
\begin{equation}
\langle {T^{r}}_{t}(r\rightarrow \infty)\rangle = a_{o} = -N^{r}_{t}(r_{H}) = \frac{1}{192\pi} f'(r_{H})h'(r_{H})= \frac{\pi}{12}T_{H}^{2}
\label{2.1.18}
\end{equation}
where $T_{H}$ is the Hawking temperature given by \cite{ghs}
\begin{equation}
T_{H} = \frac{1}{8\pi Me^{\phi_{0}}}~.
\label{2.1.20}
\end{equation}
Equation (\ref{2.1.18}) represents the energy-momentum flux of  
Hawking radiation coming from the GHS black hole \cite{ghs}. 
Moreover, the flux (\ref{2.1.18}) which is obtained by the present approach
 is compatible with that calculated by using the 
 consistent anomaly \cite{vega}.\\ 
{\it Extremal limit:}\\
At the extremality, i.e. when $Q^{2}=2e^{-2\phi_{0}}M^{2}$, 
the GHS black hole solution (\ref{2.1.2}, \ref{2.1.3}) becomes
\begin{eqnarray}
ds^{2}&=& dt^{2}-\left(1-\frac{2Me^{\phi_{0}}}{r}\right)^{-2}dr^2
-r^{2}d\Omega~.
\label{2.1.21}
\end{eqnarray}
It is easy to check that in this case the Hawking temperature
vanishes. Indeed, by substituting
\begin{eqnarray}
 f(r) &=& 1 \nonumber \\
 h(r) &=& \left(1-\frac{2Me^{\phi_{0}}}{r}\right)^{-2} \label{2.1.22}
\end{eqnarray}
in (\ref{2.1.18}), we see that the energy flux vanishes.

\subsection{\label{coanomaly5.2}Hawking radiation from D1-D5 non-extremal black hole }

\noindent As another example of the covariant anomaly approach,
we consider a non-extremal five dimensional black hole which originates
as a brane configuration in Type IIB superstring
theory compactified on $S^1 \times T^4$\cite{D1D5-1}. 
The configuration relevant to the present case is
composed of D1-branes wrapping $S^1$, D5-branes wrapping $S^1 \times
T^4$ and momentum modes along $S^1$.  The solution of the Type IIB
supergravity corresponding to this configuration is a supersymmetric
background known as the extremal five-dimensional D1-D5 black hole
having zero Hawking temperature. Hence, in order to consider Hawking radiation
we study the non-extremal D1-D5 black hole. 

   The ten-dimensional supergravity background
corresponding to the non-extremal D1-D5 black hole has the following
form in the string frame \cite{D1D5-1}:
\begin{eqnarray}
ds^2_{10}&=& f_1^{-1/2} f_5^{-1/2}( - h f_n^{-1} dt^2 
               + f_n ( dx_5 + (1-\tilde{f}_n^{-1}) dt)^2 )\nonumber \\
&&+ f_1^{1/2} f_5^{-1/2} (dx_6^2+ \cdots + dx_9^2)
+ f_1^{1/2} f_5^{1/2} ( h^{-1} dr^2 + r^2 d \Omega_3^2)\nonumber \\
e^{-2 \phi} &=& f_1^{-1} f_5 \quad, \quad
C_{05} = \tilde{f}_1^{-1} -1 \label{2.2.1}\\
F_{ijk} &=& \frac{1}{2} \epsilon_{ijkl} 
 \partial_l \tilde{f}_5 \quad,\quad
i,j,k,l=1,2,3,4\nonumber
\end{eqnarray}
 $x_5$ is the periodic coordinate along $S^{1}$ with period $2\pi R_{5}$
while $x_6,\cdots, x_9$ are periodic coordinates on 
$T^4$. Each of $x_{6},\cdots, x_{9}$ is periodically identified with $2\pi V^{1/4}$,
 where $V$ is the volume of $T^4$. $F$ is the three-form field strength 
of the RR 2-form gauge potential $C$, $F = dC$. 
Also various functions appearing in the above background
are functions of coordinates $x_1, \dots, x_4$ given by
\begin{eqnarray}
h &=& 1 - \frac{r_0^2}{r^2} \quad, \quad
f_{1,5,n} = 1+ \frac{r_{1,5,n}^2}{r^2} \nonumber \\
\tilde{f}_{1,n}^{-1} &=& 1 - \frac{r_0^2 \sinh \alpha_{1,n} 
 \cosh \alpha_{1,n}}{r^2} f_{1,n}^{-1}  \label{2.2.2} \\
r^2_{1,5,n} &=& r_0^2 \sinh^2 \alpha_{1,5,n} \quad, \quad
r^2= x_1^2 + \cdots + x_4^2~. \nonumber
\end{eqnarray}
This black hole solution is parameterized by six independent quantities
 $\alpha_{1,5,n}, r_{0}, R_{5}$ and $V$. Functions $h$, $f_{1,5,n}$,
are harmonic functions representing the non-extremality and the
presence of D1, D5, and momentum modes respectively.

   Dimensional reduction of (\ref{2.2.1}) along $S^1 \times
T^4$ following the procedure of \cite{D1D5-5} yields the Einstein metric
of the non-extremal five-dimensional black hole as \footnote{This is 
the dimensional reduction of a background metric and it is different 
from the the dimensional reduction of the scalar field action in the near
horizon limit, outlined in the appendix of this chapter.}
\begin{equation}
ds^2_5 = -\lambda^{-2/3} h~dt^2 + 
\lambda^{1/3} (h^{-1} dr^2 + r^2 d \Omega_3^2) 
\label{2.2.3}
\end{equation}
where $\lambda$ is defined by
\begin{equation}			
\lambda = f_1 f_5 f_n ~.\label{2.2.4}
\end{equation}
The event horizon $r_H$ of this black hole geometry is located at
\begin{equation}
r_H = r_0 ~.
\label{2.2.5}
\end{equation}

   Apart from the metric, 
the dimensional reduction gives us three kinds
of gauge fields.  The first one is the Kaluza-Klein gauge field
$A^{(K)}_\mu$ coming from the metric and the second one, say
$A^{(1)}_\mu$, basically stems from $C_{\mu 5}$ (here $\mu =
0,1,2,3,4$). From the background (\ref{2.2.1}), the first two gauge
fields are obtained as
\begin{eqnarray}
A^{(K)} &=& -(\tilde{f}_n^{-1} - 1 )dt \quad, \quad
A^{(1)} = ( \tilde{f}_1^{-1} - 1 )dt ~.
\label{2.2.6}
\end{eqnarray}
Unlike these gauge fields which are one-form in nature, the last one
is the two-form gauge field $A_{\mu\nu}$, originating from
$C_{\mu\nu}$ whose field strength is given by the expression of $F$
in (\ref{2.2.1}).

    Now if we consider a free complex scalar field in the black
hole background (\ref{2.2.3}) and (\ref{2.2.6}) and
perform a partial wave decomposition of $\Phi$
in terms of the spherical harmonics, then it can be shown
that the action near the horizon becomes (see the Appendix-\ref{appendix1A})
\begin{eqnarray}
S[\Phi]&=&-\sum_l \int dtdr~ r^3~\lambda^{1/2}~
  \Phi^{*}_{l}(t, r) 
  \left( - \frac{1}{f} (\partial_t -i A_t)^2 
          + \partial_r f \partial_r 
  \right)
  \Phi_{l}(t, r) 
\label{2.2.7}
\end{eqnarray}
where, $a$ is the collection of angular quantum numbers
of the spherical harmonics, $A_t = e_1 A^{(1)}_t + e_K A^{(K)}_t$ 
and 
\begin{equation}
 f(r) = \frac{\lambda^{1/2}}{h}~. \label{2.2.7a}
\end{equation}
This action describes an infinite set of massless
two-dimensional complex scalar fields in the following effective background : 
\begin{equation}
ds^2 = -f(r)dt^2 + \frac{1}{f(r)}dr^2\quad,
\quad \phi = r^3 \lambda^{1/2}
\label{2.2.8}
\end{equation} 
\begin{equation}
A_{t}(r) = - \frac{e_{1}r_{o}^2 \sinh \alpha_{1}
\cosh\alpha_{1}}{r^2 + r_{1}^2}
+ \frac{e_{k}r_{o}^2 \sinh \alpha_{n}
\cosh\alpha_{n}}{r^2 + r_{n}^2}
\label{2.2.9}
\end{equation}
where $\phi$ is the two-dimensional dilaton field. 

Once we mapped the theory from $3+1$ to $1+1$ dimensions, the computation of 
 the charge and energy-momentum flux will follow exactly in a similar way  
 illustrated in section-\ref{coanomaly2}. To calculate the charge flux we
 first note that since there are two kinds of $U(1)$ gauge symmetries, we have 
two $U(1)$ gauge currents $J_{\mu}^{(1)}$ and $J_{\mu}^{(K)}$ corresponding
to $A_{\mu}^{(1)}$ and $A_{\mu}^{(K)}$ respectively. The form for
covariant gauge anomaly for these two currents are identical
in nature and given by (\ref{1.2.15}) (but with two different 
coupling constants $e_{1}$ and $e_{K}$ corresponding to $A_{\mu}^{(1)}$
 and $A_{\mu}^{(K)}$, respectively). Solving the covariant anomalous gauge
 Ward identities (\ref{1.2.15}) for two different gauge currents
 $\langle J_{\mu(H)}^{(1)}\rangle$ and $\langle J_{\mu(H)}^{(K)}\rangle$, yields
\begin{eqnarray}
\langle J^{(1)r}_{(H)}\rangle &=& c_{H}^{(1)} +\frac{e_1}{2\pi}
\left[A_{t}(r) - A_{t}(r_{H})\right]\label{2.2.13}\\
\langle J^{(K)r}_{(H)}\rangle &=& c_{H}^{(K)} +\frac{e_K}{2\pi}
\left[A_{t}(r) - A_{t}(r_{H})\right]~. \label{2.2.13a}
\end{eqnarray}
 While, in the region away from the horizon, solving the usual conservation
 equations for $\langle J_{\mu(o)}^{(1)}\rangle$ and $\langle J_{\mu(o)}^{(K)}\rangle$, we get
 \begin{eqnarray}
\langle J^{(1)r}_{(o)}\rangle &=& c_{o}^{(1)}\label{2.2.12}\\
\langle J^{(K)r}_{(o)}\rangle &=& c_{o}^{(K)}\label{2.2.12a}~.
\end{eqnarray}
The relations among the integration constants $c^{(1)}_{o},c^{(1)}_{H}$ and 
$c^{(K)}_{o},c^{(K)}_{H}$ are obtain by splitting the total currents $\langle J^{(1)}_{\mu}\rangle$ 
 and $\langle J^{(K)}_{\mu}\rangle$ and demanding them to be anomaly free. After doing this, we get
\begin{equation}
c^{(1)}_{o} = c^{(1)}_{H} - \frac{e_1}{2\pi}A_{t}(r_{H}) \ ; \  
c^{(K)}_{o} = c^{(K)}_{H} - \frac{e_K}{2\pi}A_{t}(r_{H})~. 
\label{2.2.16}
\end{equation}
As before, the integration constants $c^{(1)}_{H}$ and $c^{(1)}_{H}$ are fixed by 
imposing the covariant boundary condition, leading to $c^{(1)}_{H}=c^{(K)}_{H} = 0$.
Finally, the Hawking charge flux corresponding to $\langle J^{(1)r} \rangle$
 and $\langle J^{(K)r} \rangle$ is given by 
\begin{eqnarray}
c^{(1)}_{o} &=& -\frac{e_1}{2\pi}A_{t}(r_{H}) = 
\frac{e_1}{2\pi}(e_1 \tanh \alpha_1 - e_K \tanh \alpha_n) \label{2.2.19}\\
c^{(K)}_{o} &=& -\frac{e_K}{2\pi}A_{t}(r_{H}) = 
\frac{e_K}{2\pi}(e_1 \tanh \alpha_1 - e_K \tanh \alpha_n)~. \label{2.2.19a}
\end{eqnarray}
Hence total charge flux is given by
\begin{equation}
c_{o} = c^{(1)}+c^{(K)}_{o} = -\frac{e}{2\pi}A_{t}(r_{H}) 
= \frac{e}{2\pi}(e_1 \tanh \alpha_1 - e_K \tanh \alpha_n)
\label{2.2.20}
\end{equation}
where, $e = e_1 + e_K$.

Now consider the  energy-momentum flux. Since we have an external gauge field,
the energy-momentum tensor will not satisfy the conservation law 
even at classical level. Rather
it gives rise to the Lorentz force law,
$\nabla_{\mu}\langle{T^{\mu}}_{\nu}\rangle = F_{\mu\nu}\langle J^{\mu}\rangle$
 \footnote{Note that here $\langle J^{\mu}\rangle =\langle J^{(1)\mu}\rangle + \langle J^{(K)\mu}\rangle$.}.
Hence the corresponding expression for the anomalous 
Ward identity for covariantly
regularized quantities is given by (\ref{1.2.35})
\begin{equation}
\nabla_{\mu}\langle {T^{\mu}}_{\nu(H)}\rangle = F_{\mu\nu}\langle J^{\mu}\rangle + \mathcal{A}_{\nu}
\label{2.2.21}
\end{equation} 
where, $\mathcal{A}_{\nu}$ is the two-dimensional 
gravitational covariant anomaly (\ref{1.2.34}).
In the region outside the horizon, 
there is no anomaly and hence the Ward identity reads 
\begin{equation}
\nabla_{\mu}\langle {T^{\mu}}_{\nu(o)}\rangle = \partial_{r}\langle {T^{r}}_{t(o)}\rangle 
= F_{rt}\langle J^{r}_{(o)}\rangle
\label{2.2.22}
\end{equation}
Using (\ref{2.2.12}, \ref{2.2.12a}), the above equation can be solved as
\begin{equation}
\langle {T^{r}}_{t(o)}\rangle = a_{o} + c_{o}A_{t}(r)
\label{2.2.23}
\end{equation}
where, $a_{o}$ is an integration constant.

 In the region near to the horizon, the Ward identity (\ref{2.2.21}) reads
\begin{equation}
\partial_{r}\langle{T^{r}}_{t(H)}\rangle = F_{rt}\langle J^{r}_{(H)}\rangle + \partial_{r}N^{r}_{t} \label{2.2.24}
\end{equation}
where, $N^{r}_{t}$ is given by (\ref{1.2.39}). 
Now substituting $\langle J^{r}_{(H)}\rangle = \langle J^{(1)r}_{(H)}\rangle + 
\langle J^{(K)r}_{(H)}\rangle$ and using (\ref{2.2.13}, \ref{2.2.13a}), we get
\begin{equation}
\langle{T^{r}}_{t(H)}\rangle = a_{H} + \int^{r}_{r_{H}} dr 
~\partial_{r}\left[c_o A_{t}+\frac{e}{4\pi}A_{t}^2 +  N^{r}_{t}\right]. 
\label{2.2.25}
\end{equation}
Following the same procedure as given in the gauge part, we arrive at
the relation
\begin{equation}
a_o = a_H + \frac{e}{4\pi}A_{t}^2(r_{H}) - N^{r}_{t}(r_{H})~. \label{2.2.26}
\end{equation} 
Implementing (as before) the boundary condition that covariant energy momentum 
tensor vanishes at the horizon, fixes $a_{H}$ to be zero.
Therefore, $a_{o}$ is given by
\begin{equation}
a_{(o)} =  \frac{e}{4\pi}A_{t}^2 (r_{H}) - N^{r}_{t}(r_{H})~.\label{2.2.27}
\end{equation}
Finally, substituting $a_{o}$ (\ref{2.2.27}) in (\ref{2.2.23}) and then
 taking its asymptotic infinity limit, we get the expression for the energy-momentum flux
\begin{eqnarray}
\langle {T^{r}}_{t}(r\rightarrow \infty)\rangle &=& \frac{e}{4\pi}A_t^2(r_H) - N^r_t(r_+) \nonumber\\
&=& \frac{e}{4\pi} (e_1 \tanh \alpha_1 - e_K \tanh \alpha_n)^2
 +\frac{\pi}{12} T_H^2 
\label{2.2.28}
\end{eqnarray}
where,
\begin{eqnarray}
T_H&=&\frac{1}{2\pi r_{0}\cosh\alpha_{1}\cosh\alpha_{5}\cosh\alpha_{n}}
\label{2.2.29}
\end{eqnarray}
is the Hawking temperature. 
Expression (\ref{2.2.28}) agrees with the energy-momentum flux
 from black body radiation with two chemical
potentials $\mu_{1} = e_{1}\tan\alpha_{1}$ and 
$\mu_{2}=e_{2}\tan\alpha_{2}$ \cite{shin}. 

\section{\label{coanomaly6}Discussions}
Using  only the expressions of covariant gauge and gravitational anomalies we have
 given a derivation of Hawking radiation from the Reissner-Nordstrom
 black hole. The quantum field theory in the region near the event horizon is anomalous.
 In this region the expressions for covariant anomalous current and energy-momentum tensor 
 were obtained by solving the covariant anomalous gauge and gravitational Ward identities.
  On the other hand, far away from the horizon the theory is anomaly free. The 
 corresponding expressions for current and energy-momentum tensor were then
 computed by solving the usual conservation laws. Both, the anomalous as well as the
 anomaly free expressions for currents/energy-momentum
 tensors contains the arbitrary constants of integration. A relation among these constants
 were obtained by demanding that the total current/energy-momentum tensor, which is 
 a sum of two combinations comming from the regions near to and away from the event horizon,
 must be anomaly free. This condition fixes one of the two integration constants.
 The remaining integration constant were then fixed by imposing the covariant boundary condition;
 namely, the vanishing of the covariant anomalous current and energy-momentum tensor at event
 horizon. This condition together with the fact that the total current/energy-momentum
 tensor is anomaly free, fixes the form for current/energy-momentum tensor away from the event
 horizon i.e $\langle J^{r}_{(o)}\rangle/ \langle {T^{r}}_{t(o)}\rangle$, completely. Finally,
 by extrapolating this anomaly free current/energy-momentum tensor gives the charge/energy-momentum
 fluxes of Hawking radiation. 

  Our approach of deriving the Hawking fluxes, uses only
 covariant expressions.  Neither the consistent anomaly nor the counterterm
 relating the different currents, which were essential inputs in \cite{isowilczek},
 were required. Consequently our analysis was economical and, we feel,
 also conceptually clean. It should be pointed out that the charge (energy) flux
 is identified with $\langle J^{r}_{(o)}\rangle$ ($\langle {T^{r}}_{t(o)} \rangle$) 
which are the expressions for the currents (energy-momentum tensors)
 exterior to the horizon. Here these currents are anomaly free, implying that
 there is no difference between the covariant and consistent expressions.
 Actually the germ of the anomaly lies in this difference \cite{rabin1,rabin2}.
 Hence it
 becomes essential, and not just desirable, to obtain the same flux in terms
 of the covariant expressions. In other words, the Hawking flux must yield  
identical results
 whether one uses the consistent or the covariant anomalies. But
the boundary condition must be covariant.
This is consistent with the universality
 of the Hawking radiation and gives further credibility to the
 anomaly based approach.

   It was shown \cite{robwilczek, isowilczek}, performing a partial wave decomposition, that 
physics near the horizon is described by an infinite collection of massless $(1+1)$ dimensional fields,
each partial wave propagating in spacetime with a metric given by the $`r- t'$ sector of the complete
spacetime metric (\ref{1.2.5}). This simplification, which effects a dimensional reduction from $d$-dimensions
to $d=2$ is also exploited here. It is however noted that greybody factors have not been included. In that 
case dimensional reduction will not yield the real Hawking radiation for $d>2$. For instance, it is known
\cite{zelnikov} that in case of $d=4$ reduction to $d=2$ and keeping only the $s$-wave $(i.e. \  l=0)$ reduces the 
Hawking flux with respect to its $2-D$ value. 

    There are distinct advantages of the covariant anomaly approach. First,
 all the expressions are manifestly covariant.
 Also, the functional forms for the covariant anomalies are unique, being governed solely
 by the gauge (diffeomorphism) transformation properties. This is
 not so for consistent anomalies. They can and do have normal parity
 terms, apart from the odd parity ones. In fact, the special
 property (\ref{1.10}) of two dimensions yields a natural form for this
 anomaly which has normal parity terms. Our observation, that the
 results of \cite{robwilczek,isowilczek} are still valid, lend further credibility 
 to this scheme of deriving Hawking radiation. 

    Further, we apply the covariant anomaly approach to compute the fluxes
 of Hawking radiation from stringy black holes. In particular, 
 we discuss the Hawking radiation
from GHS and five dimensional non-extremal D1-D5 blackhole.
For GHS blackhole, the energy-momentum flux 
was obtained when $Q^2 < 2Me^{-2\phi_{o}}$.
At extremality there is no energy flux and hence Hawking temperature
is zero. 
In the case of $D1-D5$ blackhole, fluxes of electric charge flow 
and energy-momentum tensor were obtained. The resulting fluxes 
are the same as that of the two dimensional black body 
radiation at the Hawking temperature. The present approach based on covariant
 anomalies has also been applied to various other black hole geometries (\cite{ourprdcite1}-\cite{ourprdcite4}).
       



\begin{subappendices}
\chapter*{Appendix}
\section{\label{appendix1A}Dimensional reduction}
\renewcommand{\theequation}{3A.\arabic{equation}}
\setcounter{equation}{0}  

Consider matter fields moving on the $3+1$ dimensional static spherically symmetric black hole background. In general, equations of motion governing the matter fields are complicated to solve. However,
if we consider the theory near the event horizon, the action and hence the equations of motion, get 
simplified. The point is that in the near horizon region, matter field can be decomposed into 
an infinite collection of free, massless fields propagating on the $r-t$ sector of the original
$3+1$ dimensional metric. Consequently, in the near horizon limit the matter field action
becomes conformally invariant. In this appendix, we explicitly show the dimensional reduction 
procedure for the neutral and charged (complex) scalar fields. 
 
      Let us first consider the scalar field $\Phi$ moving on the $3+1$ dimensional static Schwarzschild
black hole background represented by the metric 
\begin{equation}
 ds^2 = \gamma_{ab}dx^{a}dx^{b}=f(r)dt^2 - \frac{1}{f(r)}dr^{2} - r^2(d\theta^{2} + \sin^2 \theta d\phi^2) \label{A1}
\end{equation}
 where $f(r)$ is the metric function and, for the Schwarzschild black hole, is 
given in  terms of mass $M$ of the black hole as
\begin{equation}
 f(r) = 1 - \frac{2M}{r}~. \label{A2} 
\end{equation}
The event horizon $r_{h}$ is defined by $f(r=r_{h})=0$. 
The action for scalar field moving on the background (\ref{A1}) is given by
\begin{equation}
S = - \int \ d^{4}x \sqrt{-\gamma} \left[\Phi \Box \Phi + m^{2}\Phi^{2}\right] \label{A3}   
\end{equation}
where $m$ is the mass of scalar field $\Phi$ while  $\sqrt{-\gamma} = \sqrt{-det \gamma_{ab}}$. 
In order to study the near horizon behavior of the theory, it is convenient to use the tortoise
coordinate $r_{*}$, defined as
\begin{eqnarray}
 \frac{dr}{dr_{*}} = f(r)~. \label{A4}  
\end{eqnarray}
Then, for the Schwarzschild metric (\ref{A1}), we have
\begin{equation}
 r_{*} = \int \ dr \frac{1}{\left(1-\frac{2M}{r}\right)} = r + 2M \ln \vert \frac{r}{2M} -1\vert ~. \label{A5}
\end{equation}
Thus, in the tortoise coordinate event horizon is located at $r_{*} = -\infty$. The metric (\ref {A1}),
 in $(t,r_{*},\theta,\phi)$ coordinates now reads
\begin{equation}
 ds^2 = f(r(r_{*}))(dt^{2} - dr_{*}^2) - r^{2}(r_{*})(d\theta^2 + \sin^2\theta d\phi^2)~. \label{A6}
\end{equation}
and for the metric (\ref{A6}) we have $\sqrt{-\gamma} = f(r(r_{*}))r^2(r_{*}) \sin\theta$. 
Then by writing the scalar field action (\ref{A3}) in the tortoise coordinate, we obtain,
\begin{eqnarray}
 S &=& - \int \ dt dr_{*}d\theta d\phi \  \sin\theta \ \Phi\left[r^2(r_{*})(\partial^{2}_{t} - \partial^{2}_{r_{*}}) - 2r(r_{*})f(r(r_{*}))\partial_{r_{*}}\right]\Phi \label{A7}\\
&& + \int \ dt dr_{*}d\theta d\phi \ f(r(r_{*}))\sin\theta  \Phi \left[\frac{1}{\sin^{2}\theta}\partial^{2}_{\phi}
 + \cot\theta \partial_{\theta} + \partial^{2}_{\theta} + r^{2}(r_{*})m^{2}\right]\Phi~. \nonumber
\end{eqnarray}
Since the black hole metric is spherically symmetric, we decompose $\Phi(t,r_{*},\theta,\phi)$ in terms
of spherical harmonics 
\begin{equation}
 \Phi(t,r_{*},\theta,\phi) = \sum_{l,n} R_{l}(t,r_{*}) Y_{l,n}(\theta,\phi) \label{A8}  
\end{equation}
Substituting this ansatz in (\ref{A7}) and integrating over the angular variables, we get
\begin{eqnarray}
 S &=& - \sum_{l} \int \ dt dr_{*} \ r^{2}(r_{*}) \ R_{l}\left[\partial^{2}_{t} - \partial^{2}_{r_{*}} - \frac{1}{r(r_{*})}\partial_{r_{*}}f(r(r_{*})) \right] R_{l}\nonumber\\
&&  - \sum_{l} \int \ dt dr_{*} \ r^{2}(r_{*}) \ R_{l} f(r(r_{*}))\left(-\frac{l(l+1)}{r^{2}(r_{*})} + m^{2}\right) R_{l} ~.\label{A9}
\end{eqnarray}
Now we expand $f(r)$ about the horizon $(r=r_{h})$ as,
\begin{eqnarray}
 f(r) &=& f(r_{h}) + f'(r_{h})(r-r_{h}) +  \cdots ~. \label{A10}
\end{eqnarray}
 Keeping only the leading order term in the above expansion, the near horizon
expression for the metric coefficient $f(r)$ reads 
\begin{eqnarray}
 f(r) &\approx& f'(r_{h})(r-r_{h}) = 2\kappa (r-r_{h}) \label{A11}   
\end{eqnarray}
where $\kappa = \frac{f'(r_{h})}{2}$ is the surface gravity. Substituting (\ref{A11}) in 
(\ref{A4}) and performing the integral, we arrive at
\begin{equation}
 r \approx Ae^{2\kappa r_{*}} + r_{h}  \ ; \ A = constant ~.\label{A12}
\end{equation}
Finally, the equations (\ref{A11}) and (\ref{A12}) give
\begin{equation}
 f(r(r_{*})) \approx 2\kappa A e^{2\kappa r_{*}}~. \label{A13}
\end{equation}
Hence, near the horizon $f(r(r_{*}))$ decay exponentially fast (since $r_{*}\rightarrow -\infty$).
 By similar reasoning, the term proportional
to $\partial_{r_{*}}f(r(r_{*}))$ also vanishes exponentially. Therefore, in the near horizon
limit the action for the scalar field (\ref{A9}) becomes
\begin{equation}
 S \approx \sum_{l}\int \ dt dr_{*}  \ r^{2}_{h} \ R_{l} \left(\partial_{t}^2 - \partial_{r_{*}}^{2}\right)R_{l}~. \label{A14}
\end{equation}
Transforming back to the Schwarzschild coordinate $(t,r)$, yields
\begin{equation}
 S \approx \sum_{l}\int \ dt dr  \ r^{2}_{h} \ R_{l} \left(\frac{1}{f}\partial_{t}^2 - \partial_{r}(f\partial_{r})\right)R_{l}~. \label{A14a}
\end{equation}
The factor of $r^{2}_{h}$ in the action can be interpreted as a dilaton background
 coupled to the scalar field. 
 Thus, physics near the horizon can be effectively described by an infinite collection 
(for each value of $l$) of $(1+1)$ dimensional free massless scalar fields, each 
propagating in a $(1+1)$ dimensional spacetime given by the $t-r$ sector of the 
$3+1$ dimensional metric, that is
\begin{equation}
 ds^2 = g_{\mu\nu}dx^{\mu}dx^{\nu}=f(r)dt^2 - \frac{1}{f(r)}dr^2 \label{A15}  
\end{equation}
with ($\mu,\nu =t,r)$. 
In deriving the Hawking flux, by using the anomaly approach, we consider only
 one component among the infinite collection of scalar fields. Analysis for all other component
 is essentially the same. However, how to extract four dimensional information by summing
 over all $l$ values is still an open issue \cite{bonora1}.
 
 Next, we consider the charged (complex) scalar field moving on the $3+1$ dimensional
Reissner-Nordstrom black hole described by the metric (\ref{1.2.5}) and the gauge
 field $A_{b}$ (\ref{1.2.7}). The action for the charged scalar field is given by
\begin{equation}
 S_{cs} = \int \ d^{4}x \sqrt{-\gamma}\gamma^{ab}(\nabla_{a}-ieA_{a})\Phi(\nabla_{b}+ieA_{b})\Phi^{*}
\label{A16}
\end{equation}
 By writing (\ref{A16}) in the tortoise coordinate $r_{*}$
 appropriate for the Reissner-Nordstrom black hole, we get
\begin{eqnarray}
 S_{cs} &=& - \int \ dt dr_{*}d\theta d\phi \  \sin\theta \ \Phi^{*} \left[r^2(r_{*})(\partial^{2}_{t} - \partial^{2}_{r_{*}}) - 2r(r_{*})f(r(r_{*}))\partial_{r_{*}}\right]\Phi \label{A17}\nonumber\\
&& + \int \ dt dr_{*}d\theta d\phi \ f(r(r_{*}))\sin\theta  \Phi^{*} \left[\frac{1}{\sin^{2}\theta}\partial^{2}_{\phi}
 + \cot\theta \partial_{\theta} + \partial^{2}_{\theta} + r^{2}(r_{*})m^{2}\right]\Phi \nonumber\\
&& + \int \ dt dr_{*}d\theta d\phi \ f(r(r_{*}))\sin\theta \left[ieA_{t}[\Phi^{*}\partial_{t}\Phi
-(\partial_{t}\Phi^{*})\Phi] + e^{2}(A_{t}^2\Phi^{*}\Phi)\right] 
\end{eqnarray}
As before, by using the ansatz (\ref{A8}) (together with its complex conjugate) and  
then integrating over the angular variable, (\ref{A17}) reduces to
\begin{eqnarray}
 S_{cs} &=& \sum_{l}\int dt dr_{*} \ r^{2}(r_{*}) \ \left[|(\partial_{t} - ieA_{t})R_{l}|^2 - |\partial_{r_{*}}R_{l}|^2\right]
\nonumber\\
&& -\sum_{l}\int dt dr_{*}  \ r^{2}(r_{*}) \ f(r(r_{*}))\left(m^2 - \frac{l(l+1)}{r^{2}(r_{*})}\right)|R_{l}|^2 \label{A18}
\end{eqnarray}
In the near horizon limit, by dropping the terms proportional to $f(r(r_{*}))$, the action for the 
complex scalar field becomes
\begin{equation}
 S_{cs} \approx \sum_{l}\int dt dr_{*} \ r^{2}_{h} \ \left[|(\partial_{t} - ieA_{t})R_{l}|^2 - |\partial_{r_{*}}R_{l}|^2\right] \label{2dcomscalaction}
\end{equation}
As before, we interpret $r^{2}_{h}$ as the dilaton background coupled to the charged
 scalar field. Consequently, the $(3+1)$ dimensional charged scalar field can be 
considered as an infinite set of $d=2$ conformal fields near the horizon in $(t,r_{*})$ coordinates.
Any other field interaction terms can also be shown to be proportional to 
the damping factor $f(r(r_{*}))$ and hence they can be neglected in the near horizon limit. 
Similar analysis will also hold for the fermionic fields. Further, we would like
 to point out that the above discussion can be easily extended to the higher dimensional
 as well as non-spherical black holes \cite{dimreduction1,dimreduction2}.
          
\end{subappendices}

\chapter{\label{chap:chiralaction}Hawking Fluxes and Effective Actions}
In  chapter-\ref{chap:coanomaly} 
 we studied the relationship between the Hawking flux and  
 the covariant anomalies. An important aspect of our analysis was that 
 only covariant expressions are used throughout and consistent expressions
 were completely bypassed. The point is that since covariant boundary
 conditions, namely; the vanishing of covariant anomalous current
 (energy-momentum tensor) at the event horizon, are mandatory, it is
  rather conceptually clean and compact to discuss everything from the covariant
 point of view. The local counterterms relating the consistent and
 covariant expressions, which were essential in the approach based on
 the consistent anomalies \cite{isowilczek}, were not at all required
 in our approach. Consequently, the calculation of Hawking flux was
 simplified considerably. In both the approaches however, a splitting of space into 
 two different regions (near to and away from the horizon)
 using discontinuous step functions like $\Theta(r-r_{+}-\epsilon)$ and $H(r)$ was 
 essential to obtain the Hawking flux. This split, which enforces the use  
 of both the normal and anomalous Ward identities, also poses certain
 conceptual issues. Particularly, the definition of path integral in this
 context is not clear.

     In this chapter we provide an algorithm
 to compute the Hawking flux from a generic spherically symmetric
 black hole spacetime, based on the chiral effective actions defined  near
 to the event horizon, which only require the boundary conditions at the event horizon.
 This approach completely bypasses the use of discontinuous step functions and also solely depend on the properties 
 of the theory near the event horizon. We adopt the arguments given in 
 \cite{robwilczek,isowilczek} 
 and chapter-\ref{chap:coanomaly} which imply that effective field theories are two dimensional
 and chiral near the event horizon.   
 Then, exploiting the known structures of two dimensional 
 chiral effective action appropriately modified by
 the local counterterms \cite{leutwyler},
 the relevant expressions for the covariant currents
 and energy-momentum tensors are derived. These currents and energy-
momentum tensors are anomalous. Again, as before the 
 arbitrary constants appearing in the expressions of currents 
 and energy-momentum tensors are fixed by imposing the covariant
 boundary conditions at the event horizon. Further, we note that
 in the asymptotic limit, anomalies in the current and energy-momentum
 tensor vanish. Then the Hawking charge 
 and energy-momentum flux, which are measured at the asymptotic
 infinity, can be computed by  taking appropriately the asymptotic 
infinity limit of these chiral currents and energy-momentum tensor,
 respectively. The results obtain by this chiral effective action approach
 are in exact agreement with the that of obtained in previous chapter and also in
 \cite{isowilczek}.
    
  The chapter is arranged in the following manner.
 In section-\ref{chiralaction1} we introduce the chiral effective action.
 The currents and energy-momentum tensors following from the chiral effective action,
 suitably modified by a local counterterm, are obtained. These currents and energy-momentum
 tensors satisfy covariant gauge and gravitational anomalies, respectively
 \cite{witten,fujikawabook,bardeenzumino,bertlmannbook}. Using these expressions for 
 the covariant current and energy-momentum tensor and implementing the covariant
 boundary condition at the horizon, Hawking fluxes from the generic spherically
 symmetric charged black hole are derived in section-\ref{chiralaction2}. 
 Our results match exactly with the one obtained by solving the covariant anomalous
 Ward identities (see section-\ref{coanomaly2}).
 In section-\ref{chiralaction4} we
 implement the chiral effective action approach to study the Hawking flux
 for the Reissner-Nordstrom black hole in the presence of gravitational
 back reaction. The corrections to the Hawking charge and energy-momentum flux are
 also obtained. Finally, we conclude this chapter in section-\ref{chiralaction5}

\section{\label{chiralaction1} General setting and chiral effective action}

 Consider a general form of static spherically symmetric charged black hole represented by the metric,
\begin{equation}
ds^2 = \gamma_{ab}dx^{a}dx^{b}=f(r)dt^2 -\frac{1}{h(r)}dr^2 -r^2(d\theta^2 + \sin^2\theta d\phi^2) \label{3.1.1}
\end{equation}
where $f(r)$ and $h(r)$ are the metric coefficients. The gauge field is given by
\begin{equation}
 {\bf A} = A_{t}(r)dt~. \label{3.1.1a}
\end{equation}
 Since we are discussing static, spherically symmetric black hole solutions,
the above choice of gauge field is always possible. The outer horizon is given by   
\begin{equation}
f(r_{h}) = h(r_{h}) =0~. \label{3.1.2} 
\end{equation}
Here we consider the asymptotically Minkowski flat 
 black hole, i.e 
\begin{eqnarray}
f(r\rightarrow \infty)&=&h(r\rightarrow \infty) = 1 \ {\text {and}} \nonumber\\
f''(r\rightarrow \infty) &=& f'''(r\rightarrow\infty)= 
h''(r\rightarrow \infty)= h'''(r\rightarrow \infty)= 0~.\label{3.1.3}
\end{eqnarray}
Now consider fermionic (or complex scalar) fields propagating on this
background. It was shown in appendix-\ref{appendix1A} that,
 by using a dimensional reduction technique, 
the effective field theory near the event horizon becomes
 two dimensional with the metric given by the  $r-t$ section of (\ref{3.1.1})
\begin{equation}
ds^2=f(r)dt^2 -\frac{1}{h(r)}dr^2 ~. \label{3.1.4}
\end{equation} 
Note that $\sqrt{-g} = \sqrt{-det g_{\mu\nu}} = \sqrt{\frac{f}{h}} \ne1$
(unless $f(r) = h(r)$). On this two dimensional background, the modes
which are going in to the black hole (left moving modes) are lost and the effective theory become chiral. 

We now summarize, step by step, our methodology. For a two dimensional theory the 
expressions for the effective actions, whether anomalous (chiral) \cite{leutwyler} or
 normal \cite{polyakov,fabbri1,fabbri2,shapiro1},
 are known in the literature. 
For deriving the Hawking flux, only the form of the anomalous
 (chiral) effective action \cite{leutwyler},
 which describes the theory near the horizon, is required.
The currents and energy momentum tensors are  computed by
taking appropriate functional derivatives of this effective action. Next, the parameters appearing
in these solutions are fixed by imposing the vanishing of covariant currents (energy momentum
tensors) at the horizon. Once we have the complete form for current and
 energy-momentum tensor, the Hawking fluxes
 are  obtained from the asymptotic $(r\rightarrow {\infty})$ limits of the currents and energy momentum
tensors.   

 For the right handed Weyl fermion propagating in the presence of external
gravitational and $U(1)$ gauge field, the classical action is given by 
\begin{equation}
S[\Psi,\bar\Psi,A,g] = \int d^{2}x \ \sqrt{-g} \  \bar\Psi  \gamma^{\mu}\left(\partial_{\mu}-i\Gamma_{\mu}
-iA_{\mu}\frac{1 - \gamma_{5}}{2}\right) \Psi \label{3.1.4a}
\end{equation}
where $\Gamma_{\mu}$ is the spin connection given by \cite{leutwyler}
\begin{equation}
 \Gamma_{\mu} = -\frac{1}{2}e^{(a)}_{\mu}\bar \epsilon^{\alpha\beta}\partial_{\alpha}e_{(a)\beta}
\label{3.1.4b}
\end{equation}
and $\bar\epsilon^{\alpha\beta}$ is given by (\ref{1.9a}). The
Zweibein vectors $e^{(a)}_{\mu}$ and its inverse, defined by a relation 
$e^{\mu}_{(b)}e^{(a)}_{\mu} = \delta^{(a)}_{(b)}$,
 are connected to the metric \footnote{Here indices in the parenthesis are defined
with respect to flat spacetime.}
\begin{eqnarray}
 e^{\mu}_{(a)}e_{\mu (b)} &=& \eta_{(a)(b)}\nonumber\\
 e^{\mu}_{(a)}e^{\nu (a)} &=& g^{\mu\nu}~. \label{3.1.4c}
 \end{eqnarray}
Now we consider the quantization of $\Psi$ and $\bar\Psi$ in the presence of
 external gravitational and gauge fields. The corresponding quantum effective action   
 $\Gamma_{(H)}$ is given by \cite{leutwyler}

\begin{equation}
\Gamma_{(H)}= -\frac{1}{3} z(\omega) + z(A) \label{3.1.5}
\end{equation}
where 
\begin{equation}
z(v) = \frac{1}{4\pi}\int d^2x d^2y \epsilon^{\mu\nu}\partial_\mu v_\nu(x)
 \Delta^{-1}(x, y)\partial_\rho[(\epsilon^{\rho\sigma} +
 \sqrt{-g}g^{\rho\sigma})v_\sigma(y)]
\label{3.1.6}
\end{equation}
and $\Delta^{-1}$ is the inverse of Laplace-Beltrami operator $\partial_{\mu}(\sqrt{-g}g^{\mu\nu}\partial_{\nu})$, 
\begin{equation}
 \partial_{\mu}(\sqrt{-g}g^{\mu\nu}\partial_{\nu})\Delta^{-1}(x,y)
 = \delta(x-y)~. \label{3.1.7}
\end{equation}
Note that the effective action given in (\ref{3.1.6}) contain $\Delta^{-1}(x,y)$ and 
 hence it is non-local. The local form is obtain by introducing
  auxiliary fields. From a variation of this effective
 action the energy momentum tensor and the gauge current are computed.
 These are shown in the literature 
\cite{witten,fujikawabook,bardeenzumino,bertlmannbook,bertlmann,rabin1,rabin2}
 as consistent forms. To get their covariant forms in which we are interested,
 however, appropriate local polynomials have to be added.
 This is possible since energy momentum tensors and currents
 are only defined modulo local polynomials. Hence we have,
\begin{equation}
\delta \Gamma_{(H)} = \int d^2x  \sqrt{-g}\left( \frac{1}{2}\delta g_{\mu\nu} T^{\mu\nu} + \delta A_{\mu}J^{\mu}\right) 
+  l \label{3.1.8} 
\end{equation}
where the local polynomial is given by \cite{leutwyler},
\begin{equation}
 l = \frac{1}{4\pi}\int d^{2}x \ \epsilon^{\mu\nu}(A_{\mu}\delta A_{\nu} - 
 \frac{1}{3}w_{\mu}\delta w_{\nu} - \frac{1}{24}R \ e_{\mu}^{(a)}\delta e_{\nu(a)})~. \label{3.1.9}
\end{equation}
The covariant gauge  current $\langle J^{\mu} \rangle$ and the covariant 
energy momentum tensor $\langle T^{\mu\nu}\rangle$ 
 are read-off from the above relations as \cite{leutwyler}
\footnote{In this chapter, we have suppressed the suffix on $\langle J^{\mu} \rangle$
 and $\langle T^{\mu\nu}\rangle$ since most of the time  we will be using
 the near horizon expressions for current and energy-momentum tensor.},
 \begin{eqnarray}
\langle J^{\mu}\rangle &=& \frac{\delta \Gamma_{(H)}}{\delta A_{\mu}}= -\frac{e^2}{2\pi}D^{\mu}B \label{3.1.10}\\
\langle T_{\mu\nu}\rangle &=& \frac{\delta \Gamma_{(H)}}{\delta g_{\mu\nu}}\nonumber\\ 
 &=& \frac{e^2}{4\pi}\left(D_{\mu}B D_{\nu}B\right)
+\frac{1}{4\pi}\left(\frac{1}{48}D_{\mu}G D_{\nu}G 
-\frac{1}{24} D_{\mu} D_{\nu}G + \frac{1}{24} g_{\mu\nu}R\right)~.\label{3.1.11}
\end{eqnarray}
Note the presence of the chiral covariant derivative
 $D_{\mu}$ expressed in terms of the usual covariant
 derivative $\nabla_{\mu}$,
\begin{equation}
D_{\mu} = \nabla_{\mu} - \bar\epsilon_{\mu\nu}\nabla^{\nu} =
 -\bar\epsilon_{\mu\nu}D^{\nu}, \label{3.1.12}
\end{equation} 
The auxiliary fields $B(x)$ and $G(x)$, necessary to make the effective action (\ref{3.1.5}) local, are
defined as
\begin{eqnarray}
B(x) &=& \int \ d^2 y \ \sqrt{-g} \Delta^{-1}(x,y)\bar\epsilon^{\mu\nu}
\partial_{\mu}A_{\nu}(y) \label{3.1.13}\\
G(x) &=& \int \ d^2 y \ \sqrt{-g}\Delta^{-1}(x,y)R(y) \label{3.1.14}
\end{eqnarray}
By acting $\nabla_{\mu}\nabla^{\mu}$ on both sides of (\ref{3.1.13}) we get 
the differential equation for $B(x)$
\begin{eqnarray}
\nabla_{\mu}\nabla^{\mu}B(x) &=& \frac{1}{\sqrt{-g(x)}} \int \ d^{2}y \ 
\partial_{\mu}(\sqrt{-g}g^{\mu\nu}\partial_{\nu})\Delta^{-1}(x,y)
\epsilon^{\alpha\beta}\partial_{\alpha}A_{\beta}(y) \nonumber\\
&=& \frac{\epsilon^{\alpha\beta}}{\sqrt{-g(x)}}\int \ d^{2}y \  \delta(x-y)
\partial_{\alpha}A_{\beta}(y) \nonumber\\
&=& \bar\epsilon^{\alpha\beta}(x)\partial_{\alpha}A_{\beta}(x) \label{3.1.15}
\end{eqnarray}
where we have used (\ref{3.1.7}). Similarly, by operating 
$\nabla_{\mu}\nabla^{\mu}$ on both sides of (\ref{3.1.14}), we obtain
\begin{eqnarray}
\nabla_{\mu}\nabla^{\mu} G(x)&=& R(x)\label{3.1.16}
\end{eqnarray}    
where $R$ is the Ricci scalar, which for the metric (\ref{3.1.4}), is given by
\begin{equation}
R = \frac{f'' h}{f} + \frac{f'h'}{2f} - \frac{f'^{2}h}{2f^2}~. \label{3.1.17}
\end{equation}

Determination of the near horizon structures for the covariant current (\ref{3.1.10}) and
energy-momentum tensor (\ref{3.1.11}) hence eventually reduces to finding the 
 solutions of the differential equations (\ref{3.1.15}, \ref{3.1.16}). 

 Before proceeding further we provide some consistency checks and properties of the 
covariant current and energy-momentum tensor given in (\ref{3.1.10}, \ref{3.1.11}). 
The covariant divergence of current $\langle J^{\mu}\rangle$ and energy-momentum tensor 
$\langle T^{\mu\nu}\rangle$ satisfy the covariant Ward identities (\ref{1.2.15}, \ref{1.2.35}),
 respectively. For example, using (\ref{3.1.10}) and (\ref{3.1.13}) in (\ref{3.1.15}), we find
\begin{eqnarray}
 \nabla_{\mu}\langle J^{\mu}\rangle &=& -\frac{e^2}{2\pi}\nabla_{\mu}\nabla^{\mu}B \nonumber\\
&=& -\frac{e^2}{4\pi}\bar\epsilon^{\alpha\beta}F_{\alpha\beta}~. \label{3.1.18}
\end{eqnarray}
This is precisely the expression of covariant gauge anomaly (\ref{1.2.15}). Similarly,
 by using (\ref{3.1.11}) and (\ref{3.1.14}) in (\ref{3.1.16}) reproduces the 
 covariant anomalous gravitational Ward identity (in the presence of gauge field),
\begin{equation}
 \nabla_{\mu}\langle {T^{\mu}}_{\nu}\rangle = \langle J_{\mu} \rangle F^{\mu\nu} + \frac{1}{96\pi}
\bar\epsilon_{\nu\mu}\nabla^{\mu}R~. \label{3.1.18a}
\end{equation}
 
The trace of the covariant energy-momentum tensor is obtained by
 contracting  $\langle T^{\mu\nu} \rangle$ (\ref{3.1.11}) with 
 respect to the metric, and yields 
\begin{eqnarray}
\langle {T^{\alpha}}_{\alpha}\rangle &=&  \frac{e^2}{4\pi}\left(D^{\mu}B D_{\mu}B\right)
+\frac{1}{4\pi}\left(\frac{1}{48}D^{\mu}G D_{\mu}G 
-\frac{1}{24} D^{\mu} D_{\mu}G + \frac{1}{24}\delta^{\mu}_{\mu}R\right)~.\label{3.1.18b}  
\end{eqnarray}
Now by using (\ref{3.1.12}) we can easily see that terms like 
$D_{\mu}B D^{\mu}B, \ D_{\mu}G D^{\mu}G$ and $D^{\mu}D_{\mu}G$ drop out, leading 
to the chiral trace anomaly
\begin{equation}
 \langle {T^{\alpha}}_{\alpha}\rangle = \frac{R}{48\pi}~. \label{3.1.19}
\end{equation}
Note that the chiral theory (\ref{3.1.5}) has both a diffeomorphism 
anomaly (\ref{1.2.35}) and a trace anomaly (\ref{3.1.19}).
 This is distinct from the usual vector theory, represented by the Polyakov type action \cite{polyakov},
 where there is only trace anomaly $\langle {T^{\mu}}_{\mu}\rangle = \frac{R}{24\pi}$. No
 diffeomorphism anomaly exists. However, it is important to note that for a vector theory,
 it is possible to shift the trace anomaly to diffeomorphism anomaly by adopting
 a certain regularization prescription \cite{greenschwartz}.  
 
The covariant current (\ref{3.1.10}) and covariant energy-momentum tensor (\ref{3.1.11})
are chiral. Consequently, not all components of $\langle J^{\mu} \rangle$ and 
$\langle T^{\mu\nu}\rangle$ are independent. Let us first consider the  expression for
current (\ref{3.1.10}). Contraction on both sides of (\ref{3.1.10}) 
with respect to the antisymmetric tensor $\bar\epsilon_{\rho\mu}$ gives
\begin{eqnarray}
 \bar\epsilon_{\rho\mu}\langle J^{\mu} \rangle &=& -\frac{e^2}{2\pi}\epsilon_{\rho\mu}D^{\mu}B
=\frac{e^2}{2\pi}D_{\rho}B \nonumber\\ 
&=& -\langle J_{\rho} \rangle~. \label{3.1.20}
\end{eqnarray}
This is the chirality relation for the covariant current (\ref{3.1.10}). 
 In order to get a chirality relation for the covariant energy-momentum tensor,
 we first consider the contraction of  $\langle {T^{\rho}}_{\nu} \rangle$
  with $\bar\epsilon_{\mu\rho}$  
\begin{eqnarray}
 \bar\epsilon_{\mu\rho}\langle {T^{\rho}}_{\nu} \rangle&=& -\left[\frac{e^2}{4\pi}D_{\mu}B D_{\nu}B
+ \frac{1}{4\pi}\left(\frac{1}{48}D_{\mu}G D_{\nu}G - \frac{1}{24}D_{\mu}D_{\nu}G
-\frac{1}{24}\bar\epsilon_{\mu\nu}R\right) \right] \label{3.1.21}
\end{eqnarray}
where we have used (\ref{3.1.12}). Interchanging $\mu$ and $\nu$ in the above equation, we
 get a similar relation 
\begin{eqnarray}
 \bar\epsilon_{\nu\rho}\langle {T^{\rho}}_{\mu} \rangle &=& -\left[\frac{e^2}{4\pi}D_{\nu}B D_{\mu}B
+ \frac{1}{4\pi}\left(\frac{1}{48}D_{\nu}G D_{\mu}G - \frac{1}{24}D_{\nu}D_{\mu}G
-\frac{1}{24}\bar\epsilon_{\nu\mu}R \right)\right] \label{3.1.22}
\end{eqnarray} 
 Adding (\ref{3.1.21}) with (\ref{3.1.22}) and using the fact that
\begin{equation}
 [D_{\mu},D_{\nu}]G = 0 \ ; \ {\text {for some scalar function G} }, \label{3.1.23}
\end{equation}
we arrive at
\begin{equation}
 \bar\epsilon_{\mu\rho}\langle {T^{\rho}}_{\nu} \rangle + \bar\epsilon_{\nu\rho}\langle {T^{\rho}}_{\mu}
\rangle = - 2\langle T_{\mu\nu}\rangle
+ \frac{1}{48\pi}g_{\mu\nu}R~, \label{3.1.24}
\end{equation}
which can be further simplified by using (\ref{3.1.19}),
\begin{equation}
\langle T_{\mu\nu} \rangle = -\frac{1}{2}[\bar\epsilon_{\mu\rho} \langle {T^{\rho}}_{\nu}\rangle
 + \bar\epsilon_{\nu\rho}\langle {T^{\rho}}_{\mu}\rangle] + \frac{1}{2}g_{\mu\nu}\langle {T^{\alpha}}_{\alpha}\rangle~.\label{3.1.25}
\end{equation}
 These chirality properties (\ref{3.1.20}, \ref{3.1.25}) constrain the structure
 of the covariant current $\langle J^{\mu} \rangle$ and energy-momentum tensor
 $\langle T^{\mu\nu} \rangle$. \\

{\underbar{\it Solutions for $B$ and $G$} :}\\

Now we solve the differential equations for the auxiliary fields $B(r,t)$ and $G(r,t)$. 
First we solve (\ref{3.1.15}) for $B(r,t)$. For the metric (\ref{3.1.4}), equation
(\ref{3.1.15}) becomes
 \begin{eqnarray}
 \nabla_{\mu}\nabla^{\mu}B(r,t) = \frac{1}{\sqrt{fh}}\partial^{2}_{t}B(r,t)- \partial_{r}[\sqrt{fh}\partial_{r}B(r,t)] = -\partial_{r}A_{t} \label{3.1.26}
\end{eqnarray}
where we have used the fact that metric (\ref{3.1.4}) is static and $A_{t}(t,r)\equiv A_{t}(r)$.
The general solution for (\ref{3.1.26}) is
\begin{equation}
 B(r,t) = B_{o}(r) -a t + b~,  \  {\text {with}} \ \partial_{r}B_{o}(r) = \frac{1}{\sqrt{fh}}[A_{t}+c]~.
\label{3.1.27}
\end{equation}
Here $a,b$ and $c$ are integration constants. Similarly, the differential equation (\ref{3.1.16})
for $G(r,t)$ on using (\ref{3.1.17}), can be written as
\begin{equation}
\frac{1}{\sqrt{fh}}\partial^{2}_{t}G(r,t)
 -\partial_{r}[\sqrt{fh}\partial_{r}G(r,t)] = \sqrt{-g}R = \partial_{r}\left[\sqrt{\frac{h}{f}}f'\right]~, \label{3.1.28}
\end{equation}
which after solving, yields 
\begin{equation}
 G(r,t) = G_{o} - 4p t + q~,  \ {\text {with}} \ \partial_{r}G_{o}(r) = - \frac{1}{\sqrt{fh}}
\left[\frac{f'}{\sqrt{-g}} + z\right] \label{3.1.29}
\end{equation}
where $p, q$ and $z$ are integration constants.
\section{\label{chiralaction2} Charge and energy flux}
Now we are in a position to calculate the charge and energy-momentum flux
 from the black hole background (\ref{3.1.1}). We will see that
 the results are the same as that obtained either from the consistent 
 \cite{isowilczek} or from the covariant (section-\ref{coanomaly2}) anomaly based
  approach. 
     First we derive the charge flux. The covariant current (\ref{3.1.10}) can be written 
 in terms of ordinary partial derivative
\begin{equation}
 \langle J^{\mu} \rangle = -\frac{e^{2}}{2\pi}D^{\mu}B(r,t) = \frac{e^2}{2\pi}
[\bar\epsilon^{\mu\rho}\partial_{\rho}B(r,t) - g^{\mu\sigma}\partial_{\sigma}B(r,t)]~.\label{3.2.1}
\end{equation}
 Taking the $\mu=r$ component of above equation and then using
 $B(r,t)$ (\ref{3.1.27}), yields
\begin{equation}
\langle J^{r}(r)\rangle = \frac{e^2}{2\pi\sqrt{-g}}[\bar A_{t}(r)] \label{3.2.2}
\end{equation}
where $\bar A_{t}(r)$ is defined as
\begin{equation}
 \bar A_{t}(r) = A_{t}(r)+c+a~. \label{3.2.3}
\end{equation}
The other component $\langle J^{t}(r) \rangle$ is fixed by the chirality constraint (\ref{3.1.20})
\begin{equation}
 \langle J^{t}(r) \rangle = -\bar\epsilon^{tr}\langle J_{r}(r)\rangle =
 -\bar\epsilon^{tr}g_{rr}\langle J^{r}(r)
\rangle = \frac{\sqrt{-g}}{f}\langle J^{r}(r) \rangle~.
\label{3.2.4}
\end{equation}
Our task now is to determine the integration constants $c$ and $a$. For that, we impose the 
 covariant boundary condition (\ref{1.2.31}), i.e  vanishing of the 
 covariant current (\ref{3.2.2}) at the horizon. This leads to a relation among
 $c$ and $a$
\begin{equation}
 c + a = -A_{t}(r_{h})~.\label{3.2.5}
\end{equation}
 Hence the expression for $\langle J^{r}(r) \rangle$ takes the form
\begin{equation}
 \langle J^{r}(r)\rangle = \frac{e^2}{2\pi\sqrt{-g}}[A_{t}(r) - A_{t}(r_{h})] \label{3.2.6}
\end{equation}
Now the charge flux is given by the asymptotic $(r \rightarrow \infty)$ limit of the anomaly free current \cite{robwilczek,isowilczek}. We had observed that for the gauge fields which vanish at  
 asymptotic infinity, the covariant gauge anomaly (\ref{3.1.18}) vanishes, and hence we directly compute the
flux from (\ref{3.2.6}) by taking the $(r \rightarrow \infty)$ limit. This yields,
\begin{equation}
\langle J^{r}( r \rightarrow \infty)\rangle = - \frac{e^2}{2\pi}\left(A_{t}(r_{h}) \right) 
\label{3.2.7}
\end{equation}
This is the desired Hawking charge flux and agrees with our previous result 
given in section-\ref{coanomaly2} (see also \cite{robwilczek,isowilczek}). 

We next consider the energy momentum flux by adopting the same technique. After using the solutions
 for $B(x)$ (\ref{3.1.27}) and $G(x)$ (\ref{3.1.29}) in the general expression
 for covariant energy-momentum tensor (\ref{3.1.11}), we get
\begin{eqnarray}
\langle {T^{r}}_{t}\rangle &=& \frac{e^2}{4\pi \sqrt{-g}}\bar A^{2}_{t}(r) + \frac{1}{12\pi\sqrt{-g}}\bar P^{2}(r) + 
\frac{1}{24\pi \sqrt{-g}}[\frac{f'}{\sqrt{-g}} \bar P(r) + \bar Q(r)]\label{3.2.8}\\
\langle {T^{r}}_{r}\rangle &=& \frac{R}{96\pi} - \frac{\sqrt{-g}}{f}\langle {T^{r}}_{t}\rangle \label{3.2.9}\\
\langle {T^{t}}_{t}\rangle &=& -\langle {T^{r}}_{r}\rangle + \frac{R}{48\pi}\label{3.2.10}
\end{eqnarray}
where $\bar A(r)$ is given in (\ref{3.2.3}) and 
\begin{eqnarray}
\bar P(r) &=& p - \frac{1}{4}(\frac{f'}{\sqrt{-g}} + z)\label{3.2.12}\\
\bar Q(r) &=& \frac{1}{4}hf'' - \frac{f'}{8}(\frac{hf'}{f} - h')~.\label{3.2.13}
\end{eqnarray}
 Relation (\ref{3.2.10}) is a consequence of the trace anomaly (\ref{3.1.19}) while  
 (\ref{3.2.9}) follows from the chirality criterion (\ref{3.1.25}). 
Now we implement the boundary condition; namely the vanishing of the 
universal component of the covariant energy momentum tensor at the horizon,
 i.e $\langle {T^{r}}_t (r_{h})\rangle = 0$. Then from (\ref{3.2.5}) and (\ref{3.2.8}), we  have
\begin{eqnarray}
 \langle{T^{r}}_{t}(r=r_{h})\rangle &=& \bar P^{2}(r_{h}) + 2[\frac{f'}{\sqrt{-g}} \bar P(r_{h})
 + \bar Q(r_{h})] =0 \label{3.2.14}\\
\Rightarrow p &=& \frac{1}{4}(z \pm \sqrt{f'(r_{h})h'(r_{h})})
\label{3.2.15}
\end{eqnarray}
Substituting either of the above relations (among $p$ and $z$) in 
(\ref{3.2.8}) we get 
\begin{eqnarray}
\langle {T^{r}}_{t}(r)\rangle &=& \frac{e^2}{4\pi\sqrt{-g}}(A_{t}(r) - A_{t}(r_{h}))^2  \label{3.2.16}\\
 && + \frac{1}{192\pi\sqrt{-g}}[f'(r_{h})h'(r_{h}) - \frac{2f(r)h'(r)}{h(r)} + 2f''(r)h(r) + f'(r)h'(r)]~.
\nonumber
\end{eqnarray}
Further, by noting that for metric (\ref{3.1.4}) 
\begin{equation}
 f'(r_{h})h'(r_{h}) - \frac{2f(r)h'(r)}{h(r)} + 2f''(r)h(r) + f'(r)h'(r) = N^{r}_{t}(r) - N^{r}_{t}(r_{h})
\label{3.2.17}
\end{equation}
(see equation (\ref{2.1.8})), we can write (\ref{3.2.16}) as  
\begin{equation}
 \langle{T^{r}}_{t}(r)\rangle = \frac{e^2}{4\pi\sqrt{-g}}(A_{t}(r) - A_{t}(r_{h}))^2 +
\frac{1}{192\pi\sqrt{-g}}[N^{r}_{t}(r) - N^{r}_{t}(r_{h})]~. \label{3.2.18}
\end{equation}
This expression is in agreement with the one obtained by solving the covariant anomalous
Ward identities in the region near to the event horizon (\ref{1.2.41}). 

To obtain the energy flux, we recall that it is given by
 the asymptotic expression for the anomaly free energy momentum
tensor. As for the charge case, here too it is found
 from (\ref{3.1.18a}) that the gravitational Ward identity vanishes in this limit.
Hence the energy flux is abstracted by taking the asymptotic infinity limit of (\ref{3.2.18}). 
This yields,
\begin{equation}
\langle {T^{r}}_t(r\rightarrow\infty)\rangle =\frac{e^2}{4\pi} A^{2}_{t}(r_{+})+
 \frac{1}{192\pi}f'(r_{h})h'(r_{h})~,\label{3.2.19}
\end{equation}
which correctly reproduces the Hawking flux from the generic spherically symmetric
 charged black hole. 

 For the Reissner-Nordstrom black hole (\ref{1.2.5}), the
 charge and energy-momentum flux  can be obtain by substituting the expressions
 for metric function and the gauge potential in (\ref{3.2.7}) and (\ref{3.2.19}) 
\begin{eqnarray}
\langle J^{r}(r\rightarrow \infty)\rangle &=& \frac{e^2 Q}{2\pi r_{+}} \label{3.2.20}\\
\langle {T^{r}}_{t}(r\rightarrow \infty)\rangle &=& \frac{e^{2}Q^2}{4\pi r^{2}_{+}} + \frac{\pi T^{2}_{H}}{12}~. \label{3.2.21}
\end{eqnarray}
These expressions for the charge and energy-momentum flux are in agreement with our 
 previous results (\ref{1.2.33}, \ref{1.2.56}) obtained by using the covariant anomaly
 approach. Hence, from the above 
 analysis, we observe that only the structure of chiral effective action (\ref{3.1.5})
 and the covariant boundary conditions at the event horizon are sufficient to determine  
 the charge and energy-momentum flux completely.  
\section{\label{chiralaction4}Back reaction effect and chiral effective action}
In this section we shall implement the chiral effective action approach
 to study the Hawking flux in the presence of gravitational back reaction.
 The modified expressions for the charge and energy-momentum flux, due
 to the effect of one loop back reaction are obtained. 

 Back reaction, it might be recalled, is an effect of non-zero expectation value of the 
 energy-momentum tensor on the spacetime geometry, which acts as a source of curvature. 
  It is possible to include the effect of gravitational back reaction in the derivation of Hawking radiation. Indeed, using the conformal anomaly method
 the effect on  the spacetime geometry by one loop back reaction was computed in \cite{york, lousto}. Based on this approach, corrections to the Hawking temperature
 were obtained in \cite{lousto,fursaev}. Correction to the Hawking temperature using the back reaction equation for linearised quantum fluctuation was derived in \cite{fabbri}.
 Recently, more useful and intuitive way to understand the effect of back reaction through the  quantum tunneling formalism \cite{parikh} was developed in \cite{bibhas}. 

 We are interested in discussing the Hawking effect from the Reissner-Nordstrom
 black hole (\ref{1.2.5}) in the presence of back reaction. However, we cannot
 use the standard Reissner-Nordstrom metric directly to  compute the Hawking fluxes
  since it gets modified when we take into account the effect of one loop back reaction.
  Insteed, we shall use the modified Reissner-Nordstrom metric given in \cite{lousto}  
 \begin{equation}
 ds^2 = f(r)dt^{2} - \frac{1}{h(r)}dr^2 - r^{2}(d\theta^{2} + \sin^{2}\theta d\phi^{2})
\label{3.3.3}
\end{equation}
where
 \begin{eqnarray}
f(r) &=& 1 - \frac{2M}{r} + \frac{Q^{2}}{r^{2}} + \frac{A(\alpha)}{r}\label{3.3.1}\\
 h(r) &=&  \frac{1}{f(r)}B(\alpha)  \label{3.3.2}
\end{eqnarray}
$A(\alpha)$ and $B(\alpha)$ depend on the parameters $\alpha$
\footnote{The explicit structures  for $A$ and $B$ are given in \cite{lousto}. In our analysis,
 however, we use only the general properties of metric coefficients like $f(r_{\infty})h(r_{\infty}) =1$.
 The explicit structure of the metric is not crucial for our purpose.}. 
 The event horizon for the modified metric (\ref{3.3.3}) is now 
defined by $f(r=r_{M}) = h(r=r_{M})$ where $r_{M}$ is the modified
 horizon radius given by \cite{lousto}
\begin{equation}
 r_{M} = r_{+}\left(1+\frac{\alpha}{M^{2}}\right)^{-\frac{1}{2}}~. \label{3.3.4}
\end{equation}
Here $r_{+}$ is the radius of the outer event horizon of the original (in the absence of back reaction)
 Reissner-Nordstrom black hole (\ref{1.2.5}).
  
  Such a form is also dictated by simple scaling arguments.
 As is well known, a loop expansion is equivalent to an expansion 
in powers of the Planck constant $h$. Since, in natural units, $\sqrt{h} = M_{p}$ (the Planck mass), the one loop correction
has a form given by $\frac{\alpha}{M^2}$ . Where parameter $\alpha$ is the 
 related to the trace anomaly coefficient, taking into
account the degrees of freedom of the fields, and its explicit form is  
given by \cite{york,lousto,fursaev}.
\begin{equation}
\alpha = \frac{1}{360\pi}(-N_{o} -\frac{7}{4}N_{\frac{1}{2}} + 13N_{1} + \frac{233}{4}N_{\frac{3}{2}} - 212N_{2})~. \label{3.3.4a} 
\end{equation}
where $N_{s}$ denotes the number of fields with spin $s$ entering into the theory \cite{birrell,duff}.
 In our case, only the complex scalar field ($s=0$) exits. Therefore, we have
\begin{equation}
 \alpha = -\frac{1}{360\pi}N_{o}~.\label{3.3.4b}
\end{equation}
  
 Further, we note that the generic form for the metric (\ref{3.3.3}) in the presence of the gravitational backreaction was obtained \cite{york, lousto} by solving the semiclassical Einstein equations
\begin{equation}
R_{\mu\nu} -\frac{1}{2}g_{\mu\nu}R = \langle T^{(4)}_{\mu\nu}(g_{\mu\nu})\rangle \label{3.3.5}
\end{equation}
or in the more convenient form 
\begin{equation}
R_{\mu\nu} = \langle T^{(4)}_{\mu\nu}(g_{\mu\nu})\rangle - \frac{1}{2}\langle T^{(4)\rho}_{\rho} \rangle g_{\mu\nu}, \label{3.3.6}
\end{equation}
with the aid of confromal (trace) anomaly in $4D$ by keeping the spherical 
 symmetry intact. Here $\langle T^{(4)}_{\mu\nu}(g_{\mu\nu})\rangle$ is the
 renormalized energy-momentum tensor in $3+1$ dimensions. 
In our formalism, we consider the generic form for the $4D$ metric 
(\ref{3.3.3}), in the presence of the back reaction, as a starting point. 
 As mentioned before, by using a dimensional reduction technique
 the effective field theory near the horizon  becomes two dimensional. 
 The metric of this two dimensional theory is identical to the $r-t$
 component of the full metric (\ref{3.3.3}). On this effective $1+1$ dimensional
 background, if we omit the classically ingoing (left moving) modes, then the theory becomes 
 chiral. 

  Now we compute the modified charge and energy-momentum flux by using the chiral effective 
 action approach discussed earlier. Instead of repeating the whole analysis we just
 use the structures for the covariant current $\langle J^{r} \rangle$ and the
covariant energy-momentum tensor $\langle {T^{r}}_{t}\rangle$  given in (\ref{3.2.6})
 and (\ref{3.2.18}), respectively. Then the Hawking charge and energy-momentum
 flux, in the presence of gravitational back reaction can be obtain by appropriately taking the asymptotic infinity limit of
the covariant current and energy-momentum tensor. 

 First, let us consider the expression for the chiral covariant current as given in (\ref{3.2.6})
\begin{equation}
 \langle J^{r}(r) \rangle = \frac{e^2}{2\pi\sqrt{-g}}[A_{t}(r) - A_{t}(r_{M})] \label{3.3.7}
\end{equation}
where the gauge potential is given by
\begin{equation}
 A_{t}(r) = -\frac{Q}{r}~. \label{3.3.8}
\end{equation}
The charge flux is determined by the asymptotic infinity limit of the anomaly free current.
As is evident from the expression (\ref{3.1.18}), the covariant gauge anomaly vanishes in this limit
 and therefore we can obtain the charge flux directly from (\ref{3.3.7}) by taking its asymptotic
 limit. This gives
\begin{equation}
 \langle J^{r}(r\rightarrow \infty)\rangle = -\frac{e^2}{2\pi}A_{t}(r_{M})~. \label{3.3.8a}
\end{equation}
Finally, by using (\ref{3.3.4}) and (\ref{3.3.8}) in (\ref{3.3.8a}), we obtain 
\begin{equation}
 \langle J^{r}(r\rightarrow \infty)\rangle = \frac{e^2 Q}{2\pi r_{+}}\left(1+\frac{\alpha}{M^2}\right)^{\frac{1}{2}}~. \label{3.3.9}
\end{equation}
This is the expression for the Hawking charge flux from the Reissner-Nordstrom
 black hole in the presence of gravitational back reaction. 

 Further, by expanding $\left(1+\frac{\alpha}{M^2}\right)^{\frac{1}{2}}$ and keeping only
 leading order terms in $\alpha$, we find
\begin{equation}
 \langle J^{r}(r\rightarrow \infty)\rangle \approx \frac{e^2 Q}{2\pi r_{+}} +  \frac{e^2 Q\alpha}{4\pi r_{+}M^2}~. \label{3.3.10} 
\end{equation}
The first term in the above expression is the usual charge flux for the Reissner-Nordstrom 
 black hole while the next term represents correction to the standard value of charge flux
 due to the effect of one loop back reaction. Note that, since the trace anomaly coefficient
 $\alpha$ is negative (\ref{3.3.4b}), there is a net decrease in the Hawking charge flux
 compare to its standard value.   

 Next, we consider the expression for the chiral covariant energy-momentum tensor (\ref{3.2.18})
\begin{equation}
 \langle{T^{r}}_{t}(r)\rangle = \frac{e^2}{4\pi\sqrt{-g}}(A_{t}(r) - A_{t}(r_{h}))^2 +
\frac{1}{192\pi\sqrt{-g}}[N^{r}_{t}(r) - N^{r}_{t}(r_{M})]~. \label{3.3.11}
\end{equation}
The Hawking energy-momentum flux is given by the asymptotic limit of the
 anomaly free energy-momentum tensor. In this limit, we observe that the 
 covariant gravitational anomaly (\ref{3.1.18a}) vanishes. Hence the
 energy-momentum flux can be easily obtained by taking the asymptotic infinity
 limit of (\ref{3.3.11})
\begin{equation}
\langle {T^{r}}_t(r\rightarrow\infty) \rangle =\frac{e^2}{4\pi} A^{2}_{t}(r_{+})+
 \frac{1}{192\pi}f'(r_{h})h'(r_{M})~.\label{3.3.12}
\end{equation}
We can also write the above expression  in terms of the modified
 surface gravity $\kappa_{M} = \frac{1}{2}\sqrt{f'(r_{M})h'(r_{M})}$ as
\begin{equation}
\langle {T^{r}}_{t}(r\rightarrow \infty)\rangle = \frac{e^2 Q^2}{4\pi r_{+}^2} \left(1 + \frac{\alpha}{M^2}\right) + \frac{1}{48\pi}
\kappa_{M}^2~. \label{3.3.13} 
\end{equation}
$\kappa_{M}$ is the modification in the surface gravity due to the effect 
of one loop back reaction. Following similar arguments given below equation
(\ref{3.3.4}) we can relate $\kappa_{M}$ to the usual surface gravity
 for the Reissner-Nordstrom black hole $\kappa$ \cite{lousto,fursaev}
\begin{equation}
\kappa_{M} = \kappa\left(1 + \frac{\alpha}{M^2}\right), \label{3.3.14} 
\end{equation}
Substituting (\ref{3.3.14}) in (\ref{3.3.13}), yields 
  \begin{eqnarray}
\langle {T^{r}}_{t}(r\rightarrow \infty) \rangle &=&\frac{e^2 Q^2}{4\pi r_{+}^2} \left(1 + \frac{\alpha}{M^2}\right) + \frac{1}{48\pi}\kappa^{2}\left(1 + \frac{\alpha}{M^2}\right)^{2}\nonumber\\
 &=& \frac{e^2 Q^2}{4\pi r_{+}^2} \left(1 + \frac{\alpha}{M^2}\right)+ \frac{\pi T_{H}^2}{12}(1+\frac{\alpha}{M^2})^2, \label{3.3.15} 
\end{eqnarray} 
where $T_{H} = \frac{\kappa}{2\pi}$ is the usual Hawking temperature of the Reissner-Nordstrom
 black hole. Further,  by expanding $\left(1 + \frac{\alpha}{M^2}\right)^{2}$
 and keeping terms up to leading order
 in  $\alpha$, we arrive at
\begin{equation}
\langle{T^{r}}_{t}(r\rightarrow \infty)\rangle \approx \frac{e^2 Q^2}{4\pi r_{H}^2} + \frac{\pi T_{H}^2}{12}
 + \frac{e^2 Q^2 \alpha}{2\pi r_{H}^2 M^2} + \frac{\pi T_{H}^2 \alpha}{6M^2}. \label{3.3.16}  
\end{equation}
The first two terms in the above expression represent
 energy flux from the usual charged black hole,
 while the last two terms  are corrections due to
 the effect of one loop back reaction. Since the trace anomaly coefficient $\alpha$ is negative,
 the overall effect of gravitational back reaction, at one loop level, is to reduce
 the Hawking flux from its usual value. This feature is also valid for the fermions ($s=\frac{1}{2}$)
 and gravitons ($s=2$). On contrary, for the fields with $s=1,\frac{3}{2}, \cdots$, the situation is exactly opposite. In these cases the trace anomaly coefficient $\alpha$ becomes positive, leading to the increase
 in the Hawking charge and energy-momentum flux \cite{fursaev}. 
 
 \section{\label{chiralaction5} Discussions}
We have given a derivation of the Hawking flux from charged black holes, based on the effective action approach, which only employs the boundary conditions at the event horizon. It might be mentioned that generally such (effective action based) approaches require, apart from conditions at the horizon, some other boundary condition, as for example, the vanishing of ingoing modes at 
infinity \cite{fulling,unruh,wipf}. The latter obviously goes against the universality of the Hawking effect which should 
be determined from conditions at the horizon only. In this we have succeeded. Also, the specific structure of the effective action, which gives the Hawking radiation, is valid only at the event horizon. This is the anomalous (chiral)
effective action. Other
effective action based techniques do not categorically specify the structure of the effective action at the horizon.
Rather, they use the usual (anomaly free) form for the
effective action and are
restricted to two dimensions only \cite{fabbri2,shapiro1,wipf}.

  An important factor concerning this analysis is to realize that effective
field theories become two dimensional and chiral near the event horizon
\cite{robwilczek}. Yet another ingredient was the implementation
of a specific boundary condition: namely the vanishing of the covariant form of the current and
 energy-momentum tensor. As emphasized in 
 chapter-\ref{chap:coanomaly}, the anomaly based approach was simplified
considerably if, instead of consistent anomalies used in \cite{robwilczek, isowilczek}, covariant anomalies
were taken as the starting point. Indeed, in the present computations, we have taken that form of the effective action which yields anomalous Ward identities 
having covariant gauge and gravitational anomalies. This is distinct from the standard
 (Polyakov type) effective action \cite{isowilczek,fabbri1}. The connection these two
 as well as their correspondence with the Unruh vacuum is the topic of next chapter.

 The arbitrary constants in the covariant energy momentum tensor and the covariant current 
derived from the anomalous effective action were fixed by a boundary condition
 at the event horizon. The Hawking fluxes, which are measured at infinity, are
 then obtained by taking the asymptotic infinity limit of the covariant current
 and energy-momentum tensor. This may be compared with the anomaly based approach
  given in chapter-\ref{chap:coanomaly} and \cite{isowilczek} where the Hawking radiation
 is derived by solving  both the anomalous Ward identities as well as the usual conservation
 laws together with the imposition of covariant boundary condition at the event horizon.
 Apart from this the use of discontinuous step functions, which are essential in the
 anomaly based approach, is avoided. 
 Consequently, we have shown that aspects like covariant anomalies
 and covariant boundary conditions are not merely confined to
discussing the Hawking effect in the anomaly based approach.
Rather they have a wider applicability since our effective action
 based approach is different from (although connected with) the anomaly based approach.

  Finally, we implement the chiral effective action approach
 to compute the Hawking charge and energy-momentum flux for the
 Reissner-Nordstrom black hole taking into the account the effect
 of one loop back reaction. The point is that the $r-t$ part of usual
 charged black hole ($\sqrt{-g} =1$) gets modified to a more general
 ($\sqrt{-g}\ne 1$) due to the effect of back reaction without
 disturbing the spherical symmetry. For this general metric, the 
   expressions for the covariant current and energy momentum tensor
 were obtained. This indicates the generality of
 the chiral effective action approach. The corrections to charge and energy flux due to
 (one loop) back reaction effect were then obtained by appropriately
 taking  asymptotic limit of the current and energy momentum tensor. 

 Apart from this example, the chiral effective action approach discussed here has been
 implemented in the computation of Hawking radiation from other several black hole
 geometries \cite{plbcitedpaper1,plbcitedpaper2,plbcitedpaper3,plbcitedpaper4}.

\chapter{\label{chap:boundarycondition} Covariant boundary conditions and 
 connection with vacuum states}
The motivation of this chapter is to provide a clear understanding of
the covariant boundary condition used in the analysis of        
 chapter-\ref{chap:coanomaly}, chapter-\ref{chap:chiralaction} and also in  \cite{isowilczek},
 of deriving the Hawking flux using chiral gauge and gravitational anomalies.
 Besides this we also reveal certain new features in chiral currents and energy-momentum tensors which are useful in exhibiting their connection with the standard nonchiral expressions.

 In chapter-\ref{chap:coanomaly} we gave a method, based on the covariant gauge and
 gravitational anomalies, to compute the fluxes of Hawking radiation.
  Hawking fluxes were obtained by solving the covariant anomalous gauge/gravitational
 Ward identities (valid near the horizon) as well as the usual conservation laws (valid
 away from the horizon) together with the implementation of covariant boundary condition;
 namely, the vanishing of covariant anomalous current and energy-momentum tensor 
 at the horizon. It is important to note that the analysis of \cite{isowilczek}
 also uses the same boundary  condition, however, the expressions for gauge 
 and gravitational anomalies were taken to be consistent. Consequently, the knowledge of
 local counterterms, connecting consistent and covariant expressions,
 becomes essential in the consistent anomaly based approach \cite{isowilczek}.
 
 In another new development (see chapter-\ref{chap:chiralaction}), we obtained the Hawking
 charge and energy-momentum flux by exploiting the structure of the chiral effective action
 \cite{leutwyler}, appropriately modified by local counterterms. 
 This chiral effective action is defined in the neighborhood of the event horizon. The 
 currents and energy-momentum tensors obtained from this effective action satisfy
 the covariant gauge and gravitational Ward identities. Again, the arbitrary constants
 appearing in the expressions for covariant current and energy-momentum tensor were
 fixed by imposing the covariant boundary condition at the event horizon. Finally, the 
 Hawking charge and energy-momentum fluxes were  obtained by taking appropriately the 
 asymptotic infinity limit of these covariant currents and energy-momentum tensors,
 respectively.     
 
  Apart from these approaches, there is an alternative
 procedure to compute the
 Hawking fluxes \cite{rabinessay}.
 Like the chiral effective action approach, this
 method uses only the near horizon structures for
 the covariant current and energy-momentum
 tensor obtained by solving the covariant gauge
 and gravitational Ward identities.
 As before, the arbitrary constants appearing
 in the expressions for current
 and energy-momentum tensor were fixed by imposing the covariant boundary 
 condition. Since the gauge and gravitational anomalies vanishes
 in the asymptotic
 limit, it is expected that the asymptotic behavior of
 the chiral covariant current
 and energy-momentum tensor would be identical as that of
 anomaly free current and energy-momentum tensor. This expectation is also confirm by
 actual computation of Hawking fluxes, which were obtained by taking the asymptotic 
 infinity limit  of the covariant current and energy-momentum tensor.

  It is thus clear from the above discussion that the covariant boundary condition plays an important 
 role in the computation of Hawking fluxes. Also, the imposition of this boundary condition
 at the event horizon helps to make the whole anomaly approach consistent with the universality
 of the Hawking effect.
 
Here we give a detailed analysis for this particular boundary condition,
 clarifying its role in the computation of the Hawking flux.
 First, we compute the Hawking flux by adopting
 the approach of \cite{rabinessay}.
 It turns out that, with this choice of boundary
 condition, the components for covariant current/energy-momentum tensor $\langle J^{r} \rangle,
 \langle {T^{r}}_{t} \rangle$  
 obtained from solving the anomaly equation match exactly with the expectation values of the
 current/energy-momentum tensor, obtained from the chiral
effective action, taken by imposing the regularity condition on the outgoing
modes at the future horizon. Furthermore, we discuss the connection of our results with those found by a standard use of boundary conditions on nonchiral 
(anomaly free) currents and energy-momentum tensors. Indeed we are able to  show that our results are equivalent to the 
choice of the Unruh vacuum for a nonchiral theory. This choice, it may be recalled, is natural for discussing Hawking flux.  

In section-\ref{boundarycondition1} we provide a generalization for the approach
 discussed in \cite{rabinessay} to compute the Hawking charge and energy-momentum
 flux. Then, in section-\ref{boundarycondition2} we consider the
 near horizon structures of covariant currents and energy-momentum tensor obtained 
 from the chiral effective action (see section-\ref{chiralaction2}). However,
 this time, the arbitrary constants appearing in the expressions of these 
 currents and energy-momentum tensors are fixed by imposing appropriate
 regularity conditions on the outgoing modes at future event horizon. Here we also
  discuss the relation of the results obtained for a chiral theory, subjected
 to the regularity conditions, with those found in a usual (nonchiral) theory
 in different vacua. Our concluding remarks are contained in section-\ref{boundarycondition3}.
 Finally, there is an appendix discussing the connection between the trace anomaly 
 and gravitational anomaly for a ($1+1$) dimensional chiral theory.      
\section{\label{boundarycondition1} Charge and energy flux from covariant anomaly: A direct
 approach}
Consider a generic spherically symmetric charged black hole background
 represented by the metric 
\begin{equation}
ds^2 = \gamma_{ab}dx^{a}dx^{b}=f(r)dt^2 -\frac{1}{h(r)}dr^2 -r^2(d\theta^2 + \sin^2\theta d\phi^2) \label{4.1.1}
\end{equation}
where $f(r)$ and $h(r)$ are the metric coefficients\footnote{This metric is same as the one
 given in section-\ref{chiralaction1}. Consequently, the vector potential ${\bf A}$ and
 the metric coefficients $f(r)$ and $h(r)$ satisfy
 all the properties given in (\ref{3.1.1a}) and (\ref{3.1.3}), respectively.}. The event horizon
 for this black hole is defined by 
\begin{equation}
 f(r_{h}) = h(r_{h}) = 0 ~. \label{4.1.2}
\end{equation}
Now consider charged scalar fields propagating on this
background. As discussed earlier,  
 the effective field theory near the event horizon becomes
 two dimensional with the metric given by the  $r-t$ section of
 (\ref{4.1.1})
\begin{equation}
ds^2=f(r)dt^2 -\frac{1}{h(r)}dr^2 ~. \label{4.1.3}
\end{equation}
 On this two dimensional background, the modes
which are going in to the black hole (for example left moving modes) are lost and
 the effective theory become chiral.
 Two dimensional chiral theory possesses gravitational anomaly and,
 if gauge fields are present, also gauge anomaly
 \cite{witten,fujikawabook,bardeenzumino,bertlmannbook,bertlmann,rabin1,rabin2}.
 These anomalies are further classified in two groups - the consistent and the covariant
 \cite{fujikawabook,bardeenzumino,rabin1,rabin2}.
A derivation of Hawking flux by using the consistent gauge and gravitational anomalies
was given in \cite{isowilczek}. However, the boundary condition used to fix the 
 arbitrary constants was covariant. A complete reformulation of 
 this approach using only covariant structures was given in chapter-\ref{chap:coanomaly},
 while the corresponding effective action based approach was developed in chapter-\ref{chap:chiralaction}
  
  An efficient and quite simple way to obtain the Hawking flux was discussed in \cite{rabinessay}
 where the computation involved only the expressions for anomalous covariant Ward
 identities and the covariant boundary condition. An important advantage of this approach 
  was that the splitting of space into two different regions
 (see \cite{isowilczek} and chapter-\ref{chap:coanomaly}) is avoided. In this section
 we would first generalize this approach for the generic black hole background (\ref{4.1.1}).
 This would also help in setting up the conventions and introduce certain expressions
 that are essential for subsequent analysis. \\

{\it Charge flux:}\\
We now compute the Hawking charge flux by using only the covariant gauge anomaly
 and the covariant boundary condition. The expression for covariant gauge anomaly
 \cite{fujikawabook,bertlmannbook} for the right moving modes (\ref{1.2.15}) is 
\begin{equation}
\nabla_{\mu}\langle J^{\mu} \rangle = -\frac{e^2}{4\pi\sqrt{-g}}\epsilon^{\alpha\beta}F_{\alpha\beta}~. \label{4.1.4}
\end{equation}
 For a static background, the above equation becomes,
\begin{equation}
\partial_{r}(\sqrt{-g}\langle J^{r} \rangle) = \frac{e^{2}}{2\pi}\partial_{r}A_{t}.\label{4.1.5}
\end{equation}
Solving this equation we get 
\begin{equation}
\sqrt{-g} \langle J^{r} \rangle = c_{H} + \frac{e^2}{2\pi}[A_{t}(r) -A_{t}(r_{h})]~. \label{4.1.6}
\end{equation}
Here $c_{H}$ is an integration constant which can be fixed by 
imposing the covariant boundary condition i.e  covariant current $\langle J^{r} \rangle$
must vanish at the event horizon, 
\begin{equation}
\langle J^{r}(r=r_{h}) \rangle = 0~. \label{4.1.7}
\end{equation}
 Hence we get $c_{H} = 0$ and the expression for the current becomes,
\begin{equation}
\langle J^{r} \rangle = \frac{e^{2}}{2\pi\sqrt{-g}}[A_{t}(r) -A_{t}(r_{h})]. \label{4.1.8}
\end{equation}      
Note that the Hawking flux is measured at infinity where there is no 
anomaly. This necessitated a split of space into two distinct regions -
one near the horizon and one away from it - and the use of two Ward
identities (see chapter-\ref{chap:coanomaly} and \cite{isowilczek}) 
 This is redundant if we observe that the anomaly (\ref{4.1.4})
vanishes at asymptotic infinity. Consequently, in this approach,
the flux is directly obtained from the asymptotic infinity limit of 
(\ref{4.1.8}):  
\begin{equation}
{\text {Charge flux}} = \langle J^{r}(r\rightarrow \infty) \rangle = -\frac{e^2 A_{t}(r_{h})}{2\pi}~ . \label{4.1.9}
\end{equation}
This reproduces the familiar expression for the charge flux obtained earlier
 in chapter-\ref{chap:coanomaly}. \\

{\it Energy-momentum flux:}\\
Next, we consider the expression for the two dimensional covariant gravitational Ward identity 
 (\ref{1.2.35}) 
\begin{equation}
\nabla_{\mu}\langle T^{\mu\nu} \rangle = \langle J_{\mu}\rangle F^{\mu\nu} +
\frac{\epsilon^{\nu\mu}}{96\pi\sqrt{-g}}\nabla_{\mu}R \label{4.1.10}
\end{equation} 
where the first term is the classical contribution (Lorentz force) and 
the second is the covariant gravitational anomaly \cite{witten,bertlmann,rabinbibhas}.
Here $R$ is the Ricci scalar and for the metric (\ref{4.1.3}) it is given by
\begin{equation}
R = \frac{f'' h}{f} + \frac{f'h'}{2f} - \frac{f'^{2}h}{2f^2}. \label{4.1.11}
\end{equation}
By simplifying (\ref{4.1.10})we get, in the static background, 
\begin{equation}
\partial_{r}(\sqrt{-g}\langle{T^{r}}_{t}\rangle) = \partial_{r}N^{r}_{t}(r) 
 -\frac{e^2 A_{t}(r_{h})}{2\pi} \partial_{r}A_{t}(r) + \partial_{r}(\frac{e^2 A^{2}_{t}(r)}{4\pi})\label{4.1.12}
\end{equation}
where
\begin{equation}
N^{r}_{t} = \frac{1}{96\pi}\left( hf'' + \frac{f'h'}{2} - \frac{f'^{2}h}{f}\right). \label{4.1.13}
\end{equation}
The solution for (\ref{4.1.12}) is given by
\begin{equation}
\sqrt{-g}\langle {T^{r}}_{t}\rangle = b_{H} + [N^{r}_{t}(r) - N^{r}_{t}(r_{h})]
 + \frac{e^2}{4\pi}[A_{t}(r) -A_{t}(r_{h})]^2 ~. \label{4.1.14}
\end{equation}
Here $b_{H}$ is an integration constant . Implementing the covariant boundary condition, namely, the vanishing of covariant EM tensor at the event
horizon, 
\begin{equation}
\langle{T^{r}}_{t}(r=r_{h})\rangle = 0 \label{4.1.15}
\end{equation}
yields $b_{H} = 0$. Hence (\ref{4.1.14}) reads
\begin{equation}
\sqrt{-g}\langle{T^{r}}_{t}(r)\rangle = [N^{r}_{t}(r) - N^{r}_{t}(r_{h})]
 + \frac{e^2}{4\pi}[A_{t}(r) -A_{t}(r_{h})]^2 ~. \label{4.1.16}
\end{equation} 
Since the covariant gravitational anomaly vanishes asymptotically,
we can compute the energy flux as before by taking the asymptotic limit of (\ref{4.1.16})
\begin{equation}
{\text{energy flux}} = \langle{T^{r}}_{t}(r\rightarrow \infty)\rangle = -N^{r}_{t}(r_{h}) + \frac{e^2A_{t}^2(r_{h})}{4\pi} ~.\label{4.1.17}
\end{equation}  
This reproduces the expression for the Hawking flux found by earlier using the anomaly
based approach (chapter-\ref{chap:coanomaly}, \cite{isowilczek}) or by using the chiral
 effective action (chapter-\ref{chap:chiralaction}).

\section{\label{boundarycondition2} Covariant boundary condition and vacuum states}
It is now clear that the covariant boundary conditions play a crucial
role in the computation of Hawking fluxes using chiral
 gauge and gravitational anomalies, either in the approaches
 discussed in  \cite{isowilczek,isowilczekPRD},chapter-\ref{chap:coanomaly},
 chapter-\ref{chap:chiralaction}
 or in the more direct method \cite{rabinessay} reviewed here.
  Therefore it is worthwhile to study it in some detail.
 We adopt the following strategy. We consider the 
 expressions for the expectation values of the
 covariant current and energy-momentum
 tensor deduced from the chiral effective action \cite{leutwyler},
 suitably modified by a local counterterm.
 These are already given in section-\ref{chiralaction1}. We then
 transform the components of current and energy-momentum tensor 
 into null coordinates. The arbitrary constants appearing in the 
 expressions for current and energy-momentum tensor are now fixed by imposing
 regularity conditions on the outgoing modes at the future event horizon.
 The final results are found to match exactly with the corresponding 
expressions for the covariant current (\ref{4.1.8}) and EM tensor (\ref{4.1.16}),
 which were derived by using the covariant boundary conditions (\ref{4.1.7},\ref{4.1.15}).
 Subsequently we show that our results are consistent with
 the imposition of the Unruh vacuum on usual (nonchiral) expressions.
 
  We begin our analysis by considering the theory near the event horizon.
 An expression for the chiral covariant current obtained from the chiral
 effective action (\ref{3.1.5}) is given in (\ref{3.1.10}). Substituting
 the solution for the auxiliary field $B(r,t)$ (\ref{3.1.27}) in (\ref{3.1.10})
 and then taking $\mu=r$ component of the covariant current $\langle J^{\mu} \rangle$,
 we get 
 \begin{equation}
\langle J^{r}(r)\rangle = \frac{e^2}{2\pi\sqrt{-g}}[\bar A_{t}(r)] \label{4.2.1}
\end{equation}
 where $\bar A_{t}(r)$ is defined by (\ref{3.2.3}). The $\mu=t$ component of the covariant
 current is obtain by exploiting the chirality condition (\ref{3.1.20})
 \begin{equation}
 \langle J^{t}(r) \rangle = -\bar\epsilon^{tr}\langle J_{r}(r)\rangle =
 -\bar\epsilon^{tr}g_{rr}\langle J^{r}(r)
\rangle = \frac{\sqrt{-g}}{f}\langle J^{r}(r) \rangle~. \label{4.2.2}
\end{equation}

Next, we consider the chiral covariant energy-momentum tensor 
given in (\ref{3.1.11}). Using the solutions for the auxiliary fields $B(r,t)$ (\ref{3.1.27})
 and $G(r,t)$ (\ref{3.1.29}) in (\ref{3.1.11}),
 the expressions for the various components of $\langle {T^{\mu}}_{\nu}\rangle$
 follow from (\ref{3.2.8}-\ref{3.2.10})
\begin{eqnarray}
\langle {T^{r}}_{t} \rangle &=& \frac{e^2}{4\pi \sqrt{-g}}\bar A^{2}_{t}(r) + \frac{1}{12\pi\sqrt{-g}}\bar P^{2}(r) + 
\frac{1}{24\pi \sqrt{-g}}[\frac{f'}{\sqrt{-g}} \bar P(r) + \bar Q(r)]\label{4.2.3}\\
\langle {T^{r}}_{r} \rangle &=& \frac{R}{96\pi} - \frac{\sqrt{-g}}{f}\langle {T^{r}}_{t}\rangle \label{4.2.4}\\
\langle {T^{t}}_{t} \rangle &=& -\langle {T^{r}}_{r}\rangle+ \frac{R}{48\pi}\label{4.2.5}
\end{eqnarray}
where $\bar A(r), \bar P(r)$ and $\bar Q(r)$ are defined by the relations
(\ref{3.2.3},\ref{3.2.12}) and (\ref{3.2.13}), respectively. As illustrated
 in section-\ref{chiralaction1}, the chirality conditions (\ref{3.1.20}, \ref{3.1.25})
 imposes certain restrictions on the components of current (\ref{4.2.1}, \ref{4.2.2}) and 
 the energy-momentum tensor (\ref{4.2.3}-\ref{4.2.5}). For example the relation (\ref{4.2.4})
 is obtained by using (\ref{3.1.25}) and (\ref{4.2.3}), while the remaining component
 $\langle {T^{t}}_{t} \rangle$  can be fixed by using the expression for the
 chiral trace anomaly (\ref{3.1.19}) \footnote{For a ($1+1$) dimensional chiral theory, it is possible to derive
 a relation among the trace and gravitational anomalies. See appendix of this chapter for more
 details.}
\begin{equation}
 \langle {T^{\mu}}_{\mu} \rangle = \frac{R}{48\pi}~. \label{4.2.5a}
\end{equation}

 To further illuminates the chiral nature of the theory near the event horizon, we transform
 the various components of the covariant current and energy-momentum tensor to 
 null coordinates 
\begin{eqnarray}
v &=& t + r*  \nonumber\\ 
u &=& t -r* \label{4.2.6}  
\end{eqnarray}
where $r^{*}$ is the tortoise coordinate defined by the relation
\begin{equation}
\frac{dr}{dr*} = \sqrt{fh}~.\label{4.2.7}
\end{equation}
The metric (\ref{4.1.3}) in these coordinates looks like
\begin{equation}
ds^2 = \bar g_{\alpha\beta} dx^{\alpha}dx^{\beta}= \frac{f(r)}{2}(dudv+dvdu) \ ; \ 
\alpha,\beta = u,v ~. \label{4.2.8} 
\end{equation}
The metric coefficients $\bar g_{\alpha\beta}$ are : 
\begin{eqnarray}
 \bar g_{uu} = \bar g_{vv} = 0  \ ; \ \bar g_{uv} = \bar g_{vu} = \frac{f(r)}{2}  \label{4.2.9}
\end{eqnarray}
Now the components of the covariant current in $(u,v)$ and ($r,t$) coordinates are
 related as 
\begin{eqnarray}
\langle J_{u} \rangle &=& \frac{\partial t}{\partial u} \langle J_{t}\rangle + 
\frac{\partial r}{\partial u}\langle J_{r} \rangle\label{4.2.10}\\
\langle J_{v} \rangle &=& \frac{\partial t}{\partial v}\langle J_{t}\rangle +
 \frac{\partial r}{\partial v}\langle J_{r} \rangle ~. \label{4.2.11}
\end{eqnarray}
After using (\ref{4.2.1}, \ref{4.2.2}, \ref{4.2.6}, \ref{4.2.7}) in (\ref{4.2.10}, \ref{4.2.11}) 
 we arrive at the expressions for the components of chiral covariant current  in $u,v$ coordinates
\begin{eqnarray}
\langle J_{u}(r) \rangle &=& \frac{1}{2}[\langle J_{t}\rangle - \sqrt{fh}\langle J_{r}\rangle] =\frac{e^2}{2\pi} \bar A_{t}(r)\label{4.2.13}\\
\langle J_{v}(r) \rangle &=&  \frac{1}{2}[\langle J_{t} \rangle + \sqrt{fh}\langle J_{r}\rangle]  = 0~. \label{4.2.14}
\end{eqnarray}
 Following similar steps, the components of the chiral covariant energy-momentum
 tensor in $u,v$ coordinates are given by
\begin{eqnarray}
\langle T_{uu}(r) \rangle &=& \frac{1}{4}[f \langle {T^{t}}_{t} \rangle - f \langle {T^{r}}_{r}\rangle + 2\sqrt{-g}\langle {T^{r}}_{t}\rangle]\nonumber\\
&=& \frac{e^2}{4\pi }\bar A^{2}_{t}(r) + \frac{1}{12\pi}\bar P^2(r) + \frac{1}{24\pi }\left[\frac{f'}{\sqrt{-g}}\bar P(r) + \bar Q(r)\right] \label{4.2.15}\\
\langle T_{uv}(r)\rangle  &=& \frac{f}{4}[\langle T^{t}_{t} \rangle + \langle {T^{r}}_{r} \rangle] = \frac{1}{192\pi}fR \label{4.2.16}\\
\langle T_{vv}(r)\rangle &=&  \frac{1}{4}[f \langle {T^{t}}_{t} \rangle - f \langle {T^{r}}_{r} \rangle - 2\sqrt{-g}\langle {T^{r}}_{t}\rangle]= 0 ~.\label{4.2.17}
\end{eqnarray}
We now observe that, due to the chiral property, the $\langle J_{v} \rangle$ and $\langle T_{vv} \rangle$
components vanish everywhere. These correspond to the ingoing modes and are compatible
with the fact, stated earlier, that the near horizon theory is a two
dimensional chiral theory where the ingoing modes are lost.
Further, by rewriting  (\ref{4.2.16}) as
\begin{equation}
 \langle T_{uv}\rangle = \frac{f(r)R}{192\pi} = \frac{f(r)}{4}\langle {T^{\alpha}}_{\alpha} \rangle \label{4.2.18}
\end{equation}
we observed that the structure of $\langle T_{uv}\rangle$ is fixed by the trace anomaly (\ref{4.2.5a}).
Only the $\langle J_{u} \rangle$ and $\langle T_{uu} \rangle$ components involve the undetermined constants.
These will now be determined by considering various vacuum states.
\subsection{\label{boundarycondition2.1} Vacuum states}
In a generic spacetime three different vacua \cite{fullingvacuum} are defined
 by appropriately choosing `in' and `out' modes.  

 1) The Unruh vacuum : \\
  In this state the `in' modes are chosen as to be positive frequency
 with respect to the Schwarzschild time $`t'$. With this choice, in the asymptotic
 past the Unruh vacuum $|U\rangle$ coincides with the usual Minkowski vacuum.
 On the other hand, out modes 
 are taken to be positive frequency with respect to the Kruskal coordinate
\begin{equation}
   U = -\kappa e^{ -\kappa u}, \label{4.2.19}   \ ; \ \kappa \ {\text {is the surface gravity}}
\end{equation}
 The Kruskal coordinate $U$ acts as the affine parameter along the past horizon.
 This mimics the late time behavior of
 modes coming out of a collapsing star as its surface approaches the horizon \cite{unruh}.
 By this choice $\langle U| T_{\mu\nu}|U\rangle$ is regular on the future event horizon i.e
  a freely falling observer must see a finite amount of flux at the future event horizon
 $\mathcal{H}^{+}$. In the asymptotic future $\langle U|T_{\mu\nu}|U\rangle$ has the form of a flux
of radiation at the Hawking temperature $T_{H}$ \cite{fabbri1,fabbri2}. 
This state is the most appropriate to discuss evaporation of black holes formed
by gravitational collapse of matter.\\
 2) Hartle-Hawking vacuum :\\
 The Hartle-Hawking state $|H\rangle$ \cite{hartle} is obtained by choosing
 in modes to be positive frequency with
respect to Kruskal coordinate
\begin{equation}
 V = \kappa e^{ \kappa v} \label{4.2.20} 
\end{equation}
 the affine parameter on the future horizon,
whereas outgoing modes are defined in the same way as for Unruh vacuum.
By construction this state is regular on both the future and past event horizons.
Consequently $\langle H|T_{\mu\nu}|H\rangle$ is also regular on the future and
 past horizons. Hartle-Hawking vacuum is appropriate to describe a black hole in
 thermal equilibrium with quantum field under consideration.\\
 3) Boulware vacuum :\\
 Boulware vacuum $|B\rangle$ \cite{boulware} is obtained by choosing both `in'
 and `out' modes to be positive frequency with
respect to the Schwarzschild time coordinate $t$ . This state most closely
reproduces the familiar notion of Minkowski vacuum asymptotically. However, since
 the Schwarzschild coordinates are not well define near the horizon, the expectation
 value of the energy-momentum in the Boulware vacuum  $\langle B|T_{\mu\nu}|B\rangle$ blows up
 at the event horizon.

This general picture is modified  when dealing with a chiral theory since, as shown before, the `in' modes always vanish. Consequently this leads to a simplification and conditions are imposed only on the `out' modes. Moreover, these conditions have to be imposed on the horizon
since the chiral theory is valid only there. The natural condition, leading to the 
occurrence of Hawking flux, is that a freely falling observer must see a finite amount
of flux at the horizon. This implies that the current (EM tensor) in Kruskal 
coordinates must be regular at the future horizon. Effectively, this is the same condition on the 'out' modes in either the Unruh vacuum \cite{unruh} or the Hartle-Hawking  vacuum \cite{hartle}. As far as our analysis is concerned this is sufficient to completely determine the form
of $\langle J_{\mu}\rangle$ or $\langle T_{\mu\nu}\rangle$.
 We show that their structures are identical to those obtained in the previous section using the covariant boundary condition. 

 A more direct comparison with the conventional results obtained from Unruh or
Hartle-Hawking states is possible. In that case one has to consider the nonchiral expressions \cite{isowilczekPRD,fabbri1} containing both `in' and `out' modes.
We show that, at asymptotic infinity where the flux is measured, our expressions
agree with that calculated from Unruh vacuum only. We discuss this in some detail.
\newpage

  {\bf Regularity conditions, Unruh and Hartle-Hawking vacua :} \\

 We now fix the arbitrary constants appearing in the covariant current (\ref{4.2.13}) and energy-momentum tensor (\ref{4.2.15}) by imposing the regularity conditions, appropriate for the Unruh \cite{unruh}
 and Hartle-Hawking \cite{hartle} vacua, on outgoing modes at future
 event horizon.

 First, consider the Unruh vacuum. As mentioned earlier, in this vacuum, `out' modes are
 defined with respect to Kruskal coordinate $U$ (\ref{4.2.19}). Therefore, we first transform
 $\langle J_{u}\rangle$ to $\langle J_{U} \rangle$ defined in Kruskal coordinate. $\langle J_{u}\rangle$
 and $\langle J_{U} \rangle$ are related by
\begin{equation}
\langle J_{U} \rangle = - \frac{\langle J_{u} \rangle}{\kappa U} \label{4.2.21}
\end{equation}
Now the regularity condition tells us that a freely falling observer must see
 a finite amount of charge flux at the future event horizon.
 However, since near the future event horizon $U \rightarrow \sqrt{r-r_{h}}$ $(r\rightarrow r_{h})$,
 it implies that $\langle J_{u} \rangle$ must vanish at $r\rightarrow r_{h}$. Hence from
 (\ref{3.2.3}) and (\ref{4.2.13}) we find
\begin{equation}
c + a = -A_{t}(r_{h})~. \label{4.2.22}
\end{equation}     
Similarly, imposing the condition that $\langle T_{UU}\rangle = (\frac{1}{\kappa U})^2
 \langle T_{uu}\rangle$
must be finite at future horizon leads to $\langle T_{uu}(r\rightarrow r_{h})\rangle =0$. This yields, from (\ref{3.2.3}, \ref{3.2.12}, \ref{3.2.13}) and (\ref{4.2.15}),
\begin{equation}
p = \frac{1}{4}(z \pm \sqrt{f'(r_{h})h'(r_{h})})~. \label{4.2.23} 
\end{equation}
Substituting (\ref{4.2.22}) and (\ref{4.2.23}) in (\ref{4.2.13}) and (\ref{4.2.15}) we get 
\begin{eqnarray}
 \langle J_{u}(r) \rangle &=& \frac{e^2}{2\pi} [A_{t}(r) -A_{t}(r_{h})]\label{4.2.24}\\
\langle T_{uu}(r) \rangle &=& \frac{e^2}{4\pi}[A_{t}(r)-A_{t}(r_{h})]^2 + 
 [N^{r}_{t}(r)-N^{r}_{t}(r_{h})] \label{4.2.25}
\end{eqnarray}
where $N^{r}_{t}(r)$ is given by (\ref{4.1.13}). Finally,
 by transforming back to $r-t$ coordinates, we obtain the expressions
\begin{eqnarray}
 \langle J^{r}(r) \rangle&=& \frac{e^2}{2\pi\sqrt{-g}}[A_{t}(r) - A_{t}(r_{h})] \label{4.2.26}\\
\langle J^{t}(r) \rangle &=& \frac{\sqrt{-g}}{f}\langle J^{r}(r) \rangle \label{4.2.27}
\end{eqnarray}
for the covariant current. While the energy-momentum tensor is given by,
 \begin{eqnarray}
\sqrt{-g}\langle {T^{r}}_{t}\rangle &=& \frac{e^2}{4\pi}[A_{t}(r)-A_{t}(r_{h})]^2 + 
 [N^{r}_{t}(r)-N^{r}_{t}(r_{h})]~. \label{4.2.28}
\end{eqnarray}
Likewise, $\langle {T^{r}}_{r}\rangle$ and $\langle {T^{t}}_{t}\rangle$ follow from (\ref{4.2.4}, \ref{4.2.5})

The expressions for $\langle J^{r} \rangle$ (\ref{4.2.26}) and $\langle {T^{r}}_{t} \rangle$ (\ref{4.2.28})
agree with the corresponding ones given in (\ref{4.1.8}) and (\ref{4.1.16}).
This shows that the structures for the universal components
$\langle J^{r} \rangle, \langle {T^{r}}_{t}\rangle$ obtained by solving the anomalous Ward
identities (\ref{4.1.4},\ref{4.1.10}) subjected to the covariant boundary conditions
(\ref{4.1.7},\ref{4.1.15}) exactly coincide with the results computed by demanding
regularity at the future event horizon. 

Next, consider the Hartle-Hawking vacuum. In this case, both $\langle T_{UU} \rangle$ and $\langle T_{VV} \rangle$ are regular at the
future and past event horizons, respectively. In the null coordinates ($u,v$), the above regularity
 condition is translated into the vanishing of $\langle T_{uu} \rangle$ and $\langle T_{vv} \rangle$, at the future and past
 event horizons. In our case however, only the outgoing modes are present in the region near the horizon.
 Consequently, the regularity condition on ingoing modes i.e $\langle T_{vv}(r\rightarrow r_{h})\rangle=0$ is
 trivially satisfied (see equation (\ref{4.2.17})).
 While, the condition on outgoing modes i.e $\langle T_{uu}(r\rightarrow r_{h})\rangle=0$ is same as for the Unruh vacuum.
 Naturally, the result (\ref{4.2.28}) remain unchanged for Hartle-Hawking vacuum also. This is different
 from the results derived for the usual (nonchiral) theory \cite{fullingvacuum}. As we shall see below, in the
 conventional analysis, Hartle-Hawking vacuum corresponds to the no flux state and is suitable for describing the black hole in a thermal equilibrium with surrounding quantum fields.
 
  It is possible to compare our findings with conventional (nonchiral) computations
where the Hawking flux is obtained in the Unruh vacuum.
 We begin by considering the conservation equations
for a nonchiral theory that is valid away from the horizon.

Such equations were earlier used in \cite{robwilczek,isowilczek,isowilczekPRD} (see also chapter-\ref{chap:coanomaly}). Conservation
of the gauge current yields \footnote{We use a tilde $\langle \tilde J^{\mu}\rangle$ to distinguish nonchiral expressions from chiral ones.},
\begin{equation}
\nabla_{\mu}\langle \tilde J^{\mu} \rangle = \frac{1}{\sqrt{-g}}\partial_{\mu}(\sqrt{-g}\langle\tilde J^{\mu}\rangle) = 0 \label{4.2.29}
\end{equation}
which, in a static background, leads to,
\begin{equation}
\langle \tilde J^{r}\rangle = \frac{C_{1}}{\sqrt{-g}} \label{4.2.30}
\end{equation}
where $C_{1}$ is some constant.

   As is well know there is no regularisation that simultaneously preserves the vector as well as axial vector
gauge invariance. Indeed, for a vector gauge invariant regularisation resulting in (\ref{4.2.29}),
 the following anomaly is found in the axial current,
\begin{equation}
\nabla_{\mu}\langle \tilde J^{5\mu} \rangle = \frac{e^2}{2\pi \sqrt{-g}}\epsilon^{\mu\nu}F_{\mu\nu} \ ; \ \langle \tilde J^{5\mu}\rangle = \frac{1}{\sqrt{-g}}\epsilon^{\mu\nu}\langle \tilde J_{\nu}\rangle~. \label{4.2.31}
\end{equation}
The solution of this Ward identity is given by,
\begin{equation}
\langle \tilde J^{t} \rangle = -\frac{1}{f}[C_{2} - \frac{e^2}{\pi}A_{t}(r)] \label{4.2.32}
\end{equation} 
where $C_{2}$ is another constant.

    In the null coordinates introduced in 
(\ref{4.2.6}) the various components of the current are defined as,
\begin{eqnarray}
\langle \tilde J_{u} \rangle &=& \frac{1}{2}[C_{1} - C_{2} + \frac{e^2}{\pi}A_{t}(r)],\label{4.2.33}\\
\langle \tilde J_{v} \rangle &=& -\frac{1}{2}[C_{1} + C_{2} - \frac{e^2}{\pi}A_{t}(r)]~.\label{4.2.34}
\end{eqnarray}
The constants $C_{1}$, $C_{2}$ are now determined by using
appropriate boundary conditions corresponding to first,
the Unruh state, and then, the Hartle-Hawking state. For
the Unruh state $\langle \tilde J_{u}(r\rightarrow r_{h})\rangle = 0$ 
and $\langle \tilde J_{v}(r \rightarrow \infty)\rangle = 0$ yield,
\begin{equation}
C_{1} = - C_{2} = -\frac{e^2}{2\pi}A_{t}(r_{h})~,\label{4.2.35}
\end{equation}
so that, reverting back to $(r,t)$ coordinates, we obtain,
\begin{eqnarray}
\langle \tilde J^{r}\rangle &=& -\frac{e^2}{2\pi \sqrt{-g}}A_{t}(r_{h}),\label{4.2.36}\\
\langle \tilde J^{t} \rangle &=& \frac{e^2}{\pi f}[A_{t}(r) - \frac{1}{2}A_{t}(r_{h})]~.\label{4.2.37}
\end{eqnarray}
The Hawking charge flux, identified with $\langle \tilde J^{r}(r \rightarrow \infty)\rangle$, 
reproduces the desired result (\ref{4.1.9}). Expectedly, (\ref{4.2.36},  \ref{4.2.37}) differ from our relations (\ref{4.2.26}, \ref{4.2.27}) which are valid only near the horizon. However, at
 asymptotic infinity where the Hawking flux is measured,
 both expressions match, i.e
\begin{eqnarray}
\langle \tilde J^{r}(r\rightarrow \infty) \rangle &=& \langle J^{r}(r \rightarrow \infty)\rangle, \label{4.2.38}\\
\langle \tilde J^{t}(r\rightarrow \infty) \rangle &=& \langle J^{t}(r \rightarrow \infty)\rangle \label{4.2.39} 
\end{eqnarray}
implying the important consequence,
\begin{equation}
 \langle \tilde J^{\mu}(r\rightarrow \infty)\rangle = \langle J^{\mu}(r\rightarrow \infty)\rangle~. \label{4.2.39aa}
\end{equation}

 All the above considerations follow identically for 
the stress tensor. Now the relevant conservation law  in the presence of an external gauge field, is
\begin{equation}
\nabla_{\mu}\langle \tilde {T}^{\mu}{}_{\nu} \rangle = \langle \tilde J^{\mu}\rangle F_{\mu\nu} \label{4.2.39a}
\end{equation} 
 Also, there is a trace anomaly given by,
\begin{equation}
\langle \tilde{T}^{\mu}{}_{\mu}\rangle = \frac{R}{24\pi} \label{4.2.39b}
\end{equation}
 Taking $\nu=t$ component of (\ref{4.2.39a}) and using (\ref{4.2.30}) we get
\begin{equation}
 \partial_{r}(\sqrt{-g}\langle \tilde{T}^{r}{}_{t}\rangle) =
\langle \tilde J^{r} \rangle \partial_{r}A_{t}(r) = C_{1}\partial_{r}A_{t}~. \label{4.2.39c}
\end{equation}
which on static background, leads to 
\begin{equation}
 \langle \tilde {T}^{r}{}_{t}\rangle = \frac{1}{\sqrt{-g}}\left[D_1 + C_{1} A_{t}(r) \right] 
\label{4.2.39d}
\end{equation}
where $D_{1}$ is an integration constant. 
Similarly, by taking the $\nu=r$ component of (\ref{4.2.39a}) and using (\ref{4.2.32}), we obtain, 
\begin{equation}
 \frac{1}{\sqrt{-g}}\partial_{r}(\sqrt{-g}\langle \tilde {T}^{r}{}_{r}\rangle) = \frac{f'}{2f}\langle \tilde {T}^{t}{}_{t}\rangle-\frac{h'}{2h}\langle \tilde {T}^{r}{}_{r}\rangle + \frac{1}{f} \partial_{r}\left(C_{2}A_{t}-\frac{e^{2}}{2\pi}A_{t}^{2}\right)~.\label{4.2.39e}
\end{equation}
Now by using 
the expression for trace anomaly (\ref{4.2.39b}) we can eliminate $\langle \tilde {T}^{t}{}_{t}\rangle$ from the above equation. Then (\ref{4.2.39e}) becomes,  
\begin{equation}
\frac{1}{\sqrt{-g}}\partial_{r}(\sqrt{-g}\langle \tilde {T}^{r}{}_{r}\rangle) = \frac{f'}{2f}
\left[\frac{R}{24\pi} - \langle {T}^{r}{}_{r}\rangle \right] -\frac{h'}{2h}\langle \tilde {T}^{r}{}_{r}\rangle + \frac{1}{f} \partial_{r}\left(C_{2}A_{t}-\frac{e^{2}}{2\pi}A_{t}^{2}\right) ~.\label{4.2.39e1}
\end{equation}
Here the Ricci scalar $R$ is given in (\ref{4.1.11}). The solution for the above equation is given by,
\begin{equation}
 \langle \tilde {T}^{r}{}_{r}\rangle = \frac{f'^{2}h}{96\pi f^{2}} +
 \frac{1}{f}\left[C_{2} A_{t} - \frac{e^{2}}{2\pi}A_{t}^{2}\right] + \frac{D_{2}}{f}~.
\label{4.2.39f}
\end{equation}
 Where $D_{2}$ is a constant of integration. The remaining component $\langle \tilde {T}^{t}{}_{t}\rangle$ is determined by the trace anomaly (\ref{4.2.39b}).

After transforming the components of the energy-momentum tensor (\ref{4.2.39d}, \ref{4.2.39f})
into null coordinates (\ref{4.2.6}), we get  
\begin{eqnarray}
 \langle \tilde T_{uu} \rangle &=&  \frac{1}{4}\left(\frac{hf''}{24\pi} + \frac{f'h'}{48\pi}-
 \frac{f'^{2}h}{24\pi f} - 2\left[(C_{2}-C_{1}) A_{t} - \frac{e^{2}}{2\pi}A_{t}^{2} \right]
 + 2(D_{1}-D_{2})\right) \label{4.2.39g}\\
\langle \tilde T_{vv} \rangle &=&  \frac{1}{4}\left(\frac{hf''}{24\pi} + \frac{f'h'}{48\pi}
 -\frac{f'^{2}h}{24\pi f} - 2\left[(C_{2}+C_{1}) A_{t} - \frac{e^{2}}{2\pi}A_{t}^{2}\right]
- 2(D_{2} + D_{1})\right)  \label{4.2.39h}
\end{eqnarray}
The arbitrary constants $D_{1}$ and $D_{2}$ are now fixed by imposing the 
 boundary conditions appropriate for the  Unruh and Hartle-Hawking vacua.
Let us first consider the Unruh vacuum. In this state, the energy-momentum tensor
 in the Kruskal coordinate is regular across the horizon, leading to the condition
 $\langle \tilde T_{uu}(r\rightarrow r_{h})\rangle =0$. Using this in (\ref{4.2.39g}), and 
 noting the expressions of $C_{1}$ and $C_{2}$  (for the Unruh vacuum) given in (\ref{4.2.35}),
 we get a relation among the integration constants 
\begin{equation}
 D_{2} - D_{1} = \frac{f'(r_{h})h'(r_{h})}{96\pi} + \frac{e^2 A_{t}^{2}(r_{h})}{2\pi}\label{4.2.39i}
\end{equation}
Also, in the Unruh vacuum, there is no incoming flux at past null infinity i.e  $\langle \tilde T_{vv}(r\rightarrow \infty)\rangle=0$. This condition gives another relation
\begin{equation}
 D_{2} = -D_{1}~. \label{4.2.39j}
\end{equation}
combining (\ref{4.2.39i}) and (\ref{4.2.39j}) we have 
\begin{equation}
D_{1}= -D_{2} = \frac{1}{192\pi}f'(r_{h})h'(r_{h}) + \frac{e^2 A_{t}^{2}(r_{h})}{2\pi} ~. \label{4.2.39k}
\end{equation}
Substituting (\ref{4.2.39k}) in (\ref{4.2.39d}) we get 
\begin{eqnarray}
 \langle \tilde {T}^{r}{}_{t}\rangle &=& \frac{1}{192\pi\sqrt{-g}}f'(r_{h})h'(r_{h}) +
 \frac{e^{2}}{4\pi\sqrt{-g}}[A_{t}^{2}(r_{h})-2A_{t}(r)A_{t}(r_{h})] \label{4.2.39l}\\
\langle \tilde {T}^{r}{}_{r}\rangle &=& \frac{1}{96\pi f}\left[\frac{f'^{2}h}{f} - \frac{1}{2}f'^{2}(r_{h})h'(r_{h})\right]\nonumber\\
&& -  \frac{e^{2}}{2\pi f} \left( A_{t}^{2}(r) - A_{t}(r_{h})A_{t} + \frac{1}{2}A_{t}^{2}(r_{h})\right)
\label{4.2.39m}
\end{eqnarray}
while, $\langle \tilde{T}^{t}{}_{t}\rangle$ can be fixed from the trace anomaly (\ref{4.2.39b}). The Hawking
 energy-momentum flux, identified with $\langle \tilde{T}^{r}{}_{t}(r\rightarrow \infty)\rangle$,
 gives the desired result (\ref{4.1.17}). Once again $\langle \tilde{T}^{\mu}{}_{\nu} \rangle$ will not agree with our $\langle {T}^{\mu}{}_{\nu}\rangle$ (\ref{4.2.28}). However, at asymptotic infinity, all components agree:
\begin{equation}
\langle \tilde{T}^{\mu}{}_{\nu}(r\rightarrow\infty)\rangle=\langle{T^\mu}_{\nu}(r\rightarrow\infty)\rangle,
\label{4.2.40}
\end{equation}
leading to the identification of the Hawking flux with $\langle \tilde{T}^{r}{}_{t}(r\rightarrow\infty)\rangle$.

   The equivalences (\ref{4.2.39aa}, \ref{4.2.40}) reveal the internal consistency of our approach. They are based on two issues. First, in the asymptotic limit the covariant chiral gauge (\ref{4.1.4})
 and gravitational \ref{4.1.10}) anomalies vanish and, secondly, the boundary conditions (\ref{4.1.7}, \ref{4.1.15}) get identified with the Unruh state that is appropriate for discussing Hawking effect. It is important to note that, asymptotically,
 all the components, and not just the universal component that yields the flux, agree.  

  In the Hartle-Hawking state, the conditions
$\langle \tilde J_{u}(r \rightarrow r_{h}) \rangle=0$ and
 $\langle \tilde J_{v}(r \rightarrow r_{h})\rangle =0$ yield,
\begin{equation}
C_{1} = 0 \ ; \ C_{2} = \frac{e^2}{\pi}A_{t}(r_{h})\label{4.2.41}
\end{equation}
so that, 
\begin{eqnarray}
\langle \tilde J^{r}(r)\rangle &=& 0,\label{4.2.42} \\
\langle\tilde J^{t}(r)\rangle &=& \frac{e^2}{\pi f}(A_{t}(r) - A_{t}(r_{h}))~,\label{4.2.43}
\end{eqnarray} 
Expectedly, there is no Hawking (charge) flux now. The 
above expressions, even at asymptotic infinity, do not
agree with our expressions (\ref{4.2.26}, \ref{4.2.27}).

 Now consider the stress tensor. In the Hartle-Hawking vacuum both $\langle T_{uu} \rangle$ (\ref{4.2.39g})
 and $\langle T_{vv}\rangle$ (\ref{4.2.39h}) are regular on the future and past event horizon
 respectively i.e
\begin{equation}
 \langle \tilde T_{uu}(r\rightarrow r_{h}) \rangle = \langle \tilde T_{vv}(r\rightarrow r_{h})\rangle =0~.\label{4.2.43aa}
\end{equation}
 By evaluating (\ref{4.2.39g}, \ref{4.2.39h}) at ($r=r_{h}$) and then equating to zero, we get
 the relations among $D_{1}$ and $D_{2}$ 
 \begin{eqnarray}
  D_{1} - D_{2} &=& \frac{f'(r_{h})h'(r_{h})}{96\pi} + \frac{e^{2} A_{t}^{2}(r_{h})}{2\pi} \label{4.2.43a}\\
  D_{1} + D_{2} &=& - (D_{1}-D_{2})~. \label{4.2.43b}
 \end{eqnarray}
  Hence, for the Hartle-Hawking state, we have
 \begin{equation}
  D_{1} = 0 \ ; \ D_{2} = -\frac{f'(r_{h})h'(r_{h})}{96\pi} - \frac{e^{2} A_{t}^{2}(r_{h})}{2\pi}~.
\label{4.2.43c}
 \end{equation}
  Substituting (\ref{4.2.41}, \ref{4.2.43c}) in (\ref{4.2.39d}) and (\ref{4.2.39f}), yields the 
 expression for various components of the energy-momentum tensor:
\begin{eqnarray}
 \langle \tilde {T}^{r}{}_{t} \rangle &=& 0 \label{4.2.43d}\\
\langle \tilde {T}^{r}{}_{r}  \rangle &=& \frac{1}{96\pi f}
\left[\frac{f'^{2}h}{f}-f'(r_{h})h'(r_{h})\right] - \frac{e^2}{2\pi f}[A_{t}-A_{t}(r_{h})]^{2}~.
\label{4.2.43e}
\end{eqnarray}
There is no energy-momentum flux in the Hartle-Hawking vacuum.
 Also, the relations (\ref{4.2.43d}, \ref{4.2.43e}),
even at the asymptotic limit, do not agree with our expressions (\ref{4.2.28}).    
 
   {\bf Boulware vacuum :}\\

Apart from the Unruh and Hartle-Hawking vacua there is another
vacuum named after Boulware \cite{boulware} which closely resembles the Minkowski vacuum asymptotically. 
In this vacuum, there is no radiation in the asymptotic future. In other words this implies
 $\langle J^{r} \rangle$ and $\langle {T^{r}}_{t}\rangle$ given in (\ref{4.2.1}) and 
(\ref{4.2.3}) must vanish at $r\rightarrow \infty$ limit. Therefore, 
for the Boulware vacuum, we get
\begin{eqnarray}
 c + a &=& 0 \label{4.2.44}\\
 p &=& \frac{1}{4}z \label{4.2.45} 
\end{eqnarray} 
By substituting (\ref{4.2.44}) in (\ref{4.2.1}) and (\ref{4.2.2}) we have
\begin{eqnarray}
\langle J^{r}(r) \rangle &=& \frac{e^2}{2\pi\sqrt{-g}}A_{t}(r)  \label{4.2.46}\\
\langle J^{t}(r) \rangle &=& \frac{e^2}{2\pi f} A_{t}(r). \label{4.2.47}  
\end{eqnarray}
Similarly, by substituting (\ref{4.2.44}) and (\ref{4.2.45}) in equations (\ref{4.2.3} -\ref{4.2.5}), we get   \begin{eqnarray}
\langle {T^{r}}_{t}\rangle &=& \frac{e^2 A_{t}^2(r)}{4\pi\sqrt{-g}} + \frac{1}{\sqrt{-g}}N^{r}_{t}(r)\label{4.2.48}\\
\langle {T^{r}}_{r} \rangle &=& \frac{-e^2 A_{t}^2(r)}{4\pi f} - \frac{1}{f}N^{r}_{t}(r) + \frac{R}{96\pi} \label{4.2.49}\\
\langle {T^{t}}_{t}\rangle &=& \frac{e^2 A_{t}^2(r)}{4\pi f} +\frac{1}{f}N^{r}_{t}(r) + \frac{R}{96\pi}\label{4.2.50} 
\end{eqnarray}
Observe that there is no radiation in the asymptotic region in the Boulware
vacuum. Also, the trace anomaly (\ref{4.2.5a}) is reproduced since this
is independent of the choice of quantum state.\\
Further, we note that, in the Kruskal coordinates, $\langle J_{U}\rangle$ and
 $\langle T_{UU} \rangle$
components of current and energy-momentum tensors diverge at the horizon. This can be 
seen by substituting equations (\ref{4.2.46}-\ref{4.2.47}) in (\ref{4.2.13}). Then
the expression for $\langle J_{u} \rangle$ in Boulware vacuum becomes,
\begin{equation}
\langle J_{u} \rangle = \frac{e^2}{2\pi}A_{t}(r) \label{4.2.51}
\end{equation}
while, by putting (\ref{4.2.48}-\ref{4.2.50}) in (\ref{4.2.15}), we 
obtain, for $\langle T_{uu} \rangle$
\begin{equation}
\langle T_{uu} \rangle = \frac{e^2 A_{t}^2(r)}{4\pi} + N^{r}_{t}(r)~. \label{g}
\end{equation}
Note that in the limit ($r\rightarrow r_{h}$) $\langle J_{u}\rangle$ and $\langle T_{uu} \rangle$ do
not vanish. Hence, in the Kruskal coordinates, the  current and EM tensor diverge.
 This is expected since the Boulware vacuum is not regular near the horizon. 
\section{\label{boundarycondition3}Discussions} 
We have analyzed in details a method, briefly introduced in
\cite{rabinessay}, of computing the Hawking flux using covariant gauge
and gravitational anomalies. Contrary to earlier approaches discussed in
 chapter-\ref{chap:coanomaly} and \cite{robwilczek,isowilczek}, a split
of space into distinct regions (near to and away from horizon) using
step functions was avoided. This method is different from the one
 given in chapter-\ref{chap:coanomaly} and \cite{robwilczek,isowilczek}
 where the fluxes of Hawking radiation were obtained by demanding that 
 the complete theory composed from contributions from
 inside the horizon, near the horizon and away from the horizon
 must be anomaly free. However, the present approach uses identical (covariant)
 boundary conditions. It reinforces the crucial role of these boundary conditions,
 the study of which has been the principal objective of this paper.

 In order to get a clean understanding of these boundary conditions
we first computed the explicit structures of the covariant current
$\langle J_{\mu}\rangle$ and the covariant energy-momentum tensor $\langle T_{\mu\nu}\rangle$
from the chiral (anomalous) effective action, appropriately modified by
adding a local counterterm \cite{leutwyler}. The chiral nature 
of these structures became more transparent by passing to the null coordinates.
 In these coordinates the contribution from the ingoing (left moving) modes was manifestly seen to vanish. The outgoing (right moving)
 modes involved arbitrary parameters which were fixed by imposing regularity conditions at the future horizon. 
No condition on the ingoing (left moving) modes was required
as these were absent as a result of chirality. These findings by themselves are new. They are also different
from the corresponding expressions for $\langle J_{\mu} \rangle$, $\langle T_{\mu\nu} \rangle$, 
obtained from the  standard nonanomalous (Polyakov type) action \cite{polyakov},
satisfying $\nabla_{\mu}\langle J^{\mu}\rangle = 0, \nabla_{\mu}\langle T^{\mu\nu}\rangle = \langle J_{\mu}\rangle F^{\mu\nu}$ and $\langle {T^{\mu}}_{\mu}\rangle = \frac{R}{24\pi}$, implying the absence of any gauge or gravitational (diffeomorphism) anomaly.
 Only the trace anomaly is present. Details of the latter computation may be found in \cite{isowilczekPRD,fabbri2,shapiro1}.

We have then established a direct connection of these results
(obtained from the chiral currents)  with
the choice of the covariant boundary condition used in determining
the Hawking flux from chiral consistent \cite{isowilczek,isowilczekPRD}
 or covariant (see chapter-\ref{chap:coanomaly}) gauge and gravitational anomalies
 and also from the near horizon chiral effective action given in chapter-\ref{chap:chiralaction}.
The relevant universal component $\langle J^{r}\rangle$ or $\langle {T^{r}}_{t}\rangle$ obtained by
solving the anomaly equation subject to the covariant boundary condition
(\ref{4.1.7}, \ref{4.1.15}) agrees exactly with the result derived from
imposing regularity condition on the outgoing modes at the 
future horizon: namely, a free falling observer sees a finite amount of flux at outer horizon
indicating the possibility of Hawking radiation. Our findings, therefore,
 provide a clear justification of the covariant boundary condition.

 Finally, we put our computations in a proper perspective by comparing our findings with the standard implementation of the various vacua states on nonchiral expressions.
Specifically, we show that our results are compatible with the choice of Unruh vacuum for a nonchiral theory which eventually yields the Hawking flux. Further, we showed that, in the Unruh vacuum 
 the asymptotic forms for the components  of the covariant current and energy-momentum tensor obtained
 from the chiral effective action matches exactly with corresponding components of
 the current and energy-momentum tensor computed from the usual (nonchiral) theory. However, for
 the Hartle-Hawking vacuum there exist no such equivalence between the chiral and usual theories,
 even at the asymptotic infinity.

\begin{subappendices}
\chapter*{Appendix}
\section{\label{appendix3A} Relation between chiral trace and gravitational anomalies}
\renewcommand{\theequation}{5A.\arabic{equation}}
\setcounter{equation}{0}   
 Unlike  the case of vector theory, where the diffeomorphism invariance 
 is kept intact inspite of the presence of trace anomaly, the chiral theory
has both a diffeomorphism anomaly (gravitational anomaly) and a trace
anomaly. In $1+1$ dimensions it is possible to obtain a relation
between the coefficients of the diffeomorphism anomaly and the trace
anomaly by exploiting the chirality criterion. 

  To see this let us write the general structure of the covariant gravitational Ward
identity in the presence of an external gauge field,
\begin{equation}
\nabla_{\mu}\langle {T^{\mu}}_{\nu}\rangle = \langle J_{\mu}\rangle {F^{\mu}}_{\nu} +
N_{a}\bar \epsilon_{\nu\mu}\nabla^{\mu}R \label{3A1} 
\end{equation}  
where $N_{a}$ is an undetermined normalisation. The functional form of the
anomaly follows on grounds of dimensionality, covariance and parity.
 Likewise, the structure of the covariant trace anomaly is written as ,
\begin{equation}
\langle {T^{\mu}}_{\mu}\rangle = N_{t}R \label{3A2}  
\end{equation}
with $N_{t}$ being the normalisation.
In the null coordinates (\ref{4.2.6}, \ref{4.2.7}) and (\ref{4.2.8}) for $\nu = v$,
 the left hand side of (\ref{3A1}) becomes
\begin{eqnarray}
\nabla_{\mu}\langle {T^{\mu}}_{v}\rangle &=& \nabla_{u}\langle {T^{u}}_{v}\rangle + 
\nabla_{v}\langle {T^{v}}_{v}\rangle\nonumber\\
&& = \nabla_{u}(g^{uv}\langle T_{vv} \rangle)+ \nabla_{v}(g^{uv}\langle T_{uv} \rangle)
 = \nabla_{v}(g^{uv}\langle T_{uv}\rangle) \label{3A3} 
\end{eqnarray}  
where we have used  the fact that for a chiral theory 
$\langle T_{vv}\rangle =0$ (see equation \ref{4.2.17}). 
Also, in  null coordinates, we have, 
\begin{equation}
\langle T_{uv}\rangle = \frac{1}{2}(g_{uv}\langle {T^{v}}_{v}\rangle + g_{uv}\langle {T^{u}}_{u}\rangle) = \frac{g_{uv}}{2}\langle {T^{\mu}}_{\mu}\rangle = \frac{f}{4}\langle {T^{\mu}}_{\mu}\rangle~. \label{3A4}
\end{equation}
By using (\ref{3A2}), (\ref{3A4}) and (\ref{3A3}) we obtain,
\begin{equation}
\nabla_{\mu}\langle {T^{\mu}}_{v}\rangle = \frac{N_{t}}{2}\nabla_{v}R~. \label{3A5}
\end{equation}
where we used $g^{uv}= \frac{2}{f}$ (\ref{4.2.9}).\\

 The right hand side of (\ref{3A1}) for $\nu=v$, with the use of 
the chirality constraint $\langle J_{v}\rangle = 0$ (\ref{4.2.14}), yields
\begin{equation}
\langle J_{\mu}\rangle {F^{\mu}}_{v} + N_{a}\bar \epsilon_{v\mu}\nabla^{\mu}R = 
N_{a}\nabla_{v}R ~.\label{3A6}
\end{equation}
Hence, by equating (\ref{3A5}) and  (\ref{3A6}) we find a relationship
between $N_{a}$ and $N_{t}$
\begin{equation}
N_{a} = \frac{N_{t}}{2}\label{3A7} 
\end{equation}
which is compatible with (\ref{4.1.10}) and (\ref{4.2.5a}) with $N_{a} = \frac{N_{t}}{2} = \frac{1}{96\pi}$. It is clear that chirality enforces both
the conformal and diffeomorphism anomalies. The trivial (anomaly free)
case $N_{a} = N_{t} =0$ is ruled out because, using general arguments
based on the unidirectional property of chirality, it is possible
to prove the existence of the diffeomorphism anomaly in
$1+1$ dimensions \cite{fullinggravano}.

\end{subappendices}

\chapter{\label{chap:conclusions} Conclusions}
The motivation of this thesis was to study certain field theory aspects of
cosmology and black holes. We now summarize the results obtained in last four chapters
 and briefly comment on future prospects.

In the second chapter we studied a generalized Chaplygin gas (GCG) model
 containing a parameter $\alpha$, which is a strong contender for explaining the accelerated expansion of the 
 Universe. In particular, we gave an action formulation
 of GCG model, both in nonrelativistic as well as relativistic regimes.
 In the nonrelativistic case, we constructed a general form
 of the Lagrangian for GCG. This Lagrangian contained both the density and
 velocity fields. By using Bernoulli's equation we expressed this 
 master Lagrangian into a nonrelativistic Born-Infeld form. Further, $\alpha=1$
 limit of our model was shown to be consistent with the corresponding 
 normal Chaplygin gas model. In the relativistic domain, we proposed
 a Born-Infeld like Lagrangian for GCG. This model was manifestly Poincare
 invariant and in the nonrelativistic limit, reduced to the conventional
 GCG. We also suggested a Lagrangian for GCG, which included both 
 density and velocity fields. In order to check its Poincare invariance,
 we computed the algebra among the generators of Poincare group. 
 We observed that the Poincare algebra closed only in the large density
 limit. 

  The relativistic Lagrangian formulation for GCG, initiated here, opens up
 a host of avenues for future study. One possibility is to extend the analysis 
 \cite{bg} of the Chaplygin matter in FRW spacetime, to the case of a
 GCG, to observe its cosmological implications. Also, study of  symmetry properties of GCG (first elucidated in \cite{jac2} for usual Chaplygin gas), as well as its connection to $d$-branes, offer further scope.       
 
 In the third chapter, we provided a new mechanism, based on 
covariant gauge and gravitational anomalies, to compute the fluxes of 
 Hawking radiation. In contrast to the earlier approaches \cite{robwilczek,isowilczek},
 where the expressions for gauge/gravitational anomalies were taken to be consistent whereas 
 the boundary conditions were covariant, the analysis presented here used only covariant 
 expressions. The point was that since the covariant boundary condition was mandatory in deriving
 the Hawking flux, it was conceptually clean to discuss everything from the covariant point of view.
 There are two important reasons in favor of the covariant anomaly approach that was adopted here:
 \begin{itemize}
 \item No counterterms connecting the consistent and covariant expressions for 
 currents and energy-momentum tensors were required. 
\item Manifest covariance was preserved at all stages of the computations.
 \end{itemize}

 We also discussed applications of our covariant anomaly technique
 to the case of stringy black holes. In particular, we computed the Hawking energy-momentum
 flux from Garfinkle-Horowitz-Strominger (GHS) and D1-D5 nonextremal black holes.

 In the fourth chapter we discussed yet another way to derive the fluxes of Hawking radiation.
 This approach used only the structure of chiral effective action, which was defined in the 
 vicinity of event horizon. The current and energy-momentum tensors derived from the 
 chiral effective action, suitably modified by local counterterms, yielded the 
 covariant gauge and gravitational anomalies, respectively. The arbitrary constants
 appearing in the expressions for the chiral covariant current and energy-momentum tensor were
 then fixed by imposing the covariant boundary condition at event horizon. Once we knew the 
 forms for current and energy-momentum tensor, in the region near the event horizon, 
 the Hawking fluxes were easily obtained by taking the asymptotic infinity limit of
 the current/energy-momentum tensor. Novelty of this chiral effective action approach was that,
 unlike the previous approaches based on the consistent \cite{robwilczek,isowilczek} or
 covariant (see chapter-\ref{chap:coanomaly}) anomaly cancellation method, it used only
 the  properties of the theory near the event horizon. The structure of the chiral effective action and the 
 covariant boundary condition were the only necessary inputs and we showed that
 they were sufficient to determine the Hawking fluxes.
 Also, our method did not require the introduction of any discontinuous step functions. 
 This was consistent with the universality of Hawking effect. Next, we used this
  approach to obtain the expressions for charge and energy-flux
 from the Reissner-Nordstrom black hole in the presence of gravitational back reaction. 

  The last chapter was devoted to discussion of the covariant boundary condition used in the analysis
 of chapters-\ref{chap:coanomaly},\ref{chap:chiralaction} and also in \cite{isowilczek}. We used
 the structures of covariant current/energy-momentum tensor, derived from the 
 chiral effective action, suitably modified by a local counterterm. The arbitrary
 constants appearing in the expressions for current and energy-momentum tensor
 were fixed by imposing the regularity condition on the outgoing modes. 
 Because our theory was chiral, in the vicinity of horizon, no
 further condition on the ingoing modes was necessary.  The regularity 
 condition states that; a freely falling observer must see a finite amount
 of flux at the future horizon. This condition was sufficient to determine
 completely the forms for current and energy-momentum tensor. The expressions
 for the universal components of current and energy-momentum tensor 
 were in exact agreement with the corresponding ones obtained by solving the
 covariant gauge/gravitational anomaly and imposing the covariant boundary
 condition (see chapter-\ref{chap:coanomaly}).
 This provided a clear physical interpretation for the covariant boundary condition.

 Next, we compared our results with the standard implementation of the 
 various vacua \cite{fullingvacuum} on nonchiral expressions. In the conventional 
 analysis, the expressions for the expectation values of current and 
 energy-momentum tensor were derived by solving simultaneously
  the conservation law  and the trace anomaly \cite{fulling}.
 Similar results can also be derived by using the structure of
 Polyakov type effective action (see \cite{isowilczekPRD,fabbri1} for details).
 The arbitrary constants were fixed by imposing the conditions appropriate
 for Unruh, Hartle-Hawking and Boulware vacuum. 

 The Unruh vacuum, by construction, is appropriate for discussing the Hawking flux \cite{unruh}.
  Unruh vacuum is characterized by two properties : 
\begin{enumerate}
 \item A finite amount of flux at the future horizon. 
 This implies that the outgoing component of the current and energy-momentum tensor
 in Kruskal coordinates (i.e $\langle J_{U} \rangle$ and $\langle T_{UU}\rangle$)
 must be regular at future event horizon. However, since the outgoing components of
 the currents/energy-momentum tensors in the Kruskal coordinates $(U,V)$ are related
 to the null coordinates $(u,v)$, as $\langle J_{U}\rangle = -\frac{\langle J_{u} \rangle}{\kappa U}$ and
  $\langle T_{UU}\rangle = \left(\frac{1}{\kappa U}\right)^{2}\langle T_{uu}\rangle$, the 
 regularity condition stated above translates into vanishing of current $\langle J_{u} \rangle$
 and energy-momentum tensor $\langle T_{uu}\rangle$ in the null coordinates at future horizon.
 \item No ingoing flux at past null infinity, i.e $\langle T_{VV}(r\rightarrow \infty)\rangle =0$.
\end{enumerate} 
 In the conventional analysis, based on the trace anomaly \cite{fulling},
 both the above conditions were essential to fix the structures of
 current and energy-momentum tensor. We showed that, in the asymptotic infinity limit
 our results were compatible with the choice of Unruh vacuum for conventional (nonchiral) theory.
 For the Hartle-Hawking state no such equivalence, between the chiral and nonchiral expressions,
 was possible. Thus, we conclude that: the imposition of covariant boundary condition on the
  chiral expressions is equivalent to implementing the conditions for the Unruh vacuum on
 the nonchiral expressions.  
   
  There are certain issues which are worthwhile for future study. For example, the inclusion
 of grey body effect within the anomaly approach would be an interesting excersise. 
 The approaches given in chapter-\ref{chap:coanomaly},\ref{chap:chiralaction} and
 \cite{robwilczek,isowilczek} did not include the grey body effect. Consequently,
 the flux obtained from these approaches were compared with the fluxes associated
 with the perfect black body.  Another important issue is the computation of black hole entropy by using the
 anomaly approach. 
  There are strong reasons to believe that the black hole entropy, like  Hawking flux
  can be related to the diffeomorphism anomaly. For example, in the analysis of \cite{carlip2,carlip3}
 the counting of microstates was done by imposing the ``horizon constraints''. The algebra
 among these ``horizon constraints'' commutes only after modifying the generators for 
 diffeomorphism symmetry. This modification in the generators give rise to desired central
 charge, which ultimately leads to Bekenstein-Hawking entropy. This is roughly similar
 to the diffeomorphism anomaly mechanism, illustrated in this thesis.
 Thus, it is clear that  the covariant anomaly mechanism and the effective action approach,
  provided in this thesis, could illuminate the subject of black hole entropy.

\backmatter




\end{document}